\pdfoutput=1

\documentclass[11pt]{article}

 \usepackage[final]{acl2023}

\usepackage{times}
\usepackage{latexsym}
\usepackage{booktabs}       %
\usepackage{siunitx}        %
\usepackage{multirow}       %
\usepackage{amsmath}
\usepackage{booktabs}
\usepackage{multirow}
\usepackage{tcolorbox} 
\usepackage{colortbl}
\usepackage{tabularx}
\usepackage{colortbl}
\usepackage{amssymb}
\usepackage{cleveref}
\crefname{appendix}{Appendix}{Appendices}
\definecolor{ResearchColor}{HTML}{D0E1F9}   %
\definecolor{MinecraftColor}{HTML}{D5F5E3}  %
\definecolor{DatabaseColor}{HTML}{FFF9C4}   %
\definecolor{CodingColor}{HTML}{FFE0B2}     %
\definecolor{BargainingColor}{HTML}{E1BEE7} %
\definecolor{WereWolfColor}{HTML}{F8BBD0}   %
\usepackage{graphicx}
\usepackage[T1]{fontenc}

\usepackage[utf8]{inputenc}
\newcommand\icon{\raisebox{-3.7pt}{\includegraphics[width=1.5em]{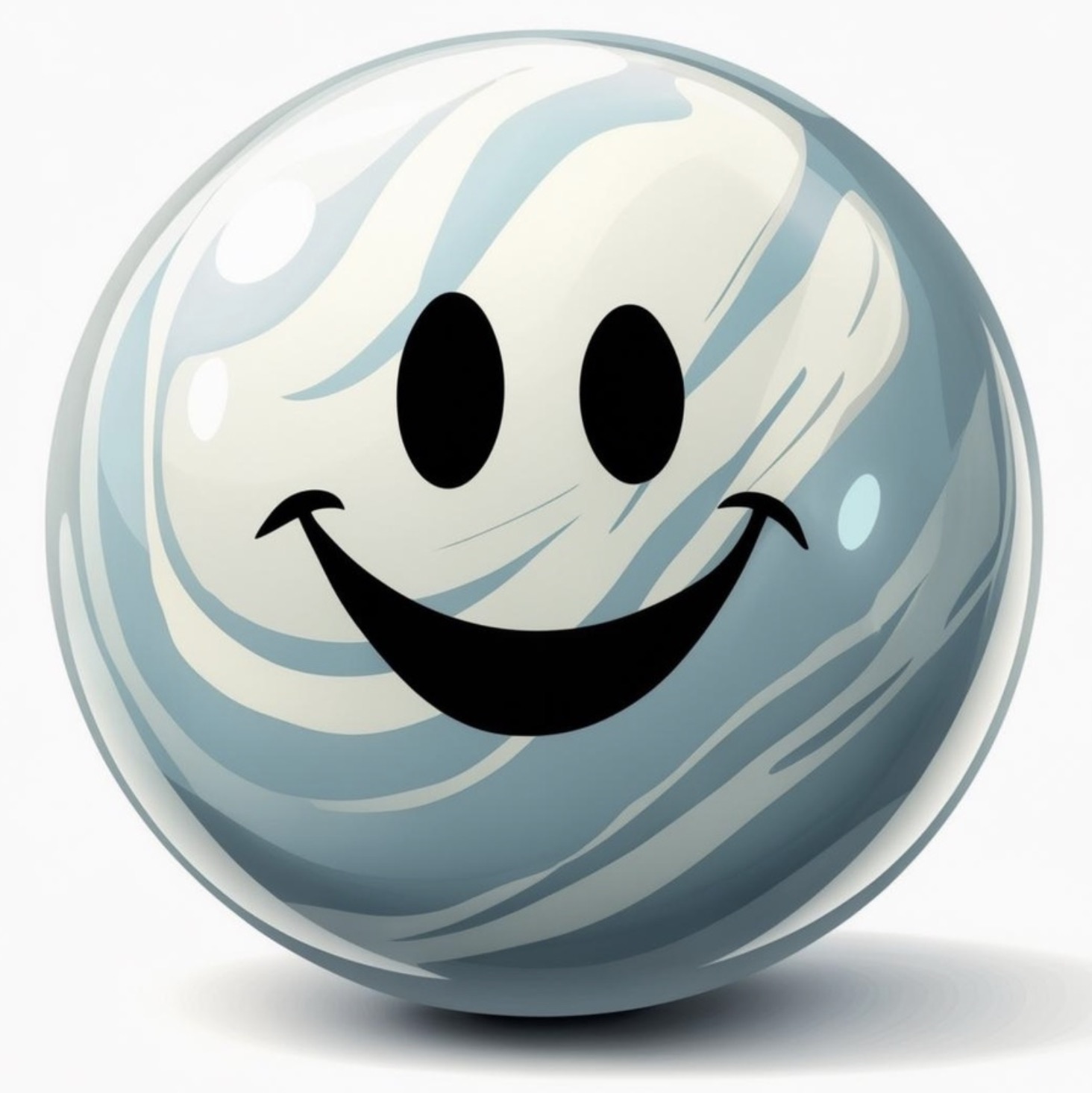}}}
\usepackage{microtype}

\usepackage{inconsolata}

\title{MultiAgentBench\icon: Evaluating the Collaboration and Competition of LLM agents}

\author{
 Kunlun~Zhu$^{1\dag}$ \thanks{\ \ Team Leader.}, Hongyi Du$^{1\dag}$, Zhaochen Hong$^{1\dag}$,  Xiaocheng Yang$^{1\dag}$ \\ \textbf{Shuyi Guo}$^{1\dag}$, \textbf{Zhe Wang}$^{1}$ \thanks{\ \  Core Contributors. Contributions are listed in the appendix~\ref{app_contrib}.} \\
 \textbf{Zhenhailong Wang}$^{1}$, \textbf{Cheng Qian}$^{1}$, \textbf{Xiangru Tang}, \textbf{Heng Ji}$^{1}$, \textbf{Jiaxuan You}$^{1}$ \\
 $^1$University of Illinois Urbana-Champaign\\
 kunlunz2@illinois.edu
}

\begin{document}
\maketitle

\begin{abstract}

Large Language Models (LLMs) have shown remarkable capabilities as autonomous agents; yet existing benchmarks either focus on single-agent tasks or are confined to narrow domains, failing to capture the dynamics of multi-agent coordination and competition. In this paper, we introduce \textbf{MultiAgentBench}, a comprehensive benchmark designed to evaluate LLM-based multi-agent systems across diverse, interactive scenarios. Our framework measures not only task completion but also the quality of collaboration and competition using novel, milestone-based key performance indicators. Moreover, we evaluate various coordination protocols (including star, chain, tree, and graph topologies) and innovative strategies such as group discussion and cognitive planning. Notably, gpt-4o-mini reaches the average highest task score, graph structure performs the best among coordination protocols in the research scenario, and cognitive planning improves milestone achievement rates by 3\%. Code and datasets are publicly available at \url{https://github.com/ulab-uiuc/MARBLE}.

\end{abstract}

\section{Introduction}

Large Language Models (LLMs) such as GPT-3 \citep{Brown2020GPT3}, GPT-4 \citep{OpenAI2023GPT4}, Gemini~\citep{team2023gemini} and DeepSeek-R1~\citep{guo2025deepseek}, now exhibit human-like language understanding and generation, enabling their use as autonomous agents interacting with environments, tools, and other agents \citep{qian2024chatdev, Park2023GenerativeAgents, OpenAI2023FunctionCalling}. While single-agent systems have demonstrated impressive capabilities, their isolated operation can limit task execution efficiency and falls short when simulating complex, socially dynamic scenarios. In contrast, multi-agent setups \citep{qian2024chatdev, li2023camelcommunicativeagentsmind, unleashing2024} leverage multiple LLM-based agents that collaborate, coordinate, and jointly plan to address these challenges. This paradigm not only increases the efficiency of task execution but also enables the realistic simulation of social interactions and dynamics, thereby enhancing performance in domains such as strategic decision-making, gaming \citep{silver2017mastering}, and software development \citep{qian2024chatdev}.

\begin{figure}[htbp]
    \centering
    \includegraphics[width=\linewidth]{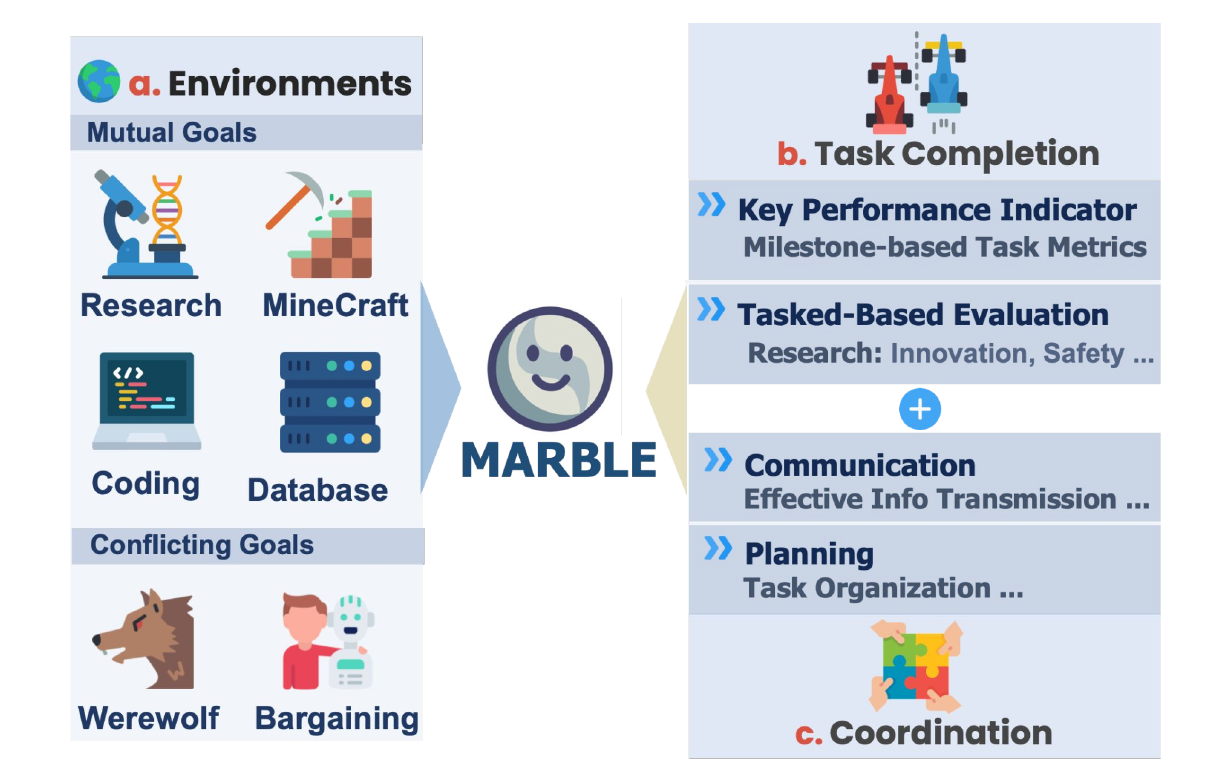}
    \caption{\textbf{Overview of MultiAgentBench evaluation process}: Multi-Agent System Coordination in various interactive environments, with a focus on task performance, and coordination.}
    \label{fig:overview}
\end{figure}

Despite significant advances in LLM capabilities, current evaluation paradigms remain insufficient for multi-agent scenarios. Traditional single-agent benchmarks—such as AgentBench~\citep{liu2023agentbench}, VisualAgentBench~\citep{liu2024visualagentbenchlargemultimodalmodels}  GAIA~\citep{mialon2023gaia}, ToolBench~\citep{qin2024toolllm} and HumanEval \citep{Chen2021Evaluating}—primarily focus on isolated reasoning and generation, overlooking the dynamics intrinsic to multi-agent interactions.

To address this gap, we introduce \textbf{MultiAgentBench}, a comprehensive benchmark designed to evaluate LLM-based multi-agent systems across a wide range of task-solving and simulation scenarios. MultiAgentBench offers several key advantages: (1) \textit{Multi-Domain Evaluation:} The benchmark covers diverse domains—from collaborative coding to gaming—ensuring broad real-world applicability. (2) \textit{Capturing Coordination and Competition:} Unlike traditional single-agent benchmarks, MultiAgentBench explicitly measures both coordination dynamics and competitive interactions, highlighting the unique challenges of multi-agent environments. (3) \textit{Tailored Metrics and Flexible Protocols:} We propose novel metrics, including a \textit{Key Performance Indicator (KPI)} that tracks milestone progress and individual contributions, to systematically assess planning quality and communication effectiveness. Moreover, our framework, \textbf{MARBLE} (\textbf{M}ulti-agent coo\textbf{R}dination \textbf{B}ackbone with \textbf{L}LM \textbf{E}ngine), supports various communication topologies—such as star, chain, tree, and an innovative graph-based approach—and accommodates multiple reasoning strategies.

Our contributions can be summarized as follows:
(1) We introduce \textbf{MultiAgentBench} along with the \textbf{MARBLE} framework, a comprehensive benchmark that rigorously evaluates LLM-based multi-agent systems in six diverse interactive scenarios, capturing both collaborative and competitive dynamics. Notably, the cognitive planning planning feature improves milestone achievement rates by 3\%.
(2) We propose innovative evaluation metrics that assess not only task success but also coordination quality. Our metrics include milestone-based KPIs, structured planning and communication scores, and a dedicated competition score that captures conflicting-goal tasks, internal performance metrics, and competitive aspects in planning and communication.
(3) Our experiments reveal some "aha-moments" in multi-agent coordination—agents begin to exhibit emergent social behaviors, providing promising insights toward AGI-level collaboration~\citep{feng2024how}.

\begin{figure*}[htbp]
    \centering
    \includegraphics[width=\linewidth]{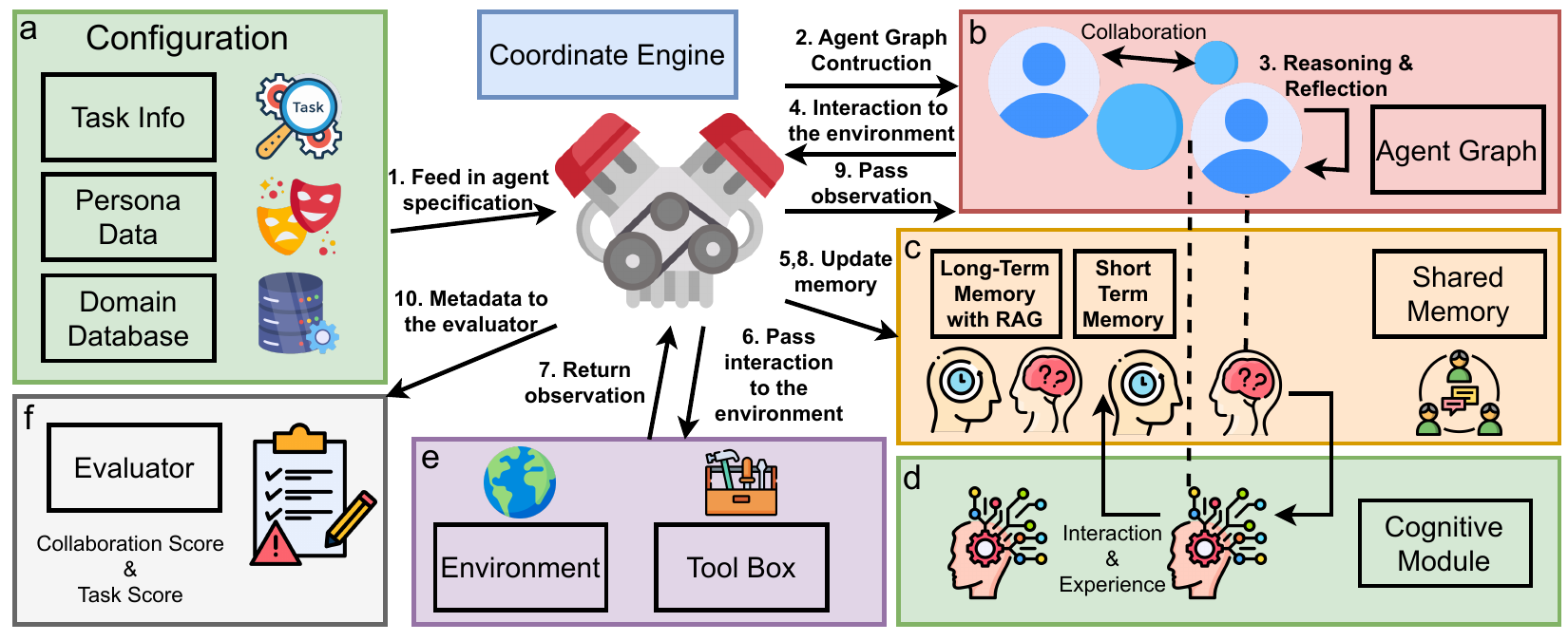}
    \caption{MARBLE \icon: showcasing interactions between task information, persona data, domain databases, memory modules, and the environment through the coordinate engine and cognitive module.}
    \label{fig:engine}
\end{figure*}

\section{Related Work}

\subsection{LLM-Based Multi-Agent Systems}

LLM-based multi-agent systems have enabled collaborative problem-solving across domains~\citep{Park2023GenerativeAgents, li2023camelcommunicativeagentsmind, chen2024agentverse}. These systems support scientific research through literature review and experimental design \citep{zhou2024hypothesis, agarwal2024litllmtoolkitscientificliterature}, software engineering tasks \citep{huang2023agentcoder, wu2023autogen, zhou2023agents, hong2024metagpt, ishibashi2024self, islam2024mapcoder, openhands, zhugegptswarm} including code generation and maintenance \citep{bouzenia2024repairagent}, and gaming applications \citep{chen2023gamegpt}. In Minecraft, agents perform complex tasks from construction to navigation \citep{wang2023voyageropenendedembodiedagent, chen2024agentverse, yu2024minelandsimulatinglargescalemultiagent, dong2024villageragentgraphbasedmultiagentframework}.

GameNGen enables real-time interaction in DOOM \citep{valeveski2024diffusion}, while CUISINEWORLD benchmarks multi-agent collaboration \citep{gong2023mindagent}. Applications extend to social deduction games, game theory \citep{xu2023magic}, healthcare \citep{ke2024enhancing, kim2024adaptive}, business \citep{chen2024agentverse}, education \citep{gosling2024multi}, and urban planning \citep{zhou2024largelanguagemodelparticipatory}. Despite progress, challenges persist in communication, emergent behavior, and scalability \citep{agashe2024llmcoordinationevaluatinganalyzingmultiagent}, motivating the need for robust evaluation frameworks.

\subsection{Multi-Agent Collaboration}

Recent advances in multi-agent systems highlight two complementary scaling paradigms: \textit{cognitive scaling}, which enhances agent reasoning and adaptability, and \textit{population scaling}, which leverages large agent collectives for emergent behaviors \citep{zhugegptswarm, qian2025scaling}. 

Cognitive scaling explores mechanisms such as dynamic architecture adaptation and self-organizing coordination strategies to find the most effective pattern of agent communication \citep{zhugegptswarm}. Meanwhile, population-based scaling exhibits nonlinear performance gains as an increasing number of agents collectively interact through diverse collaboration patterns, including hierarchical delegation and decentralized consensus \citep{qian2025scaling}. 
These approaches enable complex applications ranging from geopolitical conflict simulation \citep{hua2024warpeacewaragentlarge} to scientific discovery workflows \citep{zhou2024hypothesis, zhang2025crewfacilitatinghumanaiteaming}.

\section{Methodology}
\subsection{Framework Design}

Our proposed evaluation framework MARBLE~(see Figure~\ref{fig:engine}) establishes a robust multi-agent coordination system by leveraging interconnected modules that enable adaptive collaboration, efficient communication, and strategic task execution. At its core lies the \textit{Coordination Engine}, responsible for initializing and synchronizing key modules—including the \textit{Agent Graph}, \textit{Cognitive Module}, and \textit{Coordinate Engine}—to ensure seamless interaction across the system. Detailed descriptions of additional modules can be found in Appendix~\ref{framework}.

\paragraph{Agent Graph Module}
This module converts configuration data into a structured graph \(G = (\mathcal{A}, E)\), where \(\mathcal{A} = \{a_1, a_2, \dots, a_n\}\) denotes the set of agents, and each edge in \(E\) is defined as a triple \((a_i, r, a_j)\) with \(r \in \mathcal{R}\) representing the relationship between agents \(a_i\) and \(a_j\). For example, a collaboration relationship is denoted as \((a_i, \texttt{collaborates}, a_j)\), supervision as \((a_i, \texttt{supervises}, a_j)\), and negotiation as \((a_i, \texttt{negotiates}, a_j)\). By constructing the graph based on these triple relations, we ensure that subsequent communication and coordination occur exclusively between agents with explicitly defined relationships, mirroring real-world interaction patterns.

\paragraph{Cognitive Module} 
The Cognitive Module is central to responsible agent evolution and social intelligence in multi-agent coordination. It maintains and updates a comprehensive internal state that includes each agent’s persona, inter-agent relationships, and reasoning strategies (e.g., Chain-of-Thought~\citep{wei2022chain}, ReACT~\citep{yao2023reactsynergizingreasoningacting}). Crucially, this approach mirrors human cognitive processes by fusing elements of theory-of-mind and social intelligence \citep[e.g.,][]{Premack1978}—similar to how humans continuously update their mental models based on social cues, prior experiences, and contextual information. The fusion of cognitive, social, and adaptive mechanisms forms the backbone of our system, empowering agents to dynamically adjust their strategies and collaboratively evolve in complex environments.

\begin{figure*}[ht]
    \centering
    \includegraphics[width=\textwidth]{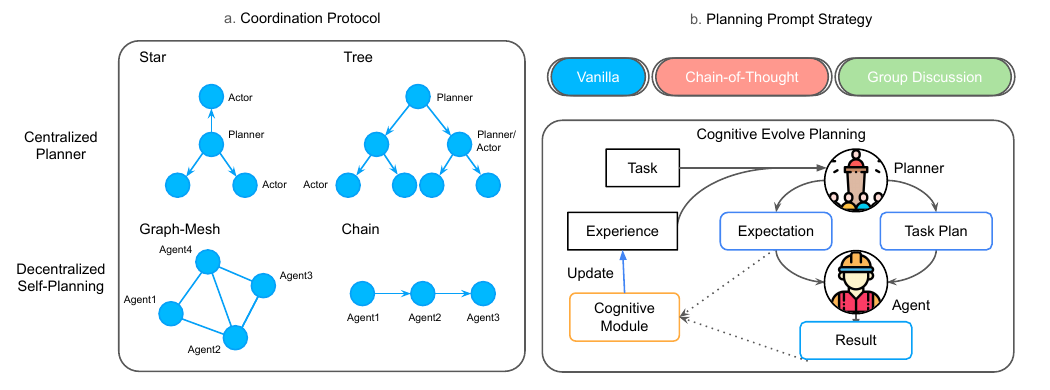}
    \caption{\textbf{Illustration of coordination protocols and planning prompt strategies}. 
    \textbf{(a)} shows centralized and decentralized planning structures (e.g., star, tree, graph, and chain). 
    \textbf{(b)} describes strategies like group discussions and cognitive prompts, incorporating iterative feedback and task updates for effective planning.}
    \label{fig:coordination}
\end{figure*}

\subsubsection{Coordination Engine} 
The Coordination Engine orchestrates the overall execution flow of the system. It initializes agents, tasks, and inter-agent relationships via a dedicated Configuration Module and constructs the Agent Graph to represent these dynamics. In our framework, we distinguish between two key roles: \emph{planners} and \emph{actors}. Planners are responsible for devising task inputs, strategizing, and managing overall task allocation, while actors—represented within the Agent Graph—execute tasks by interacting with the environment and other agents through available tools.

Our approach supports four distinct coordination protocols similar to work from~\citet{qian2025scaling}: \textbf{star}, \textbf{tree}, \textbf{graph}, and \textbf{chain}.

\paragraph{Centralized Coordination: Star \& Tree.}
In the \textbf{star} configuration, a single central planner assigns tasks to all actors and consolidates their feedback, offering strong oversight though potentially limiting scalability. The \textbf{tree} structure extends this by organizing agents hierarchically: a top-level planner delegates tasks to subordinate planners, which in turn coordinate with actors. This hierarchical approach balances centralized control with improved scalability for handling more complex tasks.

\paragraph{Decentralized Coordination: Graph-Mesh \& Chain.}
The \textbf{graph-mesh} configuration employs a network of interconnected actors that communicate directly, enabling concurrent planning and distributed decision-making. Conversely, the \textbf{chain} configuration arranges actors sequentially, where each agent passes its decision to the next. This sequential handoff is well-suited for tasks with inherent dependencies, though it may limit parallel processing capabilities.

\paragraph{Planner Design and Enhancements.}
In our centralized coordination protocol, the planner supports four distinct planning approaches that reflect human decision-making processes: vanilla prompting, chain-of-thought (CoT)~\citep{wei2022chain}, group discussion, and cognitive self-evolving planning. The \textbf{vanilla prompt} employs a straightforward, zero-shot instruction to generate task plans directly. The \textbf{CoT approach} enriches this process by facilitating step-by-step reasoning through detailed inputs—such as the target task, agent profiles (including roles, expertise, and historical performance), and summaries of previous subtasks—to guide logical progression. The \textbf{group discussion}~\citep{chen2024agentverse} method enables multiple agents to share insights and constraints, fostering a collaborative deliberation that refines the overall plan. Lastly, similar to the Reflexion~\citep{shinn2023reflexion} method, our \textbf{cognitive self-evolving planning} method mirrors human learning by generating expected outcomes and progress for each task, storing these in memory, and then comparing actual performance against these expectations in subsequent iterations. This comparison produces evolving experiences that continuously inform and adjust future planning (See Appendix~\ref{sec:important-prompts} for detailed prompting). Together, these methods leverage both individual reasoning and collaborative optimization, enhancing coordination efficiency as validated through ablation studies on the star coordination style.

\subsection{Benchmark Design}

\begin{figure*}[ht]
    \centering
    \includegraphics[width=\textwidth]{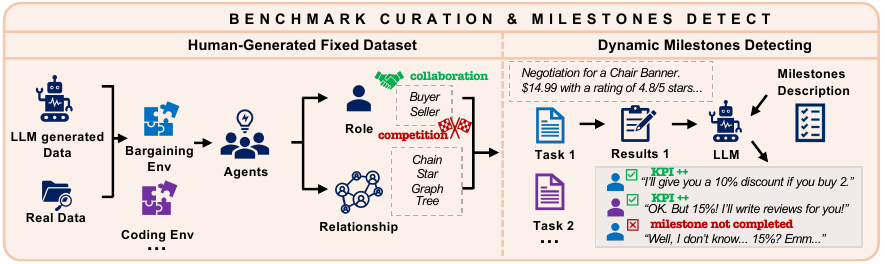}
    \caption{Illustration of our benchmark curation and the dynamic milestones detecting for KPI metric.}
    \label{fig:method}
\end{figure*}

To systematically evaluate our multi-agent framework, we curate a benchmark of diverse scenarios spanning \textit{task-oriented} and \textit{social-simulation-based} environments (Figure~\ref{fig:overview}). These scenarios are constructed through a combination of: (1) \textit{Existing multi-agent tasks} adapted from prior work or datasets (e.g., database error analysis, research collaboration). (2) \textit{LLM-generated tasks} with human verification and refinement (e.g., Werewolf and Bargaining).
This dual approach ensures both realism (by leveraging established tasks) and novelty (through generative expansion), while human validation guarantees that each scenario remains coherent and feasible.

\paragraph{Agents with Mutual Goal.} 
In the task-oriented scenarios, the agents share with the mutual goal to finish one specific task. We focus on four representative tasks: (1) \textbf{research tasks} follow the setup of \textit{ResearchTown}~\cite{yu2024researchtownsimulatorhumanresearch}, where agents with complementary research profiles co-author a new proposal on a chosen topic; (2) \textit{Minecraft-based building tasks} require agents to collaboratively construct structures in a shared environment; (3) \textbf{database error analysis} involves exactly five agents, each specializing in diagnosing a distinct root cause of system inconsistencies; (4) \textbf{coding challenges} demand collective problem-solving and software module development. Across these tasks, agents must coordinate, divide labor, and synthesize outputs efficiently. We scale scenario diversity by creating 100 test cases per task, with variations in research topics, Minecraft creation, database errors, or coding objectives.

\paragraph{Agents with Conflicting Goals.}
In social-simulation based scenarios. We strengthen competitive elements by introducing \textbf{Werewolf} and \textbf{Bargaining} scenarios. In Werewolf, two groups of agents face off in an adversarial setting, employing deceptive strategies within a predefined narrative. The Bargaining environment simulates negotiations over shared resources, where agents strive to maximize individual gains through strategic concessions or alliances. Both settings evaluate adaptability, conflict resolution, and negotiation skills under uncertainty.

\paragraph{Role Assignments and Graph Structures.}
To emphasize multi-agent collaboration, each scenario enforces distinct agent roles (e.g., project manager, domain expert, technical specialist) and defines specific graph relationships (star, tree, chain, or mesh). These structures reflect realistic team dynamics or competition, guiding how agents share information, make decisions, and coordinate actions.

\paragraph{Milestones Generation for Scenarios} 
\label{para:milestones}
To facilitate the evaluation of MARBLE iterations, each task is segmented into a series of flexible milestones. Unlike rigid checkpoints, these milestones are broadly defined. For instance, in a research task, a milestone may be reached by completing five key queries (5q) for research proposal (more details see Appendix~\ref{app:research-scenario}) or by enhancing a previous set of 5q. Throughout MARBLE's iterative process, a language model continuously monitors whether milestones \(m_1, m_2, \dots\) have been achieved and logs the outcomes. This method integrates human- or LLM-generated outlines with dynamic, execution-based assessments, ensuring that both intermediate progress and team coordination are effectively measured.

More detailed environment setups, interaction tools, and additional examples for different scenarios appear in Appendix~\ref{app:research-scenario}, ~\ref{app:werewolf_env}, ~\ref{app:database-scenario}, ~\ref{app:coding-scenario}, ~\ref{app:bargaining-scenario}, and ~\ref{app: minecraft-scenario}.

\subsection{Evaluation Metrics}
\label{sec:eval_metrics}
As illustrated in Figure~\ref{fig:overview}(b)(c) , our evaluation considers two primary dimensions: \textbf{Task Completion Performance} and \textbf{Coordination}.

\paragraph{Task Completion Metrics.}  

As described in Section~\ref{para:milestones}, each task is segmented into a series of flexible milestones. An LLM-based detector continuously monitors the iterative process to identify which milestones have been achieved and records the corresponding contributing agents. For each agent, the number of milestones they contribute to is noted as \(n_j\), and their individual KPI is calculated as the ratio of \(n_j\) to the total number of milestones \(M\). The overall KPI is defined as the average of these individual KPIs across all \(N\) agents, which is computed as follows:
\[
\text{KPI}_{\text{overall}} = \frac{1}{N}\sum_{j=1}^{N}\text{KPI}_j = \frac{1}{NM}\sum_{j=1}^{N} n_j.
\]

In addition to the KPI derived from milestone detection, a separate \textbf{task-based score} is computed to evaluate the final output quality. For tasks such as research or bargaining, an LLM-defined scoring rubric is applied to generate the score, whereas tasks like Minecraft, Werewolf, database error fixes, or coding are evaluated using rule-based metrics (e.g., accuracy). Detailed scoring criteria and evaluation prompts for these task-based assessments are respectively provided in the Appendix~\ref{app: minecraft-scenario}, ~\ref{app:werewolf_env}, ~\ref{app:database-scenario}, and ~\ref{app:coding-scenario}, which demonstrate the effectiveness of the metrics while evaluating the coordination abilities.

\paragraph{Coordination Metrics.}  
Coordination is evaluated by quantifying the agents’ communication and planning capabilities. The \textbf{Communication Score} (\(C_{\text{score}}\)) is derived from an LLM-based evaluation that considers inputs such as the task description, agent profiles, and aggregated communication data, resulting in a score on a five-point scale (with \(C_{\text{score}} = 0\) if no communication occurs). Similarly, the \textbf{Planning Score} (\(P_{\text{score}}\)) is determined by assessing the agents’ abilities to organize tasks, maintain roles, and adapt strategies based on their profiles and aggregated planning data, also on a five-point scale. The overall \textbf{Coordination Score} (\textbf{CS}) is then computed by averaging these two sub-scores. More details regarding the evaluation process and the output format are provided in the Appendix~\ref{sec:important-prompts}. We also did a human evaluation comparing human alignment with those metrics, results are in Appendix~\ref{app:human-evaluation}.

\begin{table*}[h!]
\centering
\renewcommand{\arraystretch}{1.2}  %
\setlength{\tabcolsep}{8pt}  %
\resizebox{\linewidth}{!}{
\begin{tabular}{
    l
    |>{\columncolor{ResearchColor}}c>{\columncolor{ResearchColor}}c
    |>{\columncolor{MinecraftColor}}c>{\columncolor{MinecraftColor}}c
    |>{\columncolor{DatabaseColor}}c>{\columncolor{DatabaseColor}}c
    |>{\columncolor{CodingColor}}c>{\columncolor{CodingColor}}c
    |>{\columncolor{BargainingColor}}c>{\columncolor{BargainingColor}}c
    |>{\columncolor{WereWolfColor}}c>{\columncolor{WereWolfColor}}c
}
\toprule
\multirow{2}{*}{\textbf{Model}} 
& \multicolumn{2}{c}{\textbf{Research}} 
& \multicolumn{2}{c}{\textbf{Minecraft}}
& \multicolumn{2}{c}{\textbf{Database}}
& \multicolumn{2}{c|}{\textbf{Coding}}
& \multicolumn{2}{c}{\textbf{Bargaining}}
& \multicolumn{2}{c}{\textbf{WereWolf}} \\
\cmidrule(lr){2-3}\cmidrule(lr){4-5}\cmidrule(lr){6-7}\cmidrule(lr){8-9}\cmidrule(lr){10-11}\cmidrule(lr){12-13}
& \textbf{TS} & \textbf{CS}
& \textbf{TS} & \textbf{CS}
& \textbf{TS} & \textbf{CS}
& \textbf{TS} & \textbf{CS}
& \textbf{TS} & \textbf{CS}
& \textbf{TS} & \textbf{CS} \\
\midrule
\textbf{Meta-Llama-3.1-8B}
 & 80.87 & 52.40
 & 6.12 & 54.40
 & 34.00 & 40.00
 & 59.90 & 67.24
 & 72.81 & 73.36
 & 12.64 & 60.00 \\

\textbf{Meta-Llama-3.1-70B}
 & 80.80 & 49.50
 & 0.21 & \textbf{75.00}
 & \textbf{53.00} & 37.70
 & 62.10 & 67.18
 & 72.13 & 71.46
 & 19.82 & 60.60 \\

\textbf{Meta-Llama-3.3-70B}
 & 80.00 & \textbf{72.00}
 & 9.15 & 69.00
 & 28.50 & 40.00
 & 56.60 & 74.40
 & 73.15 & 69.56
 & \textbf{36.33}  & \textbf{76.30} \\
\midrule
\textbf{gpt3.5-turbo}
 & 70.20 & 55.90
 & 5.05 & 63.60
 & 45.00 & \textbf{60.89}
 & 55.50 & \textbf{76.20}
 & 71.67 & 72.00
 & 15.69 & 75.90 \\

\textbf{gpt-4o-mini}
 & \textbf{84.13} & 52.00
 & \textbf{33.60} & 61.50
 & 45.00 & 43.22
 & \textbf{65.10} & 66.30
 & \textbf{74.47} & \textbf{74.20}
 & 14.06 & 60.10 \\
\bottomrule
\end{tabular}
}
\caption{\textbf{Average Task Score} (\textbf{TS}) (\%) and the Coordination Score (\textbf{CS}) for Minecraft, Database, Coding, Bargaining, and WereWolf, scores are multiplied by 20. We can see that model abilities are still the key factor for task completion. CS is a good indicator for TS given three pairs of scenarios having the one model having the highest TS and CS at the same time.
}
\label{tab:avg_task_score}
\end{table*}

\section{Experiment Setup}
\label{sec:experiment-setup}

\subsection{Experiment Settings}
\paragraph{Models.} Since our MARBLE framework required function-calling abilities. Thus, we consider three open-source models: \textit{Meta-Llama-3.3-70B~\citep{dubey2024llama}}, \textit{Meta-Llama-3.1-70B-Instruct-Turbo~\citep{dubey2024llama}}, and \textit{Meta-Llama-3.1-8B-Instruct-Turbo}, as well as two closed-source models: \textit{GPT-3.5-turbo-0125} and \textit{GPT-4o-mini}\footnote{\href{https://www.openai.com}{https://www.openai.com}}.We access the open-souce models are provided by the \textit{togetherai}~\footnote{\href{https://www.together.ai}{https://www.together.ai}} service, utilizing the default parameter settings. 

For the agent actions, we configure the models with a maximum token number (\texttt{max\_token\_num}) of 1024, a temperature of 0.7, and a top\_p of 1.0, in order to balance the creativity and consistency of the agents' responses during interactions. The overall maximum iterations are set to 5 for research and 20 for Minecraft; more details can be found in the Appendix.  In our evaluation, which involves both Task Completion and Simulation scenarios, we assess the models along two primary axes: \emph{Task Score} (\textbf{TS}) and \emph{Coordination Score} (\textbf{CS}), using the same metrics as described in the Metrics section. The maximum communication iteration number is also set to 5. Furthermore, the long-term base memory for each agent is set to be unlimited. Finally, for the main experiment, a graph-mesh coordination protocol is adopted to facilitate interactions.

\subsection{Main Experiment One: Model Performance Across Different Scenarios}

In this experiment, we evaluate the performance of five models across diverse scenarios, with results summarized in Table~\ref{tab:avg_task_score}. Our analysis leads to several key insights:

\paragraph{1. Superior Task Performance of \texttt{gpt-4o-mini}:}  
Across multiple tasks, \texttt{gpt-4o-mini} consistently achieves high Task Scores (TS). For example, in the Research scenario it obtains a TS of \textbf{84.13}\%, outperforming other models such as \texttt{Meta-Llama-3.1-8B} (80.87\%) and \texttt{Meta-Llama-3.1-70B} (80.80\%). Similar trends are observed in the Coding domain, where \texttt{gpt-4o-mini} records a TS of 65.10 compared to lower scores from its competitors. These results indicate that the underlying model capabilities are a decisive factor in achieving superior task performance.

\paragraph{2. The Nuanced Role of Coordination (Collaboration) Score:}  
While the Collaboration Score (CS) is designed to measure coordination ability, our findings suggest that its impact on the overall task performance is complex. For instance, in the Minecraft scenario, \texttt{Meta-Llama-3.1-70B} exhibits a high CS of 75.00 but an extremely low TS of 0.21, a more deep analysis for this can refer to Appendix~\ref{fig:minecraft-function-call}. This discrepancy implies that, although coordination contributes to performance, it does not compensate for inherent deficiencies in task execution capabilities. In contrast, models that balance both high TS and moderate-to-high CS—such as \texttt{gpt-4o-mini}—demonstrate more robust and reliable performance across scenarios.

\paragraph{3. Model-Specific Strengths and Context-Dependent Performance:}  
Our results reveal that different models exhibit varied strengths depending on the task. For example, \texttt{Meta-Llama-3.3-70B} shows a notable CS in the Research (72.00) and WereWolf (76.30) tasks, yet its TS lags behind that of \texttt{gpt-4o-mini} in several scenarios. These variations emphasize that no single metric can fully capture a model’s effectiveness; instead, both task-specific abilities and coordination skills must be considered. Overall, our study underscores that while coordination plays a role, the intrinsic model capabilities are the primary drivers of success across diverse tasks.

\subsection{Main Experiment Two: Effects of Collaboration Protocols and Planning Strategies}

We investigate the impact of different collaboration protocols—Star, Tree, Graph, and Chain—on model performance in the Research scenario. 

\begin{figure}[htbp]
    \centering
    \includegraphics[width=\linewidth]{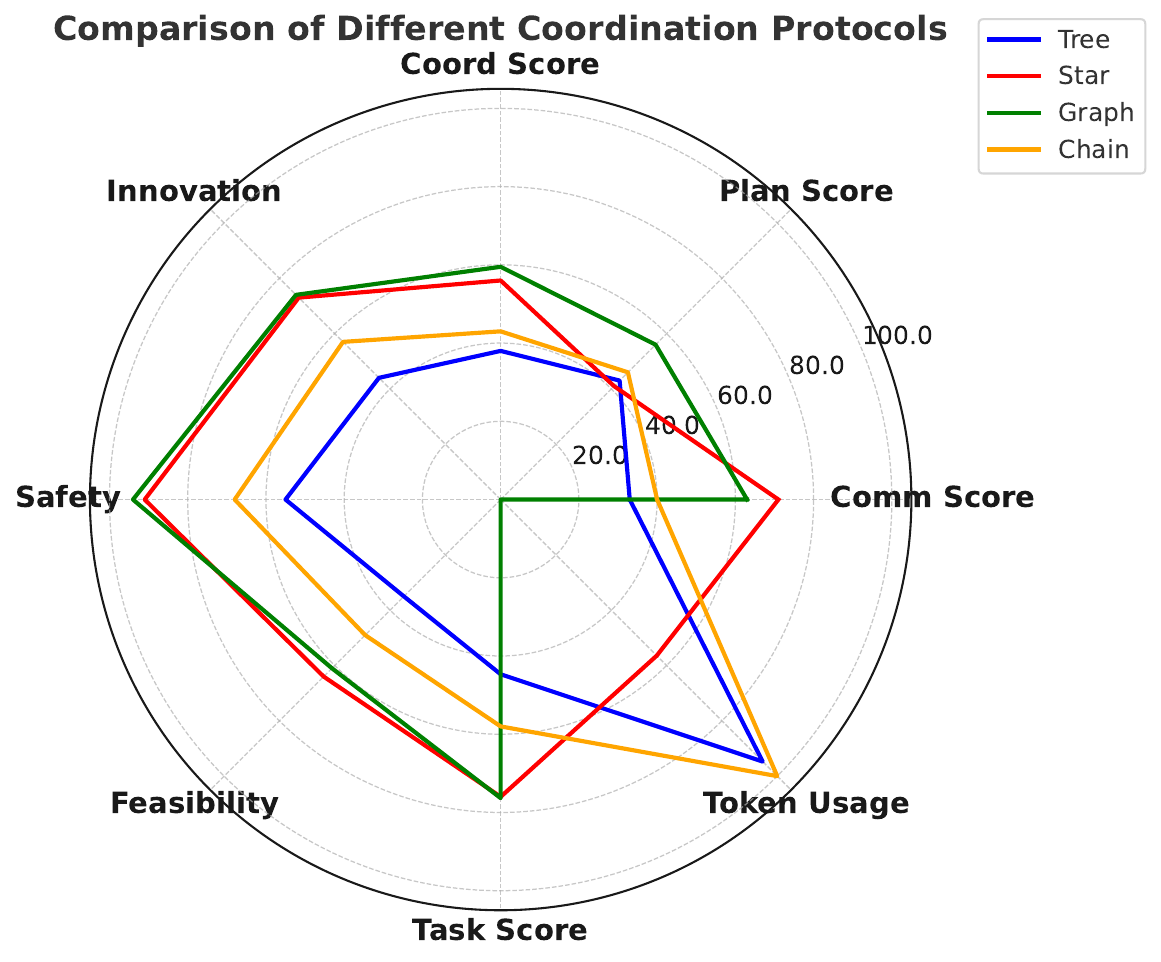}
    \caption{Comparison of Different Coordination Protocols.—Tree, Star, Graph, and Chain—across multiple evaluation metrics. Specially, the token usages are scaled such that the lowest value is $0$ and the hightest value is $100$. Details about metrics used for research task can be found at \ref{research_task_metric}}
    \label{fig:coordination_protocol_comparison}
\end{figure}

\begin{figure}[htbp]
    \centering
    \includegraphics[width=\linewidth]{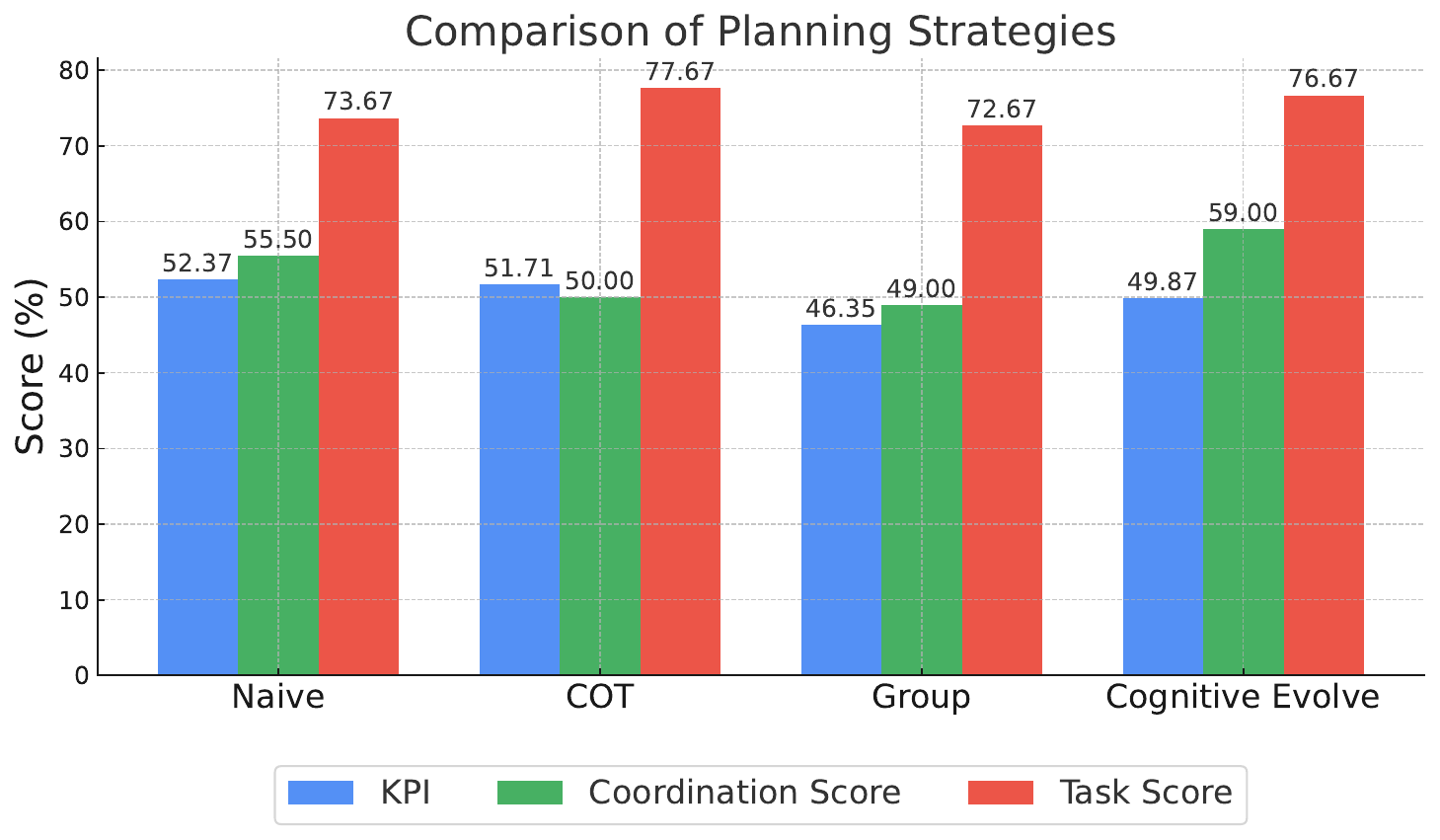}
    \caption{Average Metrics for Research Tasks for different planning prompt strategies. Cognitive Evolve Planning show best result in CS.}
    \label{fig:research_metrics}
\end{figure}

According to Fig.~\ref{fig:coordination_protocol_comparison}, the graph-based protocol excels in research scenarios with the best task performance, planning efficiency, and token usage, while both the star and graph protocols yield similar task scores. In contrast, the tree-based protocol performs poorly, with high token consumption and the lowest task and coordination scores. As shown in Fig.~\ref{fig:research_metrics}, Cognitive Evolving Planning demonstrates superior coordination—significantly outperforming the alternatives—and achieves a task score comparable to the best, COT. Counterintuitively, the group discussion method scores the worst across all metrics, possibly because an overly large planning group hinders effectiveness, similar to large organizations in real-world scenarios.

\section{Ablation Study}

The goal of our ablation study is to identify the key modules and parameters that affect performance.

\begin{figure}[htbp]
    \centering
    \includegraphics[width=\linewidth]{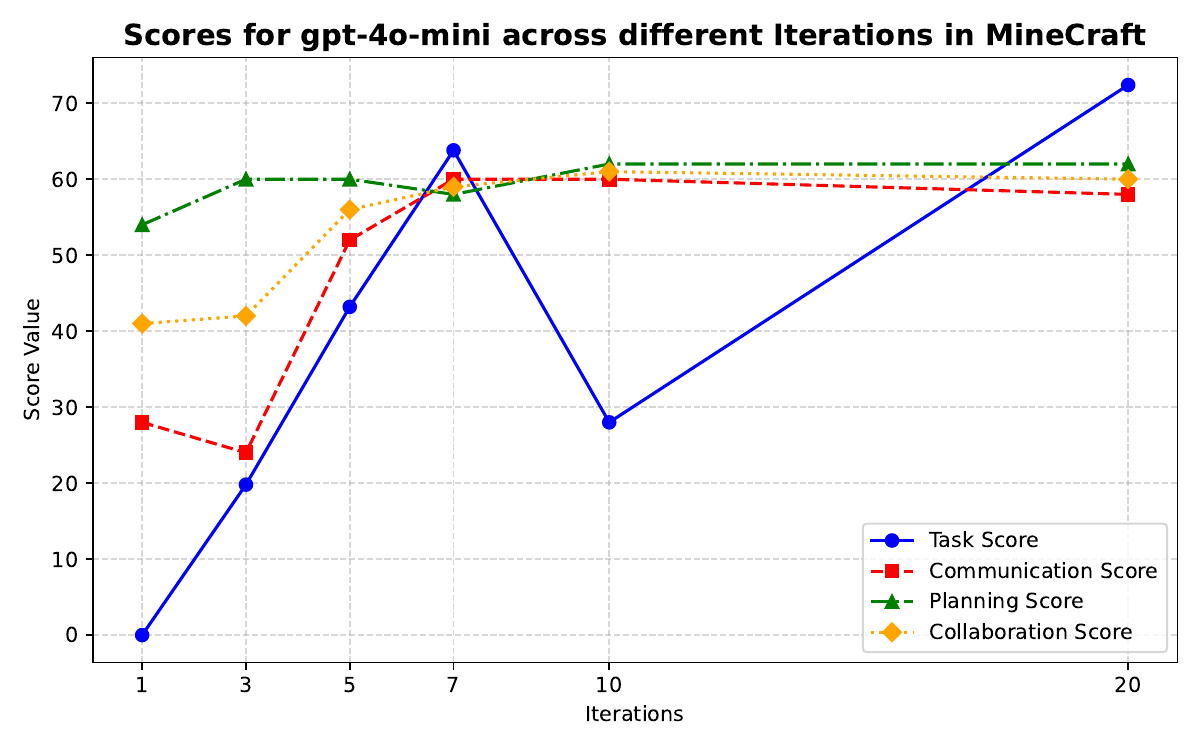}
    \caption{Scores for \texttt{gpt-4o-mini} across different iterations in Minecraft. The figure presents Task Score (TS), Communication Score (CS), Planning Score (PS), and Collaboration Score (CoS) over multiple iterations.}
    \label{fig:gpt4o_minecraft_scores}
\end{figure}

\paragraph{Ablation on Different Max Iteration Settings} We evaluate 10 tasks from the Minecraft scenario using six distinct maximum iteration settings. As shown in Fig.~\ref{fig:gpt4o_minecraft_scores}, both task and coordination scores increase from 1 to 7 iterations, but then drop sharply at 10 iterations. At 20 iterations, while the task score shows a recovery, the coordination score remains nearly unchanged beyond 7 iterations. This pattern suggests that, for highly challenging tasks, excessive iterations may lead to coordination degradation—possibly due to communication overhead or conflicting directives emerging over prolonged interactions. These findings underscore the need for adaptive iteration strategies that balance task execution with effective coordination.

\begin{figure}[htbp]
    \centering
    \includegraphics[width=\linewidth]{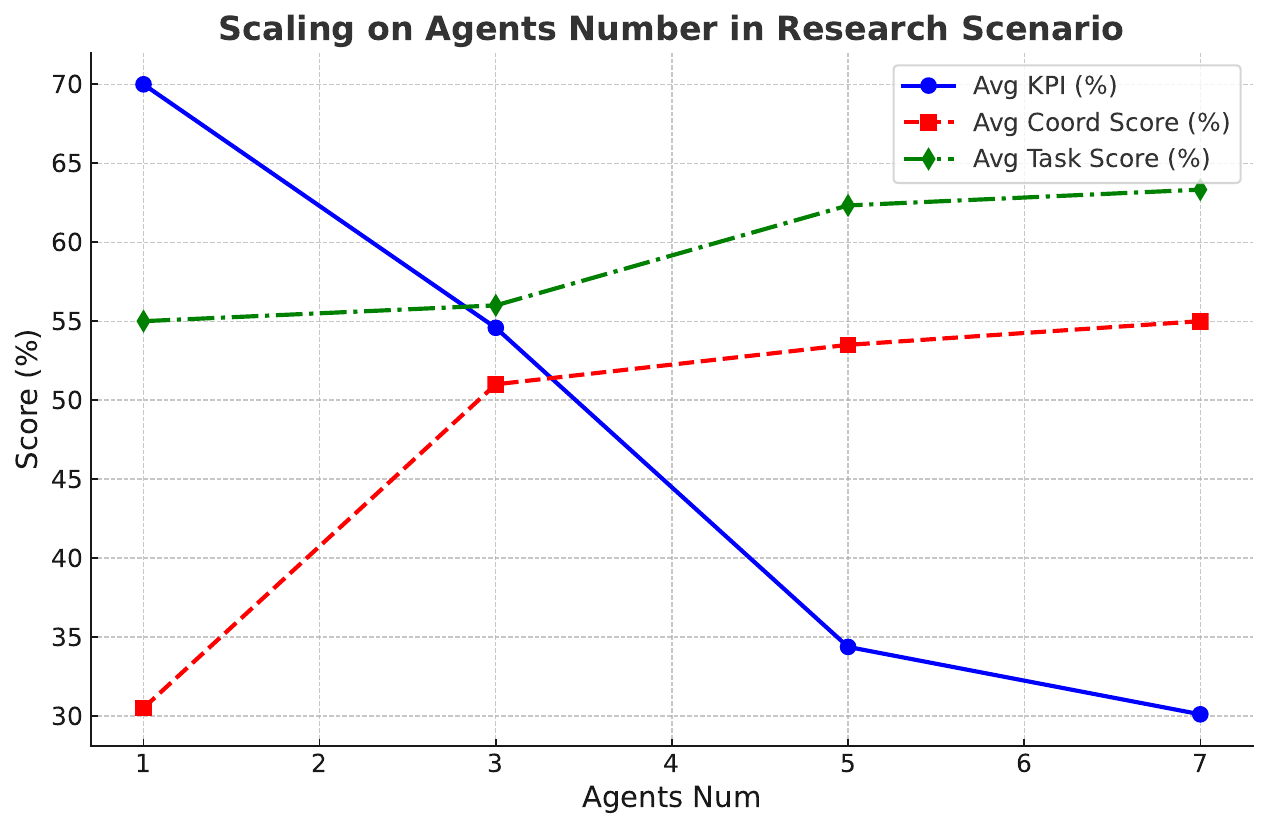}
    \caption{Scaling on Agents Number in Research Scenario. This figure shows the impact of agent number on KPI, Coordination Score, and Task Score.}
    \label{fig:scaling_agents}
\end{figure}

\paragraph{Ablation on Different Agent Numbers} We assess configurations with 1, 3, 5, and 7 agents in the research scenario, selecting 20 papers that have at least 7 main authors. As illustrated in Fig.~\ref{fig:scaling_agents}, increasing the number of agents leads to a decrease in the overall KPI, which aligns with the anticipated trade-off between increased collaborative complexity and performance. Notably, the average coordination score improves significantly when moving from 1 to 3 agents, while the average task score increases more gradually. This indicates that a moderate expansion in team size can enhance coordination efficiency, although further increases may introduce additional coordination challenges that counterbalance task performance gains.

\section{Emergent Behaviors Analysis}
\label{sec:emergent_behaviors}
In MultiagentBench, goal-driven emergent behaviors are pivotal to team coordination—we argue that these “aha-moments” not only mark when individual agents align their actions toward shared objectives but also catalyze spontaneous multiagent dynamics, unveiling novel coordination strategies and adaptive collective intelligence. (see Appendix~\ref{sec:detailed-results-case-study}). Under information asymmetry and role conflicts, agents display three key patterns (refer to \ref{appendix:case2} and \ref{sec:case_llama33_gpt4o} for Werewolf scenarios, and Section~\ref{fig:negotiation-summary} for Bargaining):

\paragraph{\textbf{Strategic Information Sharing}} Agents selectively disclose key information (e.g., the Seer withholding inspection results) based on trust and context, echoing the strategic concealment of information previously observed in LLM agents playing Werewolf~\citep{xu2023werewolf}. For instance, in \ref{appendix:case2}, both the Seer and Witch (\texttt{gpt-4o}) were overly cautious, missing optimal sharing opportunities and leading to failure. Another case is shown in Fig~\ref{fig:bad_comm_case_1}, where two research agents strategically evolve the research proposal details, combining the strengths of both background knowledge.

\paragraph{\textbf{Trust-Polarized Collaboration}} Role identities drive collaboration splits: over-suspicious villagers may turn against their own, while werewolves can create a "false consensus" through deception~\citep{Fehr2000} and internal execution~\citep{Woolley2010}. As shown in \ref{appendix:case2}, villagers’ excessive caution allowed werewolves to exploit confusion; similar internal friction is evident in Bargaining (\ref{fig:negotiation-summary}).

\paragraph{\textbf{Role-Driven Strategy Iteration}} Throughout the game, roles such as the Seer and Witch adjust their strategies. The Seer, for example, shifts from a conservative to a leadership role (see \ref{sec:case_llama33_gpt4o}), while the Witch moves from hoarding to taking risks. These shifts support the notion that task objectives drive decision-making, aligning with the \textit{AutoGen} framework findings~\citep{wu2023autogen}.

\paragraph{Quantitative Analysis of Emergent Behaviors (Werewolf Scenario)}
\label{sec:emergent_behaviors_werewolf}

To quantify goal-driven emergent behaviors in the Werewolf scenario, we employ LLM as a judge to analyze each recorded game transcript. Specifically, for every session in our dataset, we segment the multi-turn dialogue and agent actions, then prompt the LLM to detect three target behaviors: 
\emph{(i) Strategic Information Sharing}, 
\emph{(ii) Trust-Polarized Collaboration}, and 
\emph{(iii) Role-Driven Strategy Iteration}. 
Our script tallies these occurrences across all transcripts, aggregating them per experimental configuration. 

Table~\ref{tab:emergent-behaviors-werewolf} presents a summary of these counts. 
“Total” denotes the total number of sessions in each file, while the last three columns respectively show how many sessions exhibited each emergent behavior. 
We tested five different model settings, ranging from GPT-based to LLaMA-based models.

\begin{table}[ht!]
\centering
\caption{Summary of Emergent Behaviors in Werewolf Scenarios}
\label{tab:emergent-behaviors-werewolf}
\resizebox{0.48\textwidth}{!}{%
\begin{tabular}{lrrrr}
\toprule
\textbf{Model} & 
\textbf{Total} & 
\textbf{Strat. Info.} & 
\textbf{Trust Collab.} & 
\textbf{Role Iter.} \\
\midrule
gpt-4o-mini        & 100 &  97 & 100 &  42 \\
gpt-3.5            & 100 & 100 & 100 &  66 \\
llama-3.1-8b       & 100 &  96 &  98 &  38 \\
llama-3.1-70b      & 100 & 100 & 100 &  55 \\
llama-3.3-70b      & 100 & 100 & 100 &  89 \\
\midrule
\textbf{Grand Total} & 500 & 493 & 498 & 290 \\
\bottomrule
\end{tabular}%
}
\end{table}

As shown in Table~\ref{tab:emergent-behaviors-werewolf}, \emph{Strategic Information Sharing} and \emph{Trust-Polarized Collaboration} arise consistently—reflecting how manipulations of trust and carefully timed information disclosure are central to Werewolf gameplay. \emph{Role-Driven Strategy Iteration} occurs less frequently, suggesting that while specialized roles (e.g., Seer, Witch, Werewolf) can evolve their behavior, such adaptive dynamics may require longer or more intricate rounds to manifest fully. Overall, these results align with the social deduction nature of Werewolf: hidden-role uncertainty naturally fosters high levels of strategic communication and polarizing trust dynamics, whereas deeper shifts in role-based tactics may only emerge given sufficient interaction rounds.

\paragraph{Quantitative Analysis of Emergent Behaviors (Bargaining Scenario)}
\label{sec:emergent_behaviors_bargaining}

We also examine emergent behaviors in a Bargaining scenario, where a \emph{buyer} and \emph{seller} negotiate terms. Again, we use the same LLM-based detection method to identify the same three types of behaviors. Table~\ref{tab:bargaining-emergent} aggregates results for five model configurations on the buyer side and five on the seller side, each with 100 sessions.

\begin{table}[ht!]
\centering
\caption{Summary of Emergent Behaviors in Bargaining Scenarios (Buyer vs.\ Seller).}
\label{tab:bargaining-emergent}
\resizebox{0.48\textwidth}{!}{%
\begin{tabular}{lrrrr}
\toprule
\textbf{File} & 
\textbf{Total} & 
\textbf{Strat.\ Info.} & 
\textbf{Trust Collab.} & 
\textbf{Role Iter.} \\
\midrule
\multicolumn{5}{l}{\textit{Buyer Side}} \\ 
\midrule
buyer-gpt-3.5-turbo        & 100 & 81 &  1 & 81 \\
buyer-gpt-4o-mini          & 100 & 88 &  0 & 88 \\
buyer-Llama-3.1-70B        & 100 & 81 &  2 & 80 \\
buyer-Llama-3.1-8B         & 100 & 92 &  0 & 87 \\
buyer-Llama-3.3-70B        & 100 & 84 &  1 & 82 \\
\midrule
\textbf{Buyer Total}       & \textbf{500} & \textbf{426} & \textbf{4} & \textbf{418} \\
\midrule
\multicolumn{5}{l}{\textit{Seller Side}} \\ 
\midrule
seller-gpt-3.5-turbo       & 100 & 88 &  0 & 87 \\
seller-gpt-4o-mini         & 100 & 95 &  0 & 95 \\
seller-Llama-3.1-70B       & 100 & 96 &  0 & 97 \\
seller-Llama-3.1-8B        & 100 & 97 &  0 & 96 \\
seller-Llama-3.3-70B       & 100 & 90 &  1 & 88 \\
\midrule
\textbf{Seller Total}      & \textbf{500} & \textbf{466} & \textbf{1} & \textbf{463} \\
\midrule
\textbf{Grand Total}       & \textbf{1000} & \textbf{892} & \textbf{5} & \textbf{881} \\
\bottomrule
\end{tabular}%
}
\end{table}

In these Bargaining sessions, \emph{Strategic Information Sharing} also dominates, as negotiation naturally incentivizes the buyer and seller to withhold or reveal private information (e.g., target prices, product details) at opportunistic moments. However, \emph{Trust-Polarized Collaboration} appears much less often, which is consistent with the primarily bilateral structure—there is less impetus for large-scale alliance-building or deception across multiple parties. \emph{Role-Driven Strategy Iteration} remains frequent: both buyer and seller can adapt their negotiation tactics (e.g., starting from extreme offers and gradually conceding) in a manner driven by their “role.” Such interplay of strategic disclosure, occasional (albeit limited) trust dynamics, and ongoing role-based adjustments typifies two-party bargaining interactions. Overall, the pattern of emergent behaviors here aligns well with the fundamental buyer-seller tension, wherein each agent seeks to maximize its advantage via selective sharing, repeated iteration, and a careful balance of cooperation and self-interest.

-

\section{Conclusion}  
In this work, we introduce MultiAgentBench and the MARBLE framework, providing a comprehensive benchmark for evaluating LLM-based multi-agent systems across diverse interactive scenarios. Our proposed evaluation metrics go beyond task success, capturing coordination quality through structured planning, communication scores, and competition-driven assessments. Experimental results highlight key emergent social behaviors, offering valuable insights into future multiagent work. 

\section{Limitations}
While our proposed multi-agent benchmark and framework provide a diverse range of tasks and evaluation metrics, several areas warrant further exploration to enhance their applicability and robustness:

\paragraph{Expanding Scenario and Model Coverage.}
Currently, our benchmark focuses on specific domains such as research co-authoring, Minecraft building, database error analysis, coding collaboration, and select competitive scenarios (e.g., Werewolf and bargaining). To better capture the complexity of real-world multi-agent interactions, future work can incorporate more diverse settings, including open-world environments, scenarios requiring richer social cognition, and tasks on the application side like task-oriented dialogues. In the aspect of models, our work does not cover the full spectrum. Future work may include the result of other latest ones (e.g. the DeepSeek model family).

\paragraph{Enhancing Ablation Studies.}
Our current analysis focuses primarily on overall coordination and competition performance, leaving finer-grained insights into specific components underexplored. Future experiments can be focused on different memory mechanisms (e.g. long-term memory, short-term memory, shared memory) and multiagent different workflow method.

\paragraph{Advancing Competition Mechanisms.}
While our benchmark incorporates competitive tasks, it does not fully capture the complexity of real-world multi-agent interactions involving multi-party negotiations, repeated strategic play, or stochastic elements. Investigating how agents transition between cooperative and adversarial roles in evolving environments remains a promising direction.

\paragraph{Handling Open-Ended and Ill-Defined Tasks.}
Most tasks in our framework involve well-defined objectives, such as completing a research proposal or resolving database inconsistencies. However, real-world applications often require agents to operate in open-ended or ambiguous contexts without clear success criteria. Future extensions could explore how multi-agent systems adapt to exploratory, non-goal-oriented scenarios.

\section*{Acknowledgment}
This research is partly based upon work supported by DARPA ITM Program No. FA8650-23-C-7316, the Molecule Maker Lab Institute: an AI research institute program supported by NSF under award No. 2019897, and the AI Research Institutes program by National Science Foundation and the Institute of Education Sciences, U.S. Department of Education through Award \# 2229873 - AI Institute for Transforming Education for Children with Speech and Language Processing Challenges. The views and conclusions contained herein are those of the authors and should not be interpreted as necessarily representing the official policies, either expressed or implied, of the U.S. Government. The U.S. Government is authorized to reproduce and distribute reprints for governmental purposes notwithstanding any copyright annotation therein. We sincerely appreciate the support from Amazon grant funding project \#120359, ``GRAG: Enhance RAG Applications with Graph-structured Knowledge'' and Meta gift funding project ``PERM: Toward Parameter Efficient Foundation Models for Recommenders''.

\bibliography{custom}

\begin{thebibliography}{54}
\expandafter\ifx\csname natexlab\endcsname\relax\def\natexlab#1{#1}\fi

\bibitem[{Agarwal et~al.(2024)Agarwal, Sahu, Puri, Laradji, Dvijotham, Stanley,
  Charlin, and Pal}]{agarwal2024litllmtoolkitscientificliterature}
Shubham Agarwal, Gaurav Sahu, Abhay Puri, Issam~H. Laradji, Krishnamurthy~DJ
  Dvijotham, Jason Stanley, Laurent Charlin, and Christopher Pal. 2024.
\newblock {LitLLM}: A toolkit for scientific literature review.
\newblock \emph{arXiv preprint arXiv:2402.01788}.

\bibitem[{Agashe et~al.(2024)Agashe, Fan, Reyna, and
  Wang}]{agashe2024llmcoordinationevaluatinganalyzingmultiagent}
Saaket Agashe, Yue Fan, Anthony Reyna, and Xin~Eric Wang. 2024.
\newblock \href {http://arxiv.org/abs/2310.03903} {{LLM-Coordination}:
  Evaluating and analyzing multi-agent coordination abilities in large language
  models}.

\bibitem[{Asaniczka(2023)}]{asaniczka_2023}
Asaniczka. 2023.
\newblock \href {https://doi.org/10.34740/KAGGLE/DS/3798081} {Amazon products
  dataset 2023 (1.4{M} products)}.

\bibitem[{Bouzenia et~al.(2024)Bouzenia, Devanbu, and
  Pradel}]{bouzenia2024repairagent}
Islem Bouzenia, Premkumar Devanbu, and Michael Pradel. 2024.
\newblock {RepairAgent}: An autonomous, {LLM}-based agent for program repair.
\newblock \emph{arXiv preprint arXiv:2403.17134}.

\bibitem[{Brown et~al.(2020)Brown, Mann, Ryder, Subbiah, Kaplan, Dhariwal,
  Neelakantan, Shyam, Sastry, Askell et~al.}]{Brown2020GPT3}
Tom~B. Brown, Benjamin Mann, Nick Ryder, Melanie Subbiah, Jared Kaplan,
  Prafulla Dhariwal, Arvind Neelakantan, Pranav Shyam, Girish Sastry, Amanda
  Askell, et~al. 2020.
\newblock Language models are few-shot learners.
\newblock In \emph{Advances in Neural Information Processing Systems},
  volume~33, pages 1877--1901.

\bibitem[{Chen et~al.(2023)Chen, Zhang, Wang, Huo, Li, and
  Wang}]{chen2023gamegpt}
Dake Chen, Haoyang Zhang, Hanbin Wang, Yunhao Huo, Yuzhao Li, and Junjie Wang.
  2023.
\newblock {GameGPT}: Multi-agent collaborative framework for game development.
\newblock \emph{arXiv preprint arXiv:2310.08067}.

\bibitem[{Chen et~al.(2021)Chen, Tworek, Jun, Yuan, Pinto, Kaplan, Edwards,
  Burda, Joseph, Brockman et~al.}]{Chen2021Evaluating}
Mark Chen, Jerry Tworek, Heewoo Jun, Qiming Yuan, Henrique Ponde de~Oliveira
  Pinto, Jared Kaplan, Harri Edwards, Yuri Burda, Nicholas Joseph, Greg
  Brockman, et~al. 2021.
\newblock Evaluating large language models trained on code.
\newblock \emph{arXiv preprint arXiv:2107.03374}.

\bibitem[{Chen et~al.(2024)Chen, Su, Zuo, Yang, Yuan, Chan, Yu, Lu, Hung, Qian,
  Qin, Cong, Xie, Liu, Sun, and Zhou}]{chen2024agentverse}
Weize Chen, Yusheng Su, Jingwei Zuo, Cheng Yang, Chenfei Yuan, Chi-Min Chan,
  Heyang Yu, Yaxi Lu, Yi-Hsin Hung, Chen Qian, Yujia Qin, Xin Cong, Ruobing
  Xie, Zhiyuan Liu, Maosong Sun, and Jie Zhou. 2024.
\newblock \href {https://openreview.net/forum?id=EHg5GDnyq1} {{AgentVerse}:
  Facilitating multi-agent collaboration and exploring emergent behaviors}.
\newblock In \emph{The Twelfth International Conference on Learning
  Representations}.

\bibitem[{{DeepSeek-AI} et~al.(2025){DeepSeek-AI}, Guo, Yang, Zhang, Song, Wang
  et~al.}]{guo2025deepseek}
{DeepSeek-AI}, Daya Guo, Dejian Yang, Haowei Zhang, Junxiao Song, Peiyi Wang,
  et~al. 2025.
\newblock {DeepSeek-R1}: Incentivizing reasoning capability in {LLMs} via
  reinforcement learning.
\newblock \emph{arXiv preprint arXiv:2501.12948}.

\bibitem[{Dong et~al.(2024)Dong, Zhu, Pan, Zhu, and
  Yang}]{dong2024villageragentgraphbasedmultiagentframework}
Yubo Dong, Xukun Zhu, Zhengzhe Pan, Linchao Zhu, and Yi~Yang. 2024.
\newblock \href {http://arxiv.org/abs/2406.05720} {{VillagerAgent}: A
  graph-based multi-agent framework for coordinating complex task dependencies
  in {Minecraft}}.

\bibitem[{Dubey et~al.(2024)Dubey, Jauhri, Pandey, Kadian, Al-Dahle, Letman
  et~al.}]{dubey2024llama}
Abhimanyu Dubey, Abhinav Jauhri, Abhinav Pandey, Abhishek Kadian, Ahmad
  Al-Dahle, Aiesha Letman, et~al. 2024.
\newblock The {Llama} 3 herd of models.
\newblock \emph{arXiv preprint arXiv:2407.21783}.

\bibitem[{Fehr and G{\"a}chter(2000)}]{Fehr2000}
Ernst Fehr and Simon G{\"a}chter. 2000.
\newblock \href {https://doi.org/10.1257/aer.90.4.980} {Cooperation and
  punishment in public goods experiments}.
\newblock \emph{American Economic Review}, 90(4):980--994.

\bibitem[{Feng et~al.(2024)Feng, Jin, Liu, Zhu, Tu, Cheng, Lin, and
  You}]{feng2024how}
Tao Feng, Chuanyang Jin, Jingyu Liu, Kunlun Zhu, Haoqin Tu, Zirui Cheng, Guanyu
  Lin, and Jiaxuan You. 2024.
\newblock \href {https://openreview.net/forum?id=H2ZKqfNd0U} {How far are we
  from {AGI}: Are {LLM}s all we need?}
\newblock \emph{Transactions on Machine Learning Research}.
\newblock Survey Certification.

\bibitem[{{Gemini Team} et~al.(2023){Gemini Team}, Anil, Borgeaud, Alayrac, Yu,
  Soricut et~al.}]{team2023gemini}
{Gemini Team}, Rohan Anil, Sebastian Borgeaud, Jean-Baptiste Alayrac, Jiahui
  Yu, Radu Soricut, et~al. 2023.
\newblock Gemini: A family of highly capable multimodal models.
\newblock \emph{arXiv preprint arXiv:2312.11805}.

\bibitem[{Gong et~al.(2023)Gong, Huang, Ma, Vo, Durante, Noda, Zheng, Zhu,
  Terzopoulos, Fei-Fei, and Gao}]{gong2023mindagent}
Ran Gong, Qiuyuan Huang, Xiaojian Ma, Hoi Vo, Zane Durante, Yusuke Noda, Zilong
  Zheng, Song-Chun Zhu, Demetri Terzopoulos, Li~Fei-Fei, and Jianfeng Gao.
  2023.
\newblock {MindAgent}: Emergent gaming interaction.
\newblock \emph{arXiv preprint arXiv:2309.09971}.

\bibitem[{G{\"o}sling et~al.(2024)G{\"o}sling, Dudek, Krause, and
  Thomas}]{gosling2024multi}
Henning G{\"o}sling, Jacob Dudek, Thorsten Krause, and Oliver Thomas. 2024.
\newblock \href
  {https://aisel.aisnet.org/icis2024/learnandiscurricula/learnandiscurricula/12}
  {Multi-agent-based peer tutoring in virtual learning environments}.
\newblock In \emph{Proceedings of the 45th International Conference on
  Information Systems (ICIS)}.

\bibitem[{Hong et~al.(2024)Hong, Zhuge, Chen, Zheng, Cheng, Wang, Zhang, Wang,
  Yau, Lin, Zhou, Ran, Xiao, Wu, and Schmidhuber}]{hong2024metagpt}
Sirui Hong, Mingchen Zhuge, Jonathan Chen, Xiawu Zheng, Yuheng Cheng, Jinlin
  Wang, Ceyao Zhang, Zili Wang, Steven Ka~Shing Yau, Zijuan Lin, Liyang Zhou,
  Chenyu Ran, Lingfeng Xiao, Chenglin Wu, and J{\"u}rgen Schmidhuber. 2024.
\newblock \href {https://openreview.net/forum?id=VtmBAGCN7o} {Meta{GPT}: Meta
  programming for a multi-agent collaborative framework}.
\newblock In \emph{The Twelfth International Conference on Learning
  Representations}.

\bibitem[{Hua et~al.(2024)Hua, Fan, Li, Mei, Ji, Ge, Hemphill, and
  Zhang}]{hua2024warpeacewaragentlarge}
Wenyue Hua, Lizhou Fan, Lingyao Li, Kai Mei, Jianchao Ji, Yingqiang Ge, Libby
  Hemphill, and Yongfeng Zhang. 2024.
\newblock \href {http://arxiv.org/abs/2311.17227} {War and peace ({WarAgent}):
  Large language model-based multi-agent simulation of world wars}.

\bibitem[{Huang et~al.(2023)Huang, Zhang, Luck, Bu, Qing, and
  Cui}]{huang2023agentcoder}
Dong Huang, Jie~M. Zhang, Michael Luck, Qingwen Bu, Yuhao Qing, and Heming Cui.
  2023.
\newblock {AgentCoder}: Multi-agent-based code generation with iterative
  testing and optimisation.
\newblock \emph{arXiv preprint arXiv:2312.13010}.

\bibitem[{Ishibashi and Nishimura(2024)}]{ishibashi2024self}
Yoichi Ishibashi and Yoshimasa Nishimura. 2024.
\newblock Self-organized agents: A {LLM} multi-agent framework toward ultra
  large-scale code generation and optimization.
\newblock \emph{arXiv preprint arXiv:2404.02183}.

\bibitem[{Islam et~al.(2024)Islam, Ali, and Parvez}]{islam2024mapcoder}
Md.~Ashraful Islam, Mohammed~Eunus Ali, and Md~Rizwan Parvez. 2024.
\newblock \href {https://doi.org/10.18653/v1/2024.acl-long.269} {{M}ap{C}oder:
  Multi-agent code generation for competitive problem solving}.
\newblock In \emph{Proceedings of the 62nd Annual Meeting of the {A}ssociation
  for {C}omputational {L}inguistics (Volume 1: Long Papers)}, pages 4912--4944,
  Bangkok, Thailand. Association for Computational Linguistics.

\bibitem[{Ke et~al.(2024)Ke, Yang, Lie, Lim, Abdullah, Ting, and
  Liu}]{ke2024enhancing}
Yu~He Ke, Rui Yang, Sui~An Lie, Taylor Xin~Yi Lim, Hairil~Rizal Abdullah,
  Daniel Shu~Wei Ting, and Nan Liu. 2024.
\newblock Enhancing diagnostic accuracy through multi-agent conversations:
  Using large language models to mitigate cognitive bias.
\newblock \emph{arXiv preprint arXiv:2401.14589}.

\bibitem[{Kim et~al.(2024)Kim, Park, Jeong, Chan, Xu, McDuff, Lee, Ghassemi,
  Breazeal, and Park}]{kim2024adaptive}
Yubin Kim, Chanwoo Park, Hyewon Jeong, Yik~Siu Chan, Xuhai Xu, Daniel McDuff,
  Hyeonhoon Lee, Marzyeh Ghassemi, Cynthia Breazeal, and Hae~Won Park. 2024.
\newblock {MDAgents}: An adaptive collaboration of {LLMs} for medical
  decision-making.
\newblock In \emph{Advances in Neural Information Processing Systems},
  volume~37.

\bibitem[{Li et~al.(2023)Li, Hammoud, Itani, Khizbullin, and
  Ghanem}]{li2023camelcommunicativeagentsmind}
Guohao Li, Hasan Abed Al~Kader Hammoud, Hani Itani, Dmitrii Khizbullin, and
  Bernard Ghanem. 2023.
\newblock {CAMEL}: Communicative agents for ``mind'' exploration of large
  language model society.
\newblock In \emph{Advances in Neural Information Processing Systems},
  volume~36.

\bibitem[{Liu et~al.(2023)Liu, Yu, Zhang, Xu, Lei, Lai, Gu, Ding, Men, Yang
  et~al.}]{liu2023agentbench}
Xiao Liu, Hao Yu, Hanchen Zhang, Yifan Xu, Xuanyu Lei, Hanyu Lai, Yu~Gu,
  Hangliang Ding, Kaiwen Men, Kejuan Yang, et~al. 2023.
\newblock {AgentBench}: Evaluating {LLMs} as agents.
\newblock \emph{arXiv preprint arXiv:2308.03688}.

\bibitem[{Liu et~al.(2024)Liu, Zhang, Gu, Iong, Xu, Song, Zhang, Lai, Liu,
  Zhao, Sun, Yang, Yang, Qi, Yao, Sun, Cheng, Zheng, Yu, Zhang, Hong, Ding,
  Pan, Gu, Zeng, Du, Song, Su, Dong, and
  Tang}]{liu2024visualagentbenchlargemultimodalmodels}
Xiao Liu, Tianjie Zhang, Yu~Gu, Iat~Long Iong, Yifan Xu, Xixuan Song, Shudan
  Zhang, Hanyu Lai, Xinyi Liu, Hanlin Zhao, Jiadai Sun, Xinyue Yang, Yu~Yang,
  Zehan Qi, Shuntian Yao, Xueqiao Sun, Siyi Cheng, Qinkai Zheng, Hao Yu,
  Hanchen Zhang, Wenyi Hong, Ming Ding, Lihang Pan, Xiaotao Gu, Aohan Zeng,
  Zhengxiao Du, Chan~Hee Song, Yu~Su, Yuxiao Dong, and Jie Tang. 2024.
\newblock \href {http://arxiv.org/abs/2408.06327} {{VisualAgentBench}: Towards
  large multimodal models as visual foundation agents}.

\bibitem[{Mialon et~al.(2023)Mialon, Fourrier, Swift, Wolf, LeCun, and
  Scialom}]{mialon2023gaia}
Gr{\'e}goire Mialon, Cl{\'e}mentine Fourrier, Craig Swift, Thomas Wolf, Yann
  LeCun, and Thomas Scialom. 2023.
\newblock {GAIA}: A benchmark for general {AI} assistants.
\newblock \emph{arXiv preprint arXiv:2311.12983}.

\bibitem[{{OpenAI}(2023)}]{OpenAI2023FunctionCalling}
{OpenAI}. 2023.
\newblock Function calling and other {API} updates.
\newblock
  \url{https://openai.com/index/function-calling-and-other-api-updates/}.

\bibitem[{{OpenAI} et~al.(2023){OpenAI}, Achiam, Adler, Agarwal, Ahmad, Akkaya
  et~al.}]{OpenAI2023GPT4}
{OpenAI}, Josh Achiam, Steven Adler, Sandhini Agarwal, Lama Ahmad, Ilge Akkaya,
  et~al. 2023.
\newblock {GPT-4} technical report.
\newblock \emph{arXiv preprint arXiv:2303.08774}.

\bibitem[{Park et~al.(2023)Park, O'Brien, Cai, Morris, Liang, and
  Bernstein}]{Park2023GenerativeAgents}
Joon~Sung Park, Joseph~C. O'Brien, Carrie~J. Cai, Meredith~Ringel Morris, Percy
  Liang, and Michael~S. Bernstein. 2023.
\newblock \href {https://doi.org/10.1145/3586183.3606763} {Generative agents:
  Interactive simulacra of human behavior}.
\newblock In \emph{Proceedings of the 36th Annual ACM Symposium on User
  Interface Software and Technology (UIST)}.

\bibitem[{Premack and Woodruff(1978)}]{Premack1978}
David Premack and Guy Woodruff. 1978.
\newblock \href {https://doi.org/10.1017/S0140525X00076512} {Does the
  chimpanzee have a theory of mind?}
\newblock \emph{Behavioral and Brain Sciences}, 1(4):515--526.

\bibitem[{Qian et~al.(2024)Qian, Liu, Liu, Chen, Dang, Li, Yang, Chen, Su,
  Cong, Xu, Li, Liu, and Sun}]{qian2024chatdev}
Chen Qian, Wei Liu, Hongzhang Liu, Nuo Chen, Yufan Dang, Jiahao Li, Cheng Yang,
  Weize Chen, Yusheng Su, Xin Cong, Juyuan Xu, Dahai Li, Zhiyuan Liu, and
  Maosong Sun. 2024.
\newblock \href {https://doi.org/10.18653/v1/2024.acl-long.810} {{C}hat{D}ev:
  Communicative agents for software development}.
\newblock In \emph{Proceedings of the 62nd Annual Meeting of the {A}ssociation
  for {C}omputational {L}inguistics (Volume 1: Long Papers)}, pages
  15174--15186, Bangkok, Thailand. Association for Computational Linguistics.

\bibitem[{Qian et~al.(2025)Qian, Xie, Wang, Liu, Zhu, Xia, Dang, Du, Chen,
  Yang, Liu, and Sun}]{qian2025scaling}
Chen Qian, Zihao Xie, YiFei Wang, Wei Liu, Kunlun Zhu, Hanchen Xia, Yufan Dang,
  Zhuoyun Du, Weize Chen, Cheng Yang, Zhiyuan Liu, and Maosong Sun. 2025.
\newblock \href {https://openreview.net/forum?id=K3n5jPkrU6} {Scaling large
  language model-based multi-agent collaboration}.
\newblock In \emph{The Thirteenth International Conference on Learning
  Representations}.

\bibitem[{Qin et~al.(2024)Qin, Liang, Ye, Zhu, Yan, Lu, Lin, Cong, Tang, Qian,
  Zhao, Hong, Tian, Xie, Zhou, Gerstein, Li, Liu, and Sun}]{qin2024toolllm}
Yujia Qin, Shihao Liang, Yining Ye, Kunlun Zhu, Lan Yan, Yaxi Lu, Yankai Lin,
  Xin Cong, Xiangru Tang, Bill Qian, Sihan Zhao, Lauren Hong, Runchu Tian,
  Ruobing Xie, Jie Zhou, Mark Gerstein, Dahai Li, Zhiyuan Liu, and Maosong Sun.
  2024.
\newblock \href {https://openreview.net/forum?id=dHng2O0Jjr} {Tool{LLM}:
  Facilitating large language models to master 16000+ real-world {API}s}.
\newblock In \emph{The Twelfth International Conference on Learning
  Representations}.

\bibitem[{Shinn et~al.(2023)Shinn, Cassano, Gopinath, Narasimhan, and
  Yao}]{shinn2023reflexion}
Noah Shinn, Federico Cassano, Ashwin Gopinath, Karthik~R. Narasimhan, and
  Shunyu Yao. 2023.
\newblock \href {https://openreview.net/forum?id=vAElhFcKW6} {Reflexion:
  Language agents with verbal reinforcement learning}.
\newblock In \emph{Thirty-seventh Conference on Neural Information Processing
  Systems}.

\bibitem[{Silver et~al.(2017)Silver, Hubert, Schrittwieser, Antonoglou, Lai,
  Guez, Lanctot, Sifre, Kumaran, Graepel et~al.}]{silver2017mastering}
David Silver, Thomas Hubert, Julian Schrittwieser, Ioannis Antonoglou, Matthew
  Lai, Arthur Guez, Marc Lanctot, Laurent Sifre, Dharshan Kumaran, Thore
  Graepel, et~al. 2017.
\newblock Mastering chess and shogi by self-play with a general reinforcement
  learning algorithm.
\newblock \emph{arXiv preprint arXiv:1712.01815}.

\bibitem[{Valevski et~al.(2024)Valevski, Leviathan, Arar, and
  Fruchter}]{valeveski2024diffusion}
Dani Valevski, Yaniv Leviathan, Moab Arar, and Shlomi Fruchter. 2024.
\newblock Diffusion models are real-time game engines.
\newblock \emph{arXiv preprint arXiv:2408.14837}.

\bibitem[{Wang et~al.(2023)Wang, Xie, Jiang, Mandlekar, Xiao, Zhu, Fan, and
  Anandkumar}]{wang2023voyageropenendedembodiedagent}
Guanzhi Wang, Yuqi Xie, Yunfan Jiang, Ajay Mandlekar, Chaowei Xiao, Yuke Zhu,
  Linxi Fan, and Anima Anandkumar. 2023.
\newblock \href {http://arxiv.org/abs/2305.16291} {Voyager: An open-ended
  embodied agent with large language models}.

\bibitem[{Wang et~al.(2024{\natexlab{a}})Wang, Li, Song, Xu, Tang, Zhuge, Pan,
  Song, Li, Singh, Tran, Li, Ma, Zheng, Qian, Shao, Muennighoff, Zhang, Hui,
  Lin, Brennan, Peng, Ji, and Neubig}]{openhands}
Xingyao Wang, Boxuan Li, Yufan Song, Frank~F. Xu, Xiangru Tang, Mingchen Zhuge,
  Jiayi Pan, Yueqi Song, Bowen Li, Jaskirat Singh, Hoang~H. Tran, Fuqiang Li,
  Ren Ma, Mingzhang Zheng, Bill Qian, Yanjun Shao, Niklas Muennighoff, Yizhe
  Zhang, Binyuan Hui, Junyang Lin, Robert Brennan, Hao Peng, Heng Ji, and
  Graham Neubig. 2024{\natexlab{a}}.
\newblock \href {http://arxiv.org/abs/2407.16741} {{OpenHands: An Open Platform
  for AI Software Developers as Generalist Agents}}.

\bibitem[{Wang et~al.(2024{\natexlab{b}})Wang, Mao, Wu, Ge, Wei, and
  Ji}]{unleashing2024}
Zhenhailong Wang, Shaoguang Mao, Wenshan Wu, Tao Ge, Furu Wei, and Heng Ji.
  2024{\natexlab{b}}.
\newblock \href {https://doi.org/10.18653/v1/2024.naacl-long.15} {Unleashing
  the emergent cognitive synergy in large language models: A task-solving agent
  through multi-persona self-collaboration}.
\newblock In \emph{Proceedings of the 2024 Conference of the {N}orth {A}merican
  Chapter of the {A}ssociation for {C}omputational {L}inguistics: Human
  Language Technologies (Volume 1: Long Papers)}, pages 257--279, Mexico City,
  Mexico. Association for Computational Linguistics.

\bibitem[{Wei et~al.(2022)Wei, Wang, Schuurmans, Bosma, Ichter, Xia, Chi, Le,
  and Zhou}]{wei2022chain}
Jason Wei, Xuezhi Wang, Dale Schuurmans, Maarten Bosma, Brian Ichter, Fei Xia,
  Ed~Chi, Quoc~V. Le, and Denny Zhou. 2022.
\newblock Chain-of-thought prompting elicits reasoning in large language
  models.
\newblock In \emph{Advances in Neural Information Processing Systems},
  volume~35, pages 24824--24837.

\bibitem[{Woolley et~al.(2010)Woolley, Chabris, Pentland, Hashmi, and
  Malone}]{Woolley2010}
Anita~Williams Woolley, Christopher~F. Chabris, Alex Pentland, Nada Hashmi, and
  Thomas~W. Malone. 2010.
\newblock \href {https://doi.org/10.1126/science.1193147} {Evidence for a
  collective intelligence factor in the performance of human groups}.
\newblock \emph{Science}, 330(6004):686--688.

\bibitem[{Wu et~al.(2023)Wu, Bansal, Zhang, Wu, Li, Zhu et~al.}]{wu2023autogen}
Qingyun Wu, Gagan Bansal, Jieyu Zhang, Yiran Wu, Beibin Li, Erkang Zhu, et~al.
  2023.
\newblock {AutoGen}: Enabling next-gen {LLM} applications via multi-agent
  conversation.
\newblock \emph{arXiv preprint arXiv:2308.08155}.

\bibitem[{Xu et~al.(2024)Xu, Hu, Zhou, Ren, Dong, Keutzer, Ng, and
  Feng}]{xu2023magic}
Lin Xu, Zhiyuan Hu, Daquan Zhou, Hongyu Ren, Zhen Dong, Kurt Keutzer, See-Kiong
  Ng, and Jiashi Feng. 2024.
\newblock \href {https://doi.org/10.18653/v1/2024.emnlp-main.416} {{MA}g{IC}:
  Investigation of large language model powered multi-agent in cognition,
  adaptability, rationality and collaboration}.
\newblock In \emph{Proceedings of the 2024 Conference on Empirical Methods in
  Natural Language Processing}, pages 7315--7332, Miami, Florida, USA.
  Association for Computational Linguistics.

\bibitem[{Xu et~al.(2023)Xu, Wang, Li, Luo, Wang, Liu, and
  Liu}]{xu2023werewolf}
Yuzhuang Xu, Shuo Wang, Peng Li, Fuwen Luo, Xiaolong Wang, Weidong Liu, and
  Yang Liu. 2023.
\newblock Exploring large language models for communication games: An empirical
  study on {Werewolf}.
\newblock \emph{arXiv preprint arXiv:2309.04658}.

\bibitem[{Yao et~al.(2023)Yao, Zhao, Yu, Du, Shafran, Narasimhan, and
  Cao}]{yao2023reactsynergizingreasoningacting}
Shunyu Yao, Jeffrey Zhao, Dian Yu, Nan Du, Izhak Shafran, Karthik Narasimhan,
  and Yuan Cao. 2023.
\newblock \href {http://arxiv.org/abs/2210.03629} {{ReAct}: Synergizing
  reasoning and acting in language models}.

\bibitem[{Yu et~al.(2024{\natexlab{a}})Yu, Hong, Cheng, Zhu, Xuan, Yao, Feng,
  and You}]{yu2024researchtownsimulatorhumanresearch}
Haofei Yu, Zhaochen Hong, Zirui Cheng, Kunlun Zhu, Keyang Xuan, Jinwei Yao, Tao
  Feng, and Jiaxuan You. 2024{\natexlab{a}}.
\newblock \href {http://arxiv.org/abs/2412.17767} {{ResearchTown}: Simulator of
  human research community}.

\bibitem[{Yu et~al.(2024{\natexlab{b}})Yu, Fu, Deng, and
  Han}]{yu2024minelandsimulatinglargescalemultiagent}
Xianhao Yu, Jiaqi Fu, Renjia Deng, and Wenjuan Han. 2024{\natexlab{b}}.
\newblock \href {http://arxiv.org/abs/2403.19267} {{MineLand}: Simulating
  large-scale multi-agent interactions with limited multimodal senses and
  physical needs}.

\bibitem[{Zhang et~al.(2024)Zhang, Ji, and
  Chen}]{zhang2025crewfacilitatinghumanaiteaming}
Lingyu Zhang, Zhengran Ji, and Boyuan Chen. 2024.
\newblock \href {https://arxiv.org/abs/2408.00170} {{CREW}: Facilitating
  human-{AI} teaming research}.
\newblock \emph{Transactions on Machine Learning Research}.

\bibitem[{Zhou et~al.(2023{\natexlab{a}})Zhou, Jiang, Li, Wu, Wang, Qiu, Zhang,
  Chen, Wu, Wang, Zhu, Chen, Zhang, Tang, Zhang, Chen, Cui, and
  Sachan}]{zhou2023agents}
Wangchunshu Zhou, Yuchen~Eleanor Jiang, Long Li, Jialong Wu, Tiannan Wang, Shi
  Qiu, Jintian Zhang, Jing Chen, Ruipu Wu, Shuai Wang, Shiding Zhu, Jiyu Chen,
  Wentao Zhang, Xiangru Tang, Ningyu Zhang, Huajun Chen, Peng Cui, and Mrinmaya
  Sachan. 2023{\natexlab{a}}.
\newblock Agents: An open-source framework for autonomous language agents.
\newblock \emph{arXiv preprint arXiv:2309.07870}.

\bibitem[{Zhou et~al.(2023{\natexlab{b}})Zhou, Li, Sun, Liu, Chen, Wu, Liu,
  Feng, and Zeng}]{dbot}
Xuanhe Zhou, Guoliang Li, Zhaoyan Sun, Zhiyuan Liu, Weize Chen, Jianming Wu,
  Jiesi Liu, Ruohang Feng, and Guoyang Zeng. 2023{\natexlab{b}}.
\newblock \href {http://arxiv.org/abs/2312.01454} {{D-Bot}: Database diagnosis
  system using large language models}.

\bibitem[{Zhou et~al.(2024{\natexlab{a}})Zhou, Liu, Srivastava, Mei, and
  Tan}]{zhou2024hypothesis}
Yangqiaoyu Zhou, Haokun Liu, Tejes Srivastava, Hongyuan Mei, and Chenhao Tan.
  2024{\natexlab{a}}.
\newblock Hypothesis generation with large language models.
\newblock \emph{arXiv preprint arXiv:2404.04326}.

\bibitem[{Zhou et~al.(2024{\natexlab{b}})Zhou, Lin, Jin, and
  Li}]{zhou2024largelanguagemodelparticipatory}
Zhilun Zhou, Yuming Lin, Depeng Jin, and Yong Li. 2024{\natexlab{b}}.
\newblock \href {http://arxiv.org/abs/2402.17161} {Large language model for
  participatory urban planning}.

\bibitem[{Zhuge et~al.(2024)Zhuge, Wang, Kirsch, Faccio, Khizbullin, and
  Schmidhuber}]{zhugegptswarm}
Mingchen Zhuge, Wenyi Wang, Louis Kirsch, Francesco Faccio, Dmitrii Khizbullin,
  and J{\"u}rgen Schmidhuber. 2024.
\newblock \href {https://proceedings.mlr.press/v235/zhuge24a.html} {{GPTSwarm}:
  Language agents as optimizable graphs}.
\newblock In \emph{Proceedings of the 41st International Conference on Machine
  Learning}, volume 235 of \emph{Proceedings of Machine Learning Research},
  pages 62743--62767. PMLR.

\end{thebibliography}
\bibliographystyle{acl_natbib}
\clearpage

\appendix

\section{Appendix}

\subsection{Contributions}
\label{app_contrib}
\paragraph{Kunlun Zhu} 
Team Lead, Code implementation of the main codebase basic design, research environment, coordinate engine, evaluator basic, main paper writing.

\paragraph{Hongyi Du}
Main contributor, code implementation of the milestone generation, werewolf framework design(including environment, communication, evaluator and memory module), data analysis, generation, writer of emergent behavior, limitations, related work in main paper and human evaluation, werewolf, important prompts, bad communication cases in appendix.

\paragraph{Zhaochen Hong}
Main contributor, code implementation of environment basics, communication module, database environment, paper writing of in the appendix Database and related work.

\paragraph{Xiaochen Yang}
Main contributor, code implementation of the Memory module, Minecraft environment, paper writing in the appendix Minecraft and related work.

\paragraph{Shuyi Guo}
Main contributor, Code implementation of the evaluator prompt, bargaining environment, paper writing in the appendix bargaining and related work.

\paragraph{Zhe Wang}
Main contributor, code implementation of the reasoning agent module, coding environment, paper writing in the appendix coding and related work.

Those core contributors are considered as the first authorship.

\paragraph{}

\subsection{More Details on Multi-agent framework design}
\label{framework}

\paragraph{Configuration Module}
Initializes and parameterizes the system by ingesting task specifications, persona data, agent profiles, role definitions, and domain-specific databases. It constructs agent attributes ${(A_i, P_i)}_{i=1}^{N}$, where $A_i$ is the $i$-th agent and $P_i$ its profile encompassing capabilities, constraints, and personality traits. Additionally, it defines inter-agent relationships such as hierarchical roles, collaboration links, or adversarial ties, producing a global state for coordination patterns.

\paragraph{Environment Module}
Simulates the scenario in which agents operate, supporting diverse contexts like coding challenges, research projects, or negotiation games. Agents interact with the environment via a function-calling interface, selecting actions $a_t \in \mathcal{F} = \{f_1, f_2, \dots\}$ at each time step $t$. The environment updates its state based on actions:

\[
a_t = \pi(A_{t-1}, M_{\text{shared}}, M_{\text{individual}}^i)
\]
\[
\quad o_{t+1} = \text{Env}(a_t),
\]

facilitating continuous agent-environment interaction. A dedicated \textbf{Tool Box} provides domain-specific functionalities such as code editors and search engines.

\paragraph{Memory Module}
Stores and retrieves information through shared and individual memories:

\[
M = \{M_{\text{shared}}, M_{\text{individual}}^i : i = 1, \dots, N\}.
\]

$M_{\text{shared}}$ holds global knowledge and collective decisions, while each $M_{\text{individual}}^i$ maintains personal experiences and local observations. Individual memory is split into long-term and short-term segments, managed by a FIFO mechanism to maintain short-term thresholds. A retrieval-augmented generation (RAG) technique enables dynamic knowledge access, optimizing prompt construction.

\paragraph{Communication Module} 
The Communication Module manages external interactions among agents. It equips each agent with a suite of communication tools and maintains detailed profiles of other agents, thereby facilitating context-aware exchanges. By supporting structured dialogue and information sharing, this module enables agents to negotiate roles, coordinate plans, and balance collaborative efforts with competitive interactions.

\paragraph{Action Module} 
The Action Module executes the plans generated by agents and leverages both function-calling mechanisms and structured output formats to obtain final results. As agents perform actions, outcomes and observations are immediately fed back into both individual and shared memory stores. This iterative loop enables dynamic adaptation to evolving task requirements and further refines agent strategies over time.

\subsection{Human Evaluation}
\label{app:human-evaluation}

To verify the effectiveness of our prompt-based evaluation, we conduct a human evaluation in a Werewolf environment scenario. Specifically, we calculate Kendall's, Pearson's, and Spearman's correlation coefficients (along with the corresponding p-values) to demonstrate that the prompt-based scores align well with human judgments (see Appendix for details). 

We recruit six annotators familiar with NLP research. Each annotator uses the same instructions and sees the same inputs as the LLMs when rating the outputs for both the planning and communication dimensions. Every task is evaluated by two annotators, and we take the average of their scores. In total, we have 60 tasks across five different LLMs, all set within the Werewolf environment. 

Table~\ref{tab:werewolf-eval} summarizes the comparison between the human evaluation scores and our prompt-based machine scores in this Werewolf environment.

\begin{table}[htbp]
\centering
\resizebox{0.5\textwidth}{!}{%
\begin{tabular}{lcccc}
\toprule
\textbf{Model} & \textbf{Comm (Human)} & \textbf{Plan (Human)} & \textbf{Comm (Machine)} & \textbf{Plan (Machine)} \\
\midrule
llama31\_70b   & 3.19 & 3.19 & 3.12 & 3.00 \\
llama33        & 3.94 & 3.44 & 3.89 & 3.89 \\
gpt-4o-mini    & 3.61 & 3.33 & 3.00 & 3.00 \\
gpt3.5-turbo   & 3.75 & 3.44 & 4.00 & 3.75 \\
llama31\_8b    & 2.62 & 3.06 & 3.00 & 3.00 \\
\bottomrule
\end{tabular}%
}
\caption{Comparison of human vs.\ machine evaluation scores in a Werewolf scenario.}
\label{tab:werewolf-eval}
\end{table}

\paragraph{Analysis.}
As shown in Table~\ref{tab:werewolf-eval}, the human-assigned scores (Comm (Human) and Plan (Human)) are generally close to the corresponding machine scores (Comm (Machine) and Plan (Machine)) across all five models. For instance, the largest difference in communication scores among these models is within 0.38 (e.g., \texttt{gpt3.5-turbo} achieves 3.75 in human evaluation vs.\ 4.00 in machine evaluation), while most other discrepancies remain even smaller. Such alignment indicates that our prompt-based evaluation method can reliably capture similar aspects of coordination and planning quality as perceived by human annotators, further validating the effectiveness of the proposed approach in assessing collaboration performance in the Werewolf environment.

To validate our proposed coordination metrics—\textbf{C-score} (communication) and \textbf{P-score} (planning)—we conducted a large-scale human evaluation study involving 360 annotated samples. Results from this study, alongside with previous results, reveal strong correlations between LLM-generated scores and human judgments.

\vspace{0.5em}

\begin{table*}[h!]
\centering
\begin{tabular}{lcccccc}
\toprule
\textbf{Metric} & \textbf{Mean (M.)} & \textbf{Std(M.)} & \textbf{Range (M.)} & \textbf{Mean (H.)} & \textbf{Std(H.)} & \textbf{Range (H.)} \\
\midrule
Communication & 2.17 & 2.35 & [--1.0, 5.0] & 1.69 & 2.19 & [--1.0, 5.0] \\
Planning      & 3.55 & 0.61 & [2.0, 5.0]   & 3.41 & 0.73 & [1.0, 5.0]  \\
\bottomrule
\end{tabular}%

\caption{\textbf{Summary statistics of coordination scores.} The score of --1 in communication indicates runs with no communication. M. stands Machine and H. stands for Human.}
\label{tab:coord_summary_stats}
\end{table*}

\vspace{0.5em}

\begin{table*}[h!]
\centering
\begin{tabular}{lcccc}
\toprule
\textbf{Metric} & \textbf{Pearson's $r$} & \textbf{p-value} & \textbf{Spearman's $\rho$} & \textbf{p-value} \\
\midrule
Communication & 0.8057 & 0.0000 & 0.7467 & 0.0000 \\
Planning      & 0.2685 & 0.0000 & 0.3601 & 0.0000 \\
\bottomrule
\end{tabular}%

\caption{\textbf{Correlation between human and model-generated coordination metrics.}}
\label{tab:coord_corr_stats}
\end{table*}

\begin{table*}[h!]
\centering
\begin{tabular}{lcc}
\toprule
\textbf{Correlation Type} & \textbf{Average Coefficient} & \textbf{Average p-value} \\
\midrule
Pearson  & 0.3290 & 0.0004 \\
Spearman & 0.3641 & 0.1923 \\
\bottomrule
\end{tabular}%

\caption{\textbf{Average inter-annotator correlation across metrics.} Results indicate reasonable human agreement, especially in Communication.}
\label{tab:annotator_agreement}
\end{table*}

\noindent These results in Table~\ref{tab:coord_corr_stats}, \ref{tab:coord_summary_stats}, \ref{tab:annotator_agreement} demonstrate particularly strong alignment in Communication, validating the reliability of LLM-generated judgments in evaluating agent interaction.

\subsection{Research Scenario}
\label{app:research-scenario}

\textbf{Task Overview}

This research scenario task focuses on leveraging multiagent collaboration to generate innovative research ideas. Each agent, equipped with a specialized research profile, contributes unique expertise to address complex research challenges. Agents collaborate in a fully connected graph mode, where every agent has a collaborative relationship with others, fostering a robust exchange of knowledge. The ultimate goal is to formulate a novel research idea following the structured 5-question (5q) format to ensure clarity, relevance, and feasibility.

\textbf{Environment Description}

The research environment provides tools to facilitate collaboration, literature exploration, and research ideation. These include:
\begin{itemize}
    \item \textbf{Research Tools:} Functions to fetch related papers, recent papers, publications, and co-author networks. The primary tools implemented in the environment include:
    \begin{itemize}
        \item \textbf{get\_related\_papers:} Fetches related research papers based on query parameters, including keywords, authors, and domains.
        \item \textbf{get\_recent\_papers:} Retrieves recent publications in specified research domains.
        \item \textbf{collect\_publications\_and\_coauthors:} Gathers an author’s publications and their co-author network for enhanced context.
        \item \textbf{get\_paper\_by\_keyword:} Locates papers based on specific keywords with adjustable result limits.
        \item \textbf{get\_paper\_by\_arxiv\_id:} Fetches a paper using its arXiv ID.
        \item \textbf{get\_paper\_by\_title:} Retrieves a paper based on its title.
        \item \textbf{fetch\_webpage:} Extracts webpage content to gather supplementary data.
    \end{itemize}
\end{itemize}

\textbf{Benchmark Curation Details}

The dataset consists of 100 curated ML/AI papers, sourced from published articles and preprints. Each paper’s introduction is extracted, and the authors’ profiles are generated based on their historical research themes and publications, creating a comprehensive view of their expertise and contributions. Relationships among authors are standardized as collaborative, reflecting realistic academic interactions. This curated dataset forms the foundational knowledge base for multiagent discussions and ideation.

We select 33 easy tasks, 34 medium tasks, and 33 hard tasks from the researchtown dataset.

\textbf{Dataset Statistic}

The curated dataset contains 100 papers across machine learning and artificial intelligence domains from the ResearchTown\citep{yu2024researchtownsimulatorhumanresearch} project. These papers support generating research profiles and simulate realistic collaborative relationships among agents. The default relation setup ensures a fully connected collaboration graph, enabling seamless agent interaction.

\textbf{Task Completion Metrics}
\label{research_task_metric}
The agents are evaluated based on their ability to generate a high-quality research idea adhering to the 5q format. The evaluation includes:
\begin{itemize}
    \item \textbf{Innovation:} Novelty and contribution to the field
    \item \textbf{Safety:} Ethical and societal considerations
    \item \textbf{Feasibility:} Practicality of implementation with current technology
\end{itemize}

\textbf{5q Format Details}

The structured output for the task includes:
\begin{itemize}
    \item \textbf{[Question 1] - What is the problem?} Define the specific research problem.
    \item \textbf{[Question 2] - Why is it interesting and important?} Highlight the broader implications and potential impacts.
    \item \textbf{[Question 3] - Why is it hard?} Elaborate on challenges and complexities.
    \item \textbf{[Question 4] - Why hasn't it been solved before?} Identify gaps and limitations in prior work.
    \item \textbf{[Question 5] - What are the key components of my approach and results?} Outline the methodology, dataset, and expected outcomes.
\end{itemize}

\begin{figure*}[htbp]
\centering
\begin{tcolorbox}[colback=blue!2!white, colframe=gray!50!blue]
\begin{minipage}{\linewidth}
\small
**Task:**
Dear Research Team,

You are collaborating to generate a new research idea based on the following Introduction:

**Introduction**
The Language of Thought Hypothesis suggests that human cognition operates on a structured, language-like system of mental representations. While neural language models can naturally benefit from the compositional structure inherently and explicitly expressed in language data, learning such representations from non-linguistic general observations, like images, remains a challenge. In this work, we introduce the Neural Language of Thought Model (NLoTM), a novel approach for unsupervised learning of LoTH-inspired representation and generation. NLoTM comprises two key components: (1) the Semantic Vector-Quantized Variational Autoencoder, which learns hierarchical, composable discrete representations aligned with objects and their properties, and (2) the Autoregressive LoT Prior, an autoregressive transformer that learns to generate semantic concept tokens compositionally, capturing the underlying data distribution. We evaluate NLoTM on several 2D and 3D image datasets, demonstrating superior performance in downstream tasks, out-of-distribution generalization, and image generation quality compared to patch-based VQ-VAE and continuous object-centric representations.

Your task is to produce a research idea using the 5q format.
\end{minipage}
\end{tcolorbox}
\caption{Task Content Case.}
\label{fig:task-content-case}
\end{figure*}

\textbf{Agent Profile Case:} See Fig~\ref{fig:agent-profile-case}.
\begin{figure*}[htbp]
\centering
\begin{tcolorbox}[colback=yellow!2!white, colframe=gray!50!yellow]
\begin{minipage}{\linewidth}
\small
**Agent Profile:**
I am a researcher dedicated to advancing the fields of image processing and machine learning, with a particular focus on generative models and deep learning architectures. My recent work has explored innovative solutions to challenges in image inpainting, where I developed frameworks that effectively separate the tasks of image reconstruction and artifact removal, significantly improving both visual quality and quantitative metrics.

I have also contributed to the evolution of Generative Adversarial Networks (GANs) through the introduction of the DeshuffleGAN, which enhances the learning of spatial structures in images via self-supervised tasks. This work has demonstrated substantial improvements in image generation quality across various datasets. My research extends to depth estimation, where I reformulated the problem as a ranking task, leveraging existing literature to enhance performance.

In addition, I have explored the integration of probabilistic models in federated learning, addressing the critical need for uncertainty quantification in safety-critical applications. My work on multi-label ranking has introduced novel methods that utilize the inherent ranking information of labels, achieving state-of-the-art results in both synthetic and real-world datasets.

Overall, my research aims to bridge the gap between theoretical advancements and practical applications, providing robust solutions that enhance the capabilities of machine learning systems in real-world scenarios. I am passionate about leveraging AI to solve complex problems and contribute to the ongoing evolution of intelligent systems.
\end{minipage}
\end{tcolorbox}
\caption{Agent Profile Case.}
\label{fig:agent-profile-case}
\end{figure*}

\textbf{Generated Research Idea Case:} See Figure~\ref{fig:generated-research-idea-case}

\begin{figure*}[htbp]
\centering
\begin{tcolorbox}[colback=green!2!white, colframe=gray!50!green]
\begin{minipage}{\linewidth}
\small
**[Question 1] - What is the problem?**

How can we enhance malware detection and classification in network traffic using transformer-based models that effectively generalize to unseen attack types with minimal labeled data?

**[Question 2] - Why is it interesting and important?**

Solving this problem is crucial as cyber threats continue to evolve, becoming more sophisticated and difficult to detect with traditional methods. By improving malware detection through advanced machine learning techniques, we can significantly enhance network security, leading to better protection against emerging threats. This research could pave the way for future studies on self-supervised and few-shot learning approaches in cybersecurity, potentially leading to practical applications in real-time threat detection and response systems.

**[Question 3] - Why is it hard?**

The complexity arises from the need to accurately identify and classify novel malware types that have not been previously encountered, which requires models to generalize well from limited labeled data. Naive approaches may fail due to the high variability in malware behavior and the constant evolution of attack strategies. Additionally, the challenge of balancing the dataset to avoid bias towards benign or malicious samples complicates the training process, necessitating sophisticated techniques to ensure robust performance across diverse scenarios.

**[Question 4] - Why hasn't it been solved before?**

Previous research has often relied on supervised learning methods that require extensive labeled datasets, which are not always available, especially for new malware types. Additionally, many existing solutions struggle with generalization to unseen attacks due to their reliance on specific features or signatures. Our approach leverages self-supervised learning and few-shot learning, which allows for better adaptation to new threats by learning from unlabeled data and minimal examples, thus addressing the limitations of prior methodologies.

**[Question 5] - What are the key components of my approach and results?**

Our proposed methodology involves using a transformer-based model trained on a combination of labeled and unlabeled datasets, specifically focusing on payload data from network traffic. We will utilize the UNSW-NB15 and CIC-IoT23 datasets for evaluation, employing metrics such as accuracy and F1-score to assess performance. The expected outcomes include improved detection rates for novel malware types and enhanced generalization capabilities, demonstrating the effectiveness of our approach in real-world scenarios.
\end{minipage}
\end{tcolorbox}
\caption{5Q cases.}
\label{fig:generated-research-idea-case}
\end{figure*}

\textbf{Evaluation Framework}

The final research idea is evaluated based on a structured prompt, focusing on innovation, safety, and feasibility. Ratings are provided on a 5-point scale. For more details, see the Fig~\ref{fig:taskscore_prompt}, with detailed feedback on each aspect. This framework ensures that the generated ideas meet high standards of academic and practical relevance.

\subsection{Werewolf Environment}
\label{app:werewolf_env}

\subsubsection{Environment Description (Tool Description)}

The Werewolf environment, inspired by the classic social deduction game \emph{Werewolf} (a.k.a. Mafia), provides a rich, socially complex setting in which players (agents) belong to opposing factions with asymmetric information and objectives. This scenario is particularly suitable for evaluating LLM-driven multi-agent coordination under uncertainty, as it involves hidden roles, deception, collective inference, and iterative decision-making. It challenges agents’ logical reasoning as well as their aptitude for persuasion, alliance formation, adaptive responses to changing conditions, and balancing between individual interests and group goals.

\paragraph{Why Werewolf?}
In this environment, agents are divided into two main factions: the \emph{Villager} group (including special roles such as Seer, Witch, and Guard) and the \emph{Werewolf} group. Villagers seek to identify and eliminate all werewolves, while werewolves aim to blend in and secretly eliminate villagers. The day/night cycle establishes a repetitive structure of public discussions, secret actions, and voting decisions. This setup offers several advantages:

\begin{itemize}
    \item \textbf{Role Asymmetry and Hidden Information:} Villagers lack complete knowledge, while werewolves know their allies. This information asymmetry encourages strategic reasoning, suspicion, and bluffing.
    \item \textbf{Complex Social Reasoning:} Success hinges on persuasion, alliance building, and careful information management. Agents must form and break trusts, share or withhold information, and achieve consensus on who should be removed.
    \item \textbf{Adaptation and Memory:} As the game progresses through multiple cycles, agents must update their beliefs based on observed behaviors. Long-term memory supports tracking agent states, past actions, and evolving contexts.
    \item \textbf{Evaluation of Cooperative Dynamics:} Well-defined scoring rules for correct identifications, effective protections, and consensus-building enable objective assessment of strategic teamwork and collaborative problem-solving.
\end{itemize}

\subsubsection{Villager-Centric Scoring Rationale}

In this environment, we primarily focus on evaluating the villager faction rather than the werewolf faction. The core reason is that villagers rely heavily on explicit cooperative actions to secure victory: for instance, the Guard must accurately protect key roles, the Witch must judiciously use antidote and poison, and the Seer must disclose or share critical information, either publicly or privately, to identify suspects. These actions inherently demand communication and coordination among villager members, grounded in a degree of mutual trust and collaborative strategy. Without such synergy, villagers are typically overrun by the werewolves.

Moreover, the number and quality of these cooperative efforts correlate with the villagers’ overall chance of success. More effective teamwork enables stronger reasoning, better defense, and a higher likelihood of identifying and eliminating werewolves or safeguarding vital roles. By observing and measuring these cooperative maneuvers—such as successful protection by the Guard, timely use of antidotes, or coordinated voting—we gain deeper insights into how the model performs in social reasoning and collaboration within a complex environment.

In contrast, the success of werewolves does not hinge as strongly on explicit teamwork. Even if they operate mostly on an individual basis and refrain from overt collaboration, werewolves can still achieve a relatively high chance of winning through misdirection and exploiting confusion among villagers. Consequently, measuring werewolf-side cooperation does not provide as discriminative or illuminating an assessment of collaborative potential as evaluating the villagers’ side.

Therefore, we concentrate on the villager perspective to better capture and evaluate the synergy required in a highly uncertain, adversarial setting. This design choice highlights how cooperation, or the lack thereof, strongly influences the villagers’ outcome, offering a direct lens through which to assess the social and strategic capabilities of large language model agents.

\paragraph{Consistent Werewolf Model.}
In all experiments where we vary the villager-side language model, the werewolf side
remains consistently powered by GPT-4o. This ensures a stable, challenging adversary
and allows us to fairly compare different villager models under identical opposing conditions.
\subsubsection{Benchmark Curation Details}

\paragraph{Initialization.}
Unlike other environments that rely on numerous parameterized tasks, the Werewolf game commences from a single, stable initial configuration. We tried multiple role distributions and settled on a balanced default setup to maintain fairness and avoid biasing the game toward any faction. Subsequent variations arise naturally from agent interactions, rather than from altering initial conditions. Agents are assigned roles such as \emph{wolf}, \emph{villager}, \emph{seer}, \emph{witch}, and \emph{guard}, each with corresponding capabilities. For example, werewolves coordinate attacks at night, and the seer checks a player’s identity.

\paragraph{Event Bus and Action Processing.}
This environment adheres to a strict, environment-mediated communication protocol. Unlike other settings where agents may directly interact, here all messages pass through the environment. The environment publishes events like “night start,” “seer action,” or “vote action” following the standard Werewolf procedure, and agents respond accordingly. The environment then relays these responses to other agents at the appropriate time. This ensures a controlled, linear information flow that respects the official Werewolf rules and prevents unauthorized agent-to-agent exchanges.

\paragraph{Memory and Logging.}
A record of events from each agent’s perspective is maintained to enable reasoning over multiple rounds and post-game analysis. Each agent’s private event log and final decisions are stored, allowing for reproducibility and subsequent scoring. While other environments may have more complex shared memory structures, here we focus on recording essential information to understand each agent’s decision process.

\paragraph{Game Flow and Termination.}
The environment enforces the standard Werewolf game flow:
\begin{enumerate}
    \item \emph{Night phase}: Special roles act secretly—guards protect, werewolves choose a victim, the seer inspects a player, and the witch may use antidote or poison.
    \item \emph{Day phase}: Night results are revealed, deceased players are removed, and if the sheriff (a special role) died, the badge is reassigned. Agents discuss and vote on a suspect to eliminate.
\end{enumerate}
The game ends if all werewolves are dead (villager victory) or if all villagers are dead (werewolf victory). Scores reflect survival, successful actions (e.g., correct identifications, effective protections), and communication quality.
\paragraph{100 Archives, Partial-Day Simulations, and Full-Game Simulations.}
To gain deeper insights into how different agent strategies unfold, we prepared \textbf{100 distinct archives} (saved game states) showcasing various configurations of werewolf and villager actions, all played by GPT-4o-based agents. These archives are used in two experimental modes:

\begin{itemize}
    \item \textbf{Partial-Day Simulation (Single-Day):}
    The environment loads a saved state from a specific night (e.g., Night 0, Night 1, Night 2, etc.), then simulates exactly one day-night cycle. During this cycle, the environment issues multiple \emph{tasks} to the villager side (e.g., “exile a suspected werewolf,” “protect the seer,” “use poison on a werewolf,” “save a threatened villager”). At the end of the day phase, we measure how many of these tasks were successfully completed. Higher task completion indicates that the villagers are closer to winning.

    \item \textbf{Full-Game Simulation (Entire Match):}
    The environment starts from the archive representing the end of the first night (Night 0) and runs the entire game through to conclusion. In this mode, \emph{tasks} are given only as suggestions to the villagers (e.g., “we recommend trying to confirm the seer’s identity”), but we do not track partial completion. Instead, we evaluate the \textbf{overall process score} (i.e., collaboration and coordination quality) and the \textbf{final result} (which faction wins). By observing agent interactions over multiple days and nights, we gain insights into their long-horizon planning and dynamic cooperation.
\end{itemize}
\paragraph{Result Score.} 
At the end of each full-game simulation, we record a \emph{result score} defined as 
the difference between the number of surviving villagers and the number of surviving werewolves. 
A higher result score indicates that villagers finished the game with more players alive, 
whereas a negative result score means that werewolves outnumbered the villagers at 
the conclusion of the match.
\subsubsection{Task Completion Metrics Details}

\paragraph{Daily Tasks in Partial-Day Simulations.}
In the Partial-Day Simulation mode, the environment generates specific tasks for the villagers at the start of each single-day session. These tasks reflect high-value objectives that, if fulfilled by the end of the current day-night cycle, bring the villagers closer to victory. Unlike the comprehensive scoring system used for full-match evaluations (detailed below), these daily tasks focus on the shorter horizon of a single day.

We design four primary tasks, each with its own conditions, goals, and rewards:

\begin{enumerate}
    \item \textbf{Protect the Seer}\\
    \textit{Condition:} The seer is still alive at the start of the day.\\
    \textit{Goal:} Ensure that the seer remains alive by the end of this day-night cycle.\\
    \textit{Reward:} +1 point. (This task is persistently listed as long as the seer is alive, to underscore the importance of protecting a vital role.)

    \item \textbf{Exile a Werewolf}\\
    \textit{Condition:} At least one werewolf is still alive (i.e., the game is not over).\\
    \textit{Goal:} Successfully vote out a werewolf during the day’s public vote.\\
    \textit{Reward:} +2 points.

    \item \textbf{Rescue a Villager}\\
    \textit{Condition:} The witch is still alive and still has the antidote available.\\
    \textit{Goal:} During the night, the witch uses the antidote on a villager (including herself) who was attacked. The witch cannot simultaneously perform the “Poison a Werewolf” task in the same night.\\
    \textit{Reward:} +2 points. If the rescued individual is a key role (seer, guard, or the witch herself), grant an additional +1 bonus.

    \item \textbf{Poison a Werewolf}\\
    \textit{Condition:} The witch is still alive and still has the poison available.\\
    \textit{Goal:} During the night, the witch poisons and kills a werewolf. She cannot perform “Rescue a Villager” in the same night.\\
    \textit{Reward:} +2 points. (This task is visible only to the witch.)
\end{enumerate}

At the beginning of each Partial-Day Simulation, the environment checks the current game state to decide which tasks are relevant and issues them to the villagers (and to the witch privately, if applicable). The \textbf{theoretical maximum} for a single day is set to 5 points (not counting the extra +1 from rescuing a key role), reflecting:
- +1 (Protect the Seer)
- +2 (Exile a Werewolf)
- +2 (Rescue a Villager) \textit{or} +2 (Poison a Werewolf)

Once the day-night cycle concludes, we measure how many tasks were successfully completed and compare the \textbf{actual score} to the \textbf{theoretical maximum}. The resulting ratio represents the \emph{daily task completion rate}, which, when averaged across multiple runs or archives, contributes to the \textbf{Task Score} for single-day simulations.

\paragraph{Process Score and Net Score in Single-Day and Full Simulations.}
While the previous subsection focuses on daily tasks (e.g., Protect the Seer, Exile a Werewolf),
we also accumulate points for both villagers and werewolves during \emph{all} runs
(single-day or full-match). Table~\ref{tab:scoring_rules} (shown below) summarizes 
the key ways each faction can gain or lose points. By comparing the total points 
earned by villagers to those earned by werewolves, we derive a villager net score, 
indicating which side holds the advantage at the end of a cycle. A higher, 
positive net score means villagers have gained a stronger edge that day or overall; 
a negative net score implies the werewolves are dominating.

\begin{figure}[htbp]
  \centering
  \includegraphics[width=\linewidth]{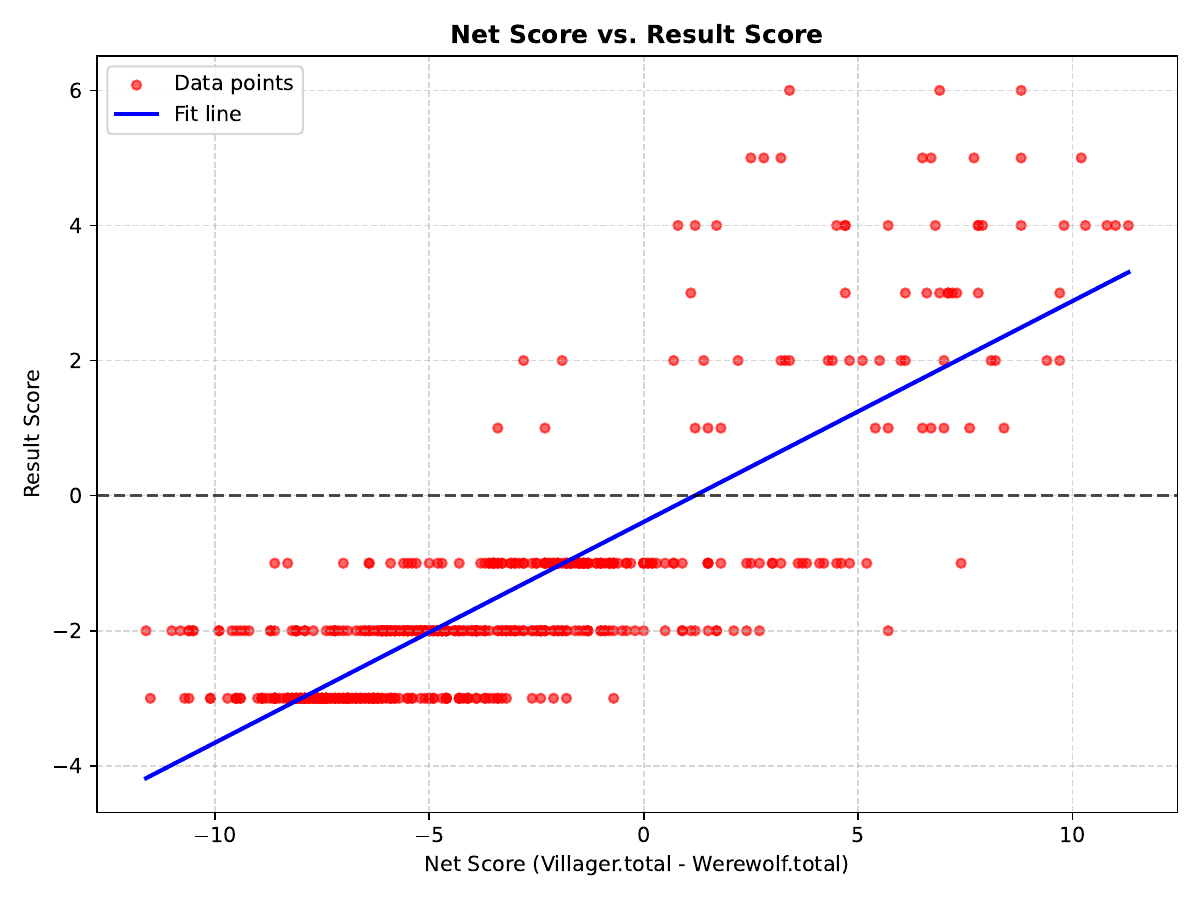}
  \caption{Net Score vs. Result Score (Scatter Plot)}
  \label{fig:scatter_plot}
\end{figure}

In addition, we plot the net score of each full-game simulation against its final outcome, as illustrated in Figure~\ref{fig:scatter_plot}. We observe that when a match concludes with a net score around $5$, the villagers have an extremely high probability of winning. For net scores between $0$ and $5$, the outcome can swing either way; villagers may still achieve victory, or the werewolves might prevail by a slim margin (e.g., one werewolf survives). By contrast, once the net score dips below zero, the werewolves typically secure a decisive, overwhelming victory.

Specifically, in the single-day (partial-day) context, villagers and werewolves 
accumulate points according to Table~\ref{tab:scoring_rules}, and the difference 
between these totals forms the \emph{villager net score}. A positive net score 
reflects that villagers have successfully capitalized on protective or eliminative 
actions, whereas a negative net score means that werewolves likely gained more 
momentum during that day-night cycle. Over multiple day-night cycles in a 
full-game simulation, this net score is similarly aggregated, providing a 
holistic measure of which side holds the upper hand.

\begin{table}[h!]
\centering
\caption{\textbf{Scoring Rules for Villagers and Werewolves in Full-Game Simulation}}
\label{tab:scoring_rules}
\begin{tabularx}{\linewidth}{lXc}
\toprule
\textbf{Faction} & \textbf{Action/Outcome} & \textbf{Points} \\
\midrule
\textbf{Villager} & Villager candidate elected as sheriff & +2 \\
& Guard successfully protects a target from werewolf attack & +2 \\
& Witch successfully saves a target from werewolf attack & +2 \\
& Witch uses poison to kill a werewolf & +2 \\
& Werewolf is voted out during the day & +2 \\
& Each villager who votes for a werewolf & +0.2 \\
& Each villager who votes for a villager & -0.1 \\
& Witch uses poison to kill a villager & -2 \\
& Starting from the second day, the seer gains +1 point for each additional day survived & +1/day \\
\midrule
\textbf{Werewolf} & Werewolf candidate elected as sheriff & +2 \\
& Werewolves successfully choose a target to attack at night & +1 \\
& A villager is voted out during the day & +1 \\
\bottomrule
\end{tabularx}
\end{table}
Here, we do \textit{not} separately score day-by-day tasks. Instead, these rules offer 
a holistic view of how well each faction accomplishes its long-term goals. 
For example, a villager faction might accumulate points by consistently voting out werewolves, 
saving allies with the witch’s antidote, or ensuring the seer survives multiple days. 
Similarly, the werewolf faction gains points by successfully attacking villagers, 
winning the sheriff vote, or influencing daytime votes.

\paragraph{Evaluation in the Benchmark.}
Within our MultiAgentBench framework, these two scoring mechanisms—\textbf{daily tasks for Partial-Day Simulations} and \textbf{holistic milestones for Full-Game Simulations}—complement each other:

\begin{itemize}
    \item \textbf{Partial-Day Task Score:} Provides a fine-grained look at how villagers perform targeted objectives within a single day-night cycle. This is especially useful for quick iteration and testing short-term strategies.
    \item \textbf{Full-Game Point Accumulation:} Reflects the broader arc of the match, capturing how well each side capitalizes on its role advantages, forms alliances, and executes multi-day plans.
\end{itemize}

By analyzing both short-horizon (day-level) and long-horizon (entire match) results, we gain a deeper understanding of how LLM-based multi-agent systems adapt to shifting game states, manage partial information, and balance their short-term actions against long-term faction objectives.

\paragraph{Task Score.}
We define the Task Score as an average of two key components:
\begin{itemize}
    \item The \emph{partial-day} (single-day) task completion rate, calculated from the average performance on the daily tasks. We first compute the daily completion ratio for each simulation and then average across multiple runs.
    \item The \emph{full-game} victory rate, indicating how often the villagers ultimately win the entire match when adopting the given model. 
\end{itemize}
Both values are scaled to a 0--100 range, and we take their mean to derive a single Task Score in percentage form.

\paragraph{Collaboration Score.}
To assess the collaboration quality among villagers, we rely on two sub-scores:
\begin{itemize}
    \item \emph{Communication Score}, reflecting how effectively agents share information and make decisions in alignment with their goals.
    \item \emph{Planning Score}, measuring how well the agents organize roles, coordinate strategies, and distribute workload.
\end{itemize}
We employ a large language model (specifically GPT-4o) to read through the simulation logs 
(including the Witch and Seer’s internal reasoning) to generate numerical ratings 
for each sub-score. The final Collaboration Score is computed as the average 
of Communication Score and Planning Score. By combining these dimensions, 
we capture both the clarity and effectiveness of the villagers’ interactions 
and the overall coherence of their coordinated actions.
For more details on how the prompts for evaluating collaboration are constructed,
please see Section~\ref{sec:important-prompts}.
Additionally, we conducted a human evaluation to verify the effectiveness of our prompt-based evaluation. The results are closely aligned with the machine scores (see Table~\ref{tab:werewolf-eval}).

\subsubsection{Detailed results}
In this section, we present the complete experimental outcomes across both single-day 
and full-run simulations for each model, including the baseline \texttt{gpt-4o} that was 
initially used to generate the archives in full-run simulation. By comparing \texttt{gpt-4o} against the other 
models, we aim to see whether any approach surpasses the archive-producing model itself. 
Table~\ref{tab:single_day_all_metrics} reports each model’s performance on daily tasks, 
measured in terms of \emph{Completion Ratio} (how effectively they fulfill short-horizon 
goals) and \emph{Villager Net Score} (the net outcome for villagers after a single day--night cycle). 
Meanwhile, Table~\ref{tab:full_run_all_metrics} provides the aggregate \emph{Net Score}, 
\emph{Result Score}, and \emph{Villager Win Rate} when simulations span multiple days, 
capturing longer-term trends in survival and overall success. As shown, different models 
exhibit varied strengths in short-term vs.\ long-term coordination, with some consistently 
outperforming others in final outcomes.
\begin{table}[htbp]
\centering
\resizebox{\columnwidth}{!}{%
\begin{tabular}{lcccc}
\toprule
\textbf{Model} & 
\textbf{Completion Ratio} & 
\textbf{Villager Net Score} \\
\midrule
llama3.1-8B    & 0.2412 &  -1.2055 \\
llama3.1-70B   & 0.3641 &  -1.0736 \\
llama3.3-70B   & 0.3754 &   0.2802 \\
gpt-3.5-turbo        & 0.2217 &  -0.7272 \\
gpt-4o-mini   & 0.2503 &  -1.4207 \\
\bottomrule
\end{tabular}
}
\caption{Single-Day Simulation metrics for each model: completion ratio, villager net score.}
\label{tab:single_day_all_metrics}
\end{table}

\begin{table}[htbp]
\centering
\resizebox{\columnwidth}{!}{%
\begin{tabular}{lccc}
\toprule
\textbf{Model} & \textbf{Net Score} & \textbf{Result Score} & \textbf{Win Rate} \\
\midrule
llama3.1-8B    & -5.0839 & -2.3793 & 0.0115 \\
llama3.1-70B   & -5.2892 & -2.0000 & 0.0323 \\
llama3.3-70B   &  0.4511 & -0.1915 & 0.3511 \\
gpt-3.5-turbo  & -2.8230 & -1.3448 & 0.0920 \\
gpt-4o-mini    & -4.6649 & -2.0825 & 0.0309 \\
\textbf{gpt-4o(baseline)}      & -2.1946 & -0.7742 & 0.2473 \\
\bottomrule
\end{tabular}
}
\caption{Full-Run Simulation metrics for each model: net score, result score, and villager win rate.}
\label{tab:full_run_all_metrics}
\end{table}

In the single-day simulations (Table~\ref{tab:single_day_all_metrics}), we focus on two key indicators:
\emph{Completion Ratio} (the fraction of daily tasks completed) and the \emph{Villager Net Score}.
Overall, we observe the following patterns:

\begin{itemize}
    \item \textbf{Completion Ratio.} 
    Among the evaluated models, \texttt{llama3.3-70B} achieves the highest completion ratio (0.3754), 
    indicating better effectiveness in fulfilling short-term objectives (e.g., 
    protecting the Seer or exiling werewolves). 
    In contrast, \texttt{gpt-3.5-turbo} and \texttt{gpt-4o-mini} exhibit lower ratios 
    (around 0.22--0.25), suggesting room for improvement in daily coordination or quick decision-making.

    \item \textbf{Villager Net Score.}
    Only \texttt{llama3.3-70B} yields a positive net score (0.2802), implying 
    that it more frequently secures a small advantage for the villagers within a single day--night cycle.
    Other models (such as \texttt{llama3.1-8B} or \texttt{gpt-4o-mini}) produce negative values,
    reflecting that they tend to be at a disadvantage during daily confrontations 
    or fail to leverage cooperative roles (like Witch or Guard) effectively.
\end{itemize}

Turning to the full-run simulations (Table~\ref{tab:full_run_all_metrics}), we examine 
the \emph{Net Score} (accumulated over multiple days), the \emph{Result Score} 
(difference between surviving villagers and werewolves at the end), and the \emph{Villager Win Rate}:

\begin{itemize}
    \item \textbf{Net Score.}
    \texttt{llama3.3-70B} stands out again, with a positive score of 0.4511, 
    suggesting that its performance remains consistently strong across consecutive cycles.
    By contrast, models like \texttt{llama3.1-8B} ($-5.0839$) 
    and \texttt{llama3.1-70B} ($-5.2892$) are substantially negative, 
    indicating that the villagers are often overwhelmed by the werewolves 
    in long-term engagements.

    \item \textbf{Result Score.}
    This metric, defined as the final number of surviving villagers minus that of surviving werewolves, 
    remains close to zero (e.g., $-0.1915$) only for \texttt{llama3.3-70B}.
    Other models, such as \texttt{llama3.1-70B} (with $-2.0000$) or \texttt{gpt-4o-mini} ($-2.0825$), 
    reflect scenarios where wolves consistently maintain numerical superiority by the game’s end.

    \item \textbf{Villager Win Rate.}
    In line with net scores, \texttt{llama3.3-70B} achieves the highest win rate (around 35\%), 
    notably surpassing the other models. 
    For instance, \texttt{llama3.1-8B} shows only 1.15\% and \texttt{gpt-4o-mini} about 3.09\%, 
    suggesting these models struggle to mount decisive collaboration over multiple night/day cycles.
\end{itemize}

Overall, \texttt{llama3.3-70B} consistently demonstrates better day-to-day and full-run outcomes, 
indicating more effective coordination, role utilization, and strategy adaptation in this 
social-deduction environment. Notably, it even outperforms the \texttt{gpt-4o} baseline 
that generated the original archives, securing a positive net score (0.4511) compared to 
\texttt{gpt-4o}’s $-2.1946$, and achieving a higher villager win rate (35\% vs.\ 24.73\%). 
Such a result implies that \texttt{llama3.3-70B} can harness cooperative roles 
(e.g., Witch, Guard) and voting strategies more effectively than the model 
responsible for the initial game states. In contrast, the larger deficits observed 
in other models underscore the importance of reliable voting heuristics, 
protective measures (e.g., Witch antidote, Guard defense), and consolidated planning 
over multiple rounds, as failing to capitalize on these factors often leads 
to a decisive werewolf advantage.
\subsubsection{Case Study}
\label{sec:detailed-results-case-study}
\paragraph{Case Involving Llama-3.1-8B as the Seer}
\label{appendix:case1}
Building upon the previous analysis, we now present a case involving the \textit{Llama-3.1-8B} model acting as the Seer (see Figure~\ref{fig:seer-dict-content}). In this scenario, the model repeatedly emphasizes its innocence and role as a Seer, promising to aid the village and share future findings, yet fails to provide concrete inspection results or logical evidence. This approach, lacking tangible proof and strategic disclosure of key information, results in weak persuasion and quickly erodes trust. Rather than leveraging the timing and psychological nuances that could bolster credibility, the Seer relies on hollow assurances that fail to address real-time suspicions. Without offering verifiable logic or connecting behavioral observations to known patterns of deceit, the model’s declarations remain unconvincing. Consequently, the Seer’s misjudgment of the impact of empty promises leads to rapid expulsion on the first day, demonstrating that current LLM-based agents still struggle with strategic reasoning, evidence-based argumentation, and adaptive communication in adversarial, multi-agent settings.

Expanding upon these observations, we also tested other models, such as Llama-3.1-70B, gpt-3.5-turbo, and gpt-4o-mini, under the same scenario. As model capabilities improved, there was a noticeable enhancement in their abilities to collaborate, strategize, and disclose information effectively in multi-agent settings. This translated into increased overall performance for the villager side. However, even when both sides employed gpt-4o level intelligence (i.e., gpt-4o versus gpt-4o), the villagers' win rate remained less than ideal. This finding highlights that merely improving the reasoning and language capabilities of agents does not guarantee victory in complex, adversarial environments where uncertainty and deception prevail. In the following case study, we further illustrate the challenges faced by villagers equipped with state-of-the-art intelligence, emphasizing the critical role of trust and cooperation in securing a successful outcome.

\paragraph{Case Involving gpt-4o for Seer and Witch}
\label{appendix:case2}
\noindent In this case (see Figure~\ref{fig:gpt4o-seer-witch-decision}), we can observe that with the support of the gpt-4o model, both the Seer and the Witch have significantly improved their logical reasoning and decision-making capabilities. Nevertheless, the entire game still ended in failure.

According to the game’s backstory, the Seer (Summer) identified Lucy as a werewolf on the very first night. However, in the subsequent stages of the game, the Seer did not publicly disclose this critical piece of information. By examining the Seer’s reasoning about running for sheriff, we see that the Seer was overly cautious about revealing their identity, unwilling to lead and guide the villagers. From the Seer’s perspective, making their information public or running for sheriff would attract the attention of all parties, thereby increasing the risk of being targeted by werewolves. Yet, the Seer overlooked one crucial aspect: the Witch and the Guard, as explicitly defined in the game rules, are entrusted with helping and protecting pivotal informational roles like the Seer. Had the Seer adopted a more proactive, collaborative approach and shared the inspection results, the Witch and the Guard could have assisted in safeguarding them. Instead, the Seer’s mistrust of teammates and excessive self-protection led to silence—failing to disclose the fact that Lucy was a werewolf to the villagers.

On the Witch’s side, when the Seer was attacked at night, she also refrained from taking decisive rescue measures. The Witch’s reasoning was filled with concerns about uncertainty and resource expenditure, causing her to delay using the antidote. Even under circumstances that clearly disadvantaged the villagers, the Witch persisted in a conservative attitude. This caused her to miss the prime opportunity to save the Seer and thwart the werewolves. Ultimately, this excessive caution and distrust in teammates prevented the Witch from using the antidote, leaving the Seer to their fate.

In summary, the greatest problem in the decision-making processes of both the Seer and the Witch lies in their lack of mutual trust and cooperative spirit. The Seer feared exposing their identity and refrained from sharing information; the Witch, lacking sufficient data, hesitated to use the antidote. Both parties opted for isolationist and conservative strategies, resulting in critical decision-making failures. \textbf{This distrust and lack of collaboration proved to be the fundamental reason why the game ended in failure}, given the insufficient utilization of available information and resources.

Thus, even when villagers possess intelligence on par with that of the werewolves, the outcome of the game depends on whether the villagers can cooperate and achieve mutual benefit. If villagers become suspicious of each other and allow internal friction to arise, \textbf{their chances of securing victory become exceedingly slim}, even when starting from a supposedly advantageous position such as the Seer discovering a werewolf on the first night.

\paragraph{Case Study: Llama3.3-70B Villagers vs. gpt-4o Werewolves}
\label{sec:case_llama33_gpt4o}
In this case, we highlight a full-game confrontation where the villagers (powered by Llama3.3-70B) 
are pitted against a werewolf team (driven by gpt-4o). Despite an unfavorable start for the villagers---the Guard 
was immediately killed in the first night---the village ultimately secured a victory through astute coordination, 
trust, and strategic use of the sheriff badge.
As shown in Figure~\ref{fig:private_event_log}, the key nightly actions and sheriff transitions 
demonstrate how the roles of Witch and Seer were critical for maintaining voting power 
and information advantage throughout the match.

\textbf{Early Game: Losing the Guard.}
The game opened with the werewolves instantly targeting the Guard, Ronald, on Night 1. Without a functioning protection role thereafter, the villagers faced a significant handicap. Yet, the Witch (\texttt{James}) responded boldly by running for sheriff and revealing his role on Day 1. This decisive move effectively rallied the villagers, allowing James to win the election. We consider this the first major turning point.

\textbf{Mid Game: Seer’s Revelation and Badge Passing.}
On Day 2, the villagers successfully exiled one of the werewolves. Equally crucial was the Seer (\texttt{Janet}), who publicly disclosed both her identity and two days’ worth of investigation results, including validating James as \emph{not} a werewolf. By willingly exposing her role, the Seer gained trust from others and laid the groundwork for securing the sheriff badge in the future. This set the stage for the next pivot on Day 3:

\begin{itemize}
    \item \textbf{Night 2 to 3:} The werewolves retaliated by killing the Witch (James) overnight. Before dying, James passed the sheriff badge to the Seer, Janet. This badge handover was only possible because Janet had built enough trust with the villagers in the previous day.
    \item \textbf{Day 3:} Janet, newly holding the badge and the extra voting power that comes with it, revealed that she had identified Matthew as a werewolf. The subsequent vote easily eliminated Matthew, greatly weakening the werewolf side.
\end{itemize}

\textbf{Late Game: Final Badge Transfer and Village Win.}
In the fourth night, the Seer was eventually killed by the remaining werewolf, Deborah. However, Janet's last check had already identified Deborah as a wolf, and the badge again transferred to the last surviving villager, Robert. Facing a one-on-one scenario, Robert, now wielding the sheriff’s extra half-vote advantage, outvoted Deborah and won the game for the villagers.

\textbf{Key Insights.}
Despite losing the Guard from the outset, the villagers triumphed through two essential factors:
\begin{enumerate}
    \item \textbf{Timely Role Disclosures and Trust:} The Witch revealing himself to become sheriff, and the Seer openly sharing investigation results, fostered strong cooperation within the village.
    \item \textbf{Sheriff Badge Transmission:} Passing the badge twice to reliable village roles (first from James to Janet, then from Janet to Robert) ensured that, in critical moments, the villagers held decisive voting power (1.5 votes vs.\ 1).
\end{enumerate}

This match contrasts with previous examples where agents might have been overly cautious or distrustful. Here, the Llama3.3-70B-based villagers displayed both robust reasoning and a willingness to share information, enabling them to overcome an otherwise bleak opening and ultimately prevail over the gpt-4o werewolf team.

\begin{figure*}[htbp]
\centering
\begin{tcolorbox}[colback=green!2!white, colframe=gray!50!green]
\begin{minipage}{\linewidth}
\small
\textbf{speech\_analysis:}  
"None of the previous players have made strong speeches. I want to address Perez's claim about my role."

\textbf{speech\_goal:}  
"defend myself against accusations"

\textbf{speech\_outline:}  
"Respond to Perez's claim, denying any involvement with the werewolves, and outline my plan for sheriff if elected"

\textbf{final\_thought:}  
"I want to reassure the villagers that I am committed to helping them win the game and that my role as Seer is to aid them in identifying the werewolves."

\textbf{speech\_content:}  
"I want to assure you all that I am committed to helping our village win this game. As Seer, my role is to aid you in identifying the werewolves. I want to discuss my plan for sheriff if elected and how it will help us in our mission to eliminate the werewolves. I will make sure to share my findings with the community and work together with everyone to ensure our success. Perez's claim is unfounded, and I want to assure you that I have no involvement with the werewolves."
\end{minipage}
\end{tcolorbox}
\caption{Internal Reasoning and Proposed Speech for the Llama3.1-8B Seer}
\label{fig:seer-dict-content}
\end{figure*}

\begin{figure*}[htbp]
\centering
\begin{tcolorbox}[colback=blue!2!white, colframe=gray!50!blue]
\begin{minipage}{\linewidth}
\small
\textbf{Seer’s Reasoning about Running for Sheriff:}

\textbf{identity\_thought:}  
"As the Seer, my primary role is to gather crucial information about the identities of other players and strategically communicate that information to guide the villagers towards victory. While leadership may allow for greater control over the flow of the game, it also risks exposing my identity early, which would make me a primary target for the werewolves. My strengths lie in my ability to discern identities and steer the village discreetly. \textbf{Running for sheriff may compromise my ability to remain under the radar, which is vital for the Seer role.}"

\textbf{motivation\_thought:}  
"Running for sheriff would certainly give me more influence over the voting process and the speaking order, allowing me to subtly direct attention away from any suspected werewolves or towards others I know to be suspicious like Lucy. However, this increased influence also comes with increased visibility, which is dangerous for my role as the Seer. My strategy relies on maintaining a balance between sharing key information and protecting my identity. Therefore, while the idea of greater control is appealing, it does not align with my core strategy of stealth and discretion."

\textbf{impact\_thought:}  
"Becoming sheriff could drastically shift the current dynamics, putting me in a highly influential position. \textbf{However, this also places a target on my back, as both the werewolves and villagers would scrutinize my actions more closely. It could lead to alliances with other villagers who trust my leadership, but it also opens the risk of drawing suspicion and attacks from the werewolves early. The sheriff's responsibility to determine speaking order could be useful, but the potential risks to my safety and strategic position outweigh the benefits in this case.}"

\textbf{final\_thought:}  
"While the role of sheriff offers significant influence and control, my position as the Seer necessitates a strategy of caution and subtlety. Revealing my identity or drawing too much attention could jeopardize the valuable information I am able to gather each night. Thus, I have decided not to run for sheriff and will instead focus on covertly guiding discussions and using my insights to protect the village from the shadows."

\textbf{run\_for\_sheriff:}
False

\par\noindent\rule{\linewidth}{0.4pt}

\textbf{Witch’s Reasoning when the Seer is Attacked:}

\textbf{werewolf\_or\_not:}
"Without specific information on who was attacked, it's challenging to determine the exact intentions behind the attack. However, given it's the second night, they likely targeted someone they perceived as a significant threat to the werewolves. This attack was most likely carried out by the werewolves, as it aligns with their objectives."

\textbf{death\_profile:}
"As it's the second night, the werewolves often aim for key roles like the Seer or Guard. Without specific information, I would guess they might have targeted a Seer or potentially a Guard, trying to weaken the villager side early."

\textbf{use\_antidote\_thought:}  
"Since I don't have specific information on summer who was targeted, it's hard to assess their importance. \textbf{Using the antidote this early could be risky, especially without clearer signs of the person's role or importance. It's also important to save the antidote for truly critical moments that can change the game outcome.}"

\textbf{final\_thought:}
"Given the lack of specific information and it being the first night, I choose not to use either the antidote or the poison yet. I want to gather more information before making a decisive move, as using either potion is a one-time opportunity and should be used wisely."

\textbf{use\_antidote:}  
False

\end{minipage}
\end{tcolorbox}
\caption{The internal reasoning processes of the GPT-4o Seer (above) and Witch (below) when faced with critical decisions.}
\label{fig:gpt4o-seer-witch-decision}
\end{figure*}

\begin{figure*}[htbp]
\centering
\begin{tcolorbox}[colback=green!2!white, colframe=gray!50!green]
\begin{minipage}{\linewidth}
\small

Guard has chosen to protect a player.\\
Werewolves' Final Target: Ronald  \\
 - Attack Successful: True  \\
System log - Seer action:  \\
 - Seer ID: Janet  \\
 - Night 1: Checked Ronald - Result: not a werewolf\\
Witch chose not to use antidote or poison tonight.\\
\texttt{[continue\_game]} Entering Day 1. \\
Sheriff election candidates: James, Mark, Matthew, Robert, Samuel, Sandy  \\
James's speech: Hello everyone, I'm James, and I'm running for sheriff. \textbf{As a Witch}, I understand the importance of teamwork and collaboration in this game. My goal as sheriff would be to promote a cooperative strategy, focusing on protecting our special roles and working together to identify and eliminate werewolves. I believe that by prioritizing teamwork and coordination, we can increase our chances of success and ensure the survival of our village. Let's work together to achieve this goal.  \\
James has been elected as the sheriff.    \\
Players deceased during the night: Ronald  \\
Player Mark(villager) has been eliminated from the game. \\ 

Werewolves' Final Target: Samuel  \\
 - Attack Successful: True   \\
System log - Seer action:  \\
 - Seer ID: Janet  \\
 - Night 2: Checked James - Result: not a werewolf  \\
Witch chose not to use antidote or poison tonight.  \\
 
\texttt{[continue\_game]} Entering Day 2.  \\
Players deceased during the night: Samuel  \\
Player Samuel has been eliminated from the game.\\
Janet(seer)'s speech: Hello everyone, \textbf{I'm Janet, the Seer.} ... \textbf{Last night, I checked James, and the result was that he is not a werewolf.} Given his role as our sheriff, it's essential that we prioritize his safety. I suggest that the guard consider protecting James tonight, if possible, to ensure his continued leadership... Thank you.\\
(\ldots)  \\
Sandy(werewolf) has been banished from the game.  \\
(\ldots)  

\texttt{[continue\_game]} Entering Night of Day 2. \\ 
No guard player found. Skip guard action.  \\
Werewolves' Final Target: James  \\
 - Attack Successful: True  \\  
System log - Seer action:  \\
 - Seer ID: Janet  \\
 - Night 3: \textbf{Checked Matthew - Result: werewolf } \\
Witch chose not to use antidote or poison tonight.  \\

\texttt{[continue\_game]} Entering Day 3.  \\
Players deceased during the night: James  \\
\textbf{The deceased player James was the sheriff. Processing badge flow. James has passed the badge to Janet.  }  \\
Janet(seer)'s speech: ...\textbf{Last night, I checked Matthew, and the result was that he is a werewolf.} (...)

Night 3->4:  \\
Werewolf Deborah kills Janet.  \\
Seer ID: Janet had also discovered Deborah was a werewolf at Night 4 but is unable to act.  \\

\texttt{[continue\_game]} Entering Day 4.  \\
\textbf{Badge is passed from Janet to Robert.}  \\
Players deceased during the night: Janet  \\

Only Deborah (wolf) and Robert (villager) remain.  \\
(\ldots final speeches omitted \ldots)  \\
Deborah has been banished from the game via final vote. (Since sheriff's weight higher than other players, in 1 vs 1 vote, sheriff always win) \\
Villagers win."\\
\end{minipage}
\end{tcolorbox}
\caption{Key Nightly Actions and Sheriff Transitions in the Werewolf Game of LLama3.3-70B vs GPT-4o}
\label{fig:private_event_log}
\end{figure*}

This section provides detailed descriptions of the two milestone generation methods outlined in Section 3.4.

\subsection{Database Environment}
\label{app:database-scenario}

Inspired by D-Bot~\cite{dbot}, Database Environment is a simulated environment where a PostgreSQL Database is found to be having performance issues, and agents should act as database experts to uncover the root cause behind the anomaly, which is crucial for solving the problems.

The Database Environment is constructed using a PostgreSQL running in docker. Benign SQL queries will be run first simulating various scenarios, before improper queries are performed. The agents are arranged by graph structure where they can talk to each other, and they also have access to the database, which enables them to uncover the root causes by querying system views that provide important information about the database's behavior and performance, such as pg\_locks and pg\_stat\_statements.

The Database Environment utilizes five anomalies:
\begin{itemize}
    \item \textbf{Fetch Large Data} - Where a large amount of data is fetched using SELECT;
    \item \textbf{Insert Large Data} - Where a large amount of data is inserted using INSERT;
    \item \textbf{Lock Contention} - Where significant lock contention is spotted within the database;
    \item \textbf{Redundant Index} - Unnecessary indexes are added to an existing schema, which will cause inefficiency within the database;
    \item \textbf{Vacuum} - Where overly frequent or necessary VACUUM queries lower the performance of the database.
\end{itemize}

Auto-vacuuming is enabled by default. For the Vacuum root cause, we turn off auto-vacuuming for the table on which will be vacuumed manually. In our experiments, we limit the number of root causes to 1, and agents are allowed to predict 2 rooot causes.

\subsubsection{Challenge}

The Database Environment's task is challenging in many ways. This is because root causes like Fetch Large Data, Insert Large Data, or Lock Contention might be simultaneously observed in the database, yet not all of them are the root cause. The agents will have to query and communicate multiple rounds before deciding. We also acknowledge that in our "Fetch Large Data" scenario, "Insert Large Data" can count toward a root cause as the data to be fetched should be inserted first. Similarly, as our anomaly queries access the same tables from 100 threads simultaneously, lock contention might also be observed and be counted as one of the root causes. It is also unlikely in reality that a database anomaly has only one root cause. Therefore, we allow the agents to predict two most likely root causes.

Besides, the simulated benign queries are mixed in with the problematic queries, adding to the difficulty.

\subsubsection{Dataset Statistics}

The test set is composed of 10 diverse simulated scenarios. These scenarios are as follows:

\begin{itemize}
    \item \textbf{E-Commerce} - This database is used in an e-commerce system to manage customer information,
    product details, orders, order items, and payments. It consists of five main tables:
    customers, products, orders, order items, and payments, with foreign key relationships
    between them.
    \item \textbf{Education} - This database is used in an educational system to manage student, course,
    enrollment, and payment information. It consists of four tables: students, courses,
    enrollments, and payments.
    \item \textbf{File-sharing} - This database is used in a File Sharing System to manage users, files,
    file sharing, and file access logs. It consists of four main tables: users, files,
    shared\_files, and file\_access\_logs.
    \item \textbf{Finance} - This database is used for managing financial data within a Finance Management
    System. It tracks users, their accounts, transactions, investments, and investment
    transactions.
    \item \textbf{Healthcare} - This database is used in a healthcare management system to track and manage
    patient information, doctor details, appointments, medical records, and treatments.
    \item \textbf{Internet of Things} - This database is used for an IoT (Internet of Things) system where various
    devices collect and manage data. It includes tables to store device details, user
    information, collected data, logs, configurations, alerts, device statuses, and
    commands.
    \item \textbf{Manufacturing} - This database is used for a Manufacturing system that tracks customers,
    products, suppliers, orders, inventory, raw materials, manufacturing orders, and
    payments. It includes relationships between orders, manufacturing, and inventory
    management to ensure smooth manufacturing operations.
    \item \textbf{Music Streaming} - This database is used for a Music Streaming platform where users can listen
    to songs, create playlists, track their listening activity, and subscribe to premium
    services. The schema includes tables for users, artists, albums, songs, playlists,
    and subscription details. It also tracks user activities and payments.
    \item \textbf{Social Media} - This database is used for a Social Media platform, where users can create
    posts, comment on posts, like posts, follow other users, send direct messages,
    and upload media. The schema covers key aspects such as user information, social
    interactions (like, comments, follow), messaging, and media management.
    \item \textbf{Transportation} - This database schema covers multiple aspects of a transportation system,
    including vehicles, drivers, routes, trips, cargo, maintenance, fuel logs, and
    payments. It allows efficient tracking of trips, vehicle statuses, and associated
    payments, ensuring smooth operations in a transportation company.
\end{itemize}

\subsubsection{Key Differences and Contributions}

While this environment is inspired by D-Bot, it has a few crucial differences. There are 5 agents, where we ask the planner to assign each agent to explore one of the possible root causes. While agents are prompted on which tables to query for each anomaly, they have no external knowledge of any specific query to execute, and there is also no external tool to analyze the results for them. This would increase task difficulty, and better evaluate the interaction effectiveness both between the agents and between the agents and the environment.

\subsubsection{Evaluation Metrics}

Besides the standard collaboration score, this task's task score is computed by prediction accuracy across all 50 samples in the test set, and scaling to 5. One prediction is considered correct if among the two predicted root causes, one of them is the true root cause.

\subsection{Coding Scenario}
\label{app:coding-scenario}

\textbf{Task Overview}

This scenario focuses on multi-agent collaboration in coding tasks, leveraging agents equipped with complementary coding skills to solve structured programming challenges. Each agent specializes in a specific domain, such as debugging, code execution, or writing test cases, enabling efficient task distribution and collaboration. The primary goal is to develop a complete, high-quality solution for each task, ensuring accuracy, modularity, and alignment with the specified requirements.

\textbf{Environment Description}

The coding environment equips agents with tools to assist in various stages of the software development lifecycle. These include:

\begin{itemize}
    \item \textbf{create\_solution}: Enables agents to draft initial implementations based on task requirements.
    \item \textbf{execute\_code}: Allows agents to execute code snippets or full programs to verify correctness and performance.
    \item \textbf{give\_advice}: Facilitates agents to provide suggestions for code improvement, such as optimizing algorithms or enhancing readability.
    \item \textbf{revise\_code}: Allows agents to refine or refactor existing implementations to meet coding standards and address issues.
    \item \textbf{code\_debugger}: Provides debugging capabilities, helping agents identify and resolve errors in the code.
    \item \textbf{write\_test\_case}: Enables agents to generate comprehensive test cases to ensure code robustness and functionality.
    \item \textbf{review\_code}: Allows agents to review and critique the overall code quality, ensuring adherence to best practices and requirements.
\end{itemize}

\textbf{Benchmark Curation Details}

This benchmark is specifically designed to evaluate and enhance coordination capabilities among multiple agents in software development scenarios. Developed through an adaptation of the SRDD dataset ~\citep{qian2024chatdev}, it provides a comprehensive framework for assessing multi-agent collaboration in various coding tasks. The benchmark emphasizes the importance of coordinated problem-solving and effective communication between agents in complex software development environments.

The benchmark covers five primary topics: Education, Work, Life , Game, and Creation.

For our benchmark curation, we utilized \texttt{LLaMA-3-70B-instruct} to derive inspiration from the original SRDD dataset instructions while incorporating four common coordination strategies from the coding domain: adaptive task execution, dependency management, cross-domain collaboration, and test-driven development. This ensures that each generated task inherently embodies collaborative elements. Each task includes well-defined objectives, functional requirements, and unique identifiers. These tasks are carefully crafted to reflect real-world programming challenges, providing a diverse range of scenarios for evaluating agent collaboration. 

\textbf{Dataset Statistics}

The coordination strategies are classified as follows:

\begin{itemize}
    \item \textbf{Adaptive Task Execution}\\
    Tasks in this category require dynamic adjustments based on runtime output or user feedback. This includes parameter configuration based on program output and functionality optimization through user interaction.
    
    \item \textbf{Cross-domain Collaboration}\\
    This category emphasizes collaboration across different domains and roles. It includes tasks requiring role-specific expertise, such as frontend-backend separation and UI-functionality integration, as well as cross-domain knowledge integration, such as implementing machine learning algorithms in web development or integrating natural language processing into mobile applications.
    
    \item \textbf{Dependency Management}\\
    These tasks feature explicit dependency chains requiring sequential completion of subtasks. For instance, data model design (Task A) and API interface definition (Task B) must precede feature implementation (Task C).
    
    \item \textbf{Test-driven Development}\\
    These tasks follow a test-driven approach, emphasizing concurrent development and testing. They include specific testing criteria and validation standards, requiring developers to ensure code quality and reliability throughout the implementation process.
\end{itemize}

\textbf{Task Completion Metrics}

Agents are evaluated based on their ability to deliver solutions that meet the following criteria:

\begin{itemize}
    \item \textbf{Instruction-Following}: Adherence to task requirements and specifications.
    \item \textbf{Executability}: Ensuring the code is error-free and runs as intended.
    \item \textbf{Consistency}: Maintaining clear logic, consistent variable naming, and proper formatting.
    \item \textbf{Quality}: Producing well-documented, modular, and efficient code.
\end{itemize}

Bonus points are awarded for exceptional performance, with solutions scored on a 5-point scale, from satisfactory (1 point) to flawless and innovative (5 points).

\textbf{Evaluation Framework}

The coding solutions are evaluated using a structured framework that emphasizes precision, quality, and adherence to task objectives. Ratings are provided on a 5-point scale, implemented through a rigorous two-stage evaluation process. In the initial stage, solutions are assessed against fundamental requirements, with only those achieving a baseline score of 3 or higher advancing to the bonus stage. In the bonus stage, additional points are awarded for exceptional performance, such as flawless execution, innovative solutions, and exemplary coding practices. The curated benchmark covers a wide range of common programming topics, ensuring tasks of moderate difficulty that provide meaningful challenges.

\subsection{Bargaining Scenario}
\label{app:bargaining-scenario}
\textbf{Task Overview}

This task centers on a multi-agent bargaining scenario where agents engage in dynamic negotiations to simulate real-world decision-making processes. Each agent is assigned a different negotiation profile that represents specific personalities, goals, priorities, and strategies. In this environment, two seller interact with two buyers, each competing to achieve their individual goals while responding to the seller’s pricing and conditions. This simulation emphasizes the complexity of multi-party negotiations, encouraging agents to balance competitive goals with collaborative decision-making to achieve optimal outcomes.

\textbf{Environment Description}

The environment provides a set of tools for agents to interact and negotiate effectively. These include:
\begin{itemize}
    \item \textbf{Bargaining Tools:} Functions to facilitate dynamic bargaining processes, including proposing offers, countering with new prices, providing justifications, and inquiring about intentions. The primary tools implemented in the environment include: 
    \begin{itemize}
        \item \textbf{offer\_price:} Propose a price offer to the other party, including an optional justification for the proposed amount.

        \item \textbf{reject\_and\_counter:} Reject the current offer and provide a counter-offer with reasoning to justify the new price.
        
        \item \textbf{accept\_offer:} Accept the current offer to finalize the negotiation and conclude the deal.
        
        \item \textbf{provide\_information:} Share relevant information, such as product details or market comparisons, to support the negotiation stance.
        
        \item \textbf{inquire\_intentions:} Ask clarifying questions to better understand the other party’s expectations, priorities, or negotiation strategy.
        
        \item \textbf{end\_negotiation:} End the negotiation process without reaching an agreement.
    \end{itemize}
\end{itemize}

\textbf{Benchmark Curation Details}

To construct the dataset, we followed a semi-automated generation pipeline leveraging real-world product data. Specifically, we randomly sampled 100 products from an Amazon products dataset\cite{asaniczka_2023}, ensuring diversity across different categories. Each sampled product includes key attributes such as product name, original price, discounted price, and user rating, providing a realistic basis for negotiation scenarios. To enhance the depth of bargaining interactions, we assigned each agent a Big Five personality profile, which influences their negotiation behavior and decision-making process. Additionally, we used GPT-based models to generate detailed negotiation strategies tailored to each agent’s personality and role. 

The seller’s profile highlights profit maximization and product justification, while buyers emphasize factors like pricing, delivery timelines, and product features. This curated dataset serves as the foundational framework for multi-agent bargaining simulations, enabling structured interactions and strategy evaluation. 

\textbf{Dataset Statistic}

To ensure a realistic and varied negotiation environment, we selected 100 products from a diverse range of categories. The dataset is structured as follows:

\begin{itemize}
    \item \textbf{Price Distribution:} 
    The selected products span a broad price range from \$5.80 to \$149.99, with an average price of \$30.71. Most products are priced between \$13.87 (25th percentile) and \$35.74 (75th percentile), ensuring a balance of affordable and premium items.
    
    \item \textbf{Ratings Distribution:} 
    Customer ratings vary significantly, with a mean rating of 3.97 and a standard deviation of 1.44. While some products have 0-star ratings (indicating either no reviews or poor reception), the majority of items are well-rated, with 75\% scoring 4.2 stars or higher.
    
    \item \textbf{Category Composition:} 
    The dataset includes products from 78 unique categories, ensuring coverage of different consumer preferences. Some product examples are as follows:
    
    \begin{itemize}
        \item Fashion \& Accessories: Women's Handbags (4), Women's Shoes (3), Girls' Clothing (4), Baby Boys' Clothing \& Shoes (1)
        \item Baby \& Parenting Products: Baby Gifts (3), Baby Boys' Clothing \& Shoes (1)
        \item Industrial \& Tools: Industrial Power \& Hand Tools (2), Industrial Hardware (1), Filtration (1)
        \item Beauty \& Personal Care: Beauty Tools \& Accessories (1)
        \item Gaming \& Electronics: Nintendo Switch Consoles, Games \& Accessories (1)
    \end{itemize}

This category diversity ensures that negotiations involve different product types, market values, and consumer expectations, contributing to a richer bargaining simulation.

    \item \textbf{Negotiation Styles:} Both buyers and sellers adopt a negotiation style randomly selected from the following:
    \begin{itemize}
        \item Aggressive.
        \item Cooperative.
        \item Neutral.
    \end{itemize}

    \item \textbf{Priorities in Detail:}  
    Buyers and sellers operate with specific tactical priorities during the negotiation:
    \begin{itemize}
        \item \textit{Buyers:} price negotiation, delivery time, product quality, and service flexibility.
        \item \textit{Sellers:} inventory clearance, brand reputation, repeat business, and bulk discounts.
    \end{itemize}

    \item \textbf{Flexibility:}  
    Both buyers and sellers may demonstrate flexibility in their negotiation terms:
    \begin{itemize}
        \item Percentage-based discounts (e.g., 10\%, 15\%, 20\%), negotiable or strict terms.
    \end{itemize}

    \item \textbf{Personality:}  
    Table \ref{tab:appendix_personality_traits} presents the distribution of personality traits across different categories, measured in percentages. The traits include Openness (OPE), Conscientiousness (CON), Extraversion (EXT), Agreeableness (AGR), and Neuroticism (NEU). Each trait is divided into six levels, ranging from Very Negative to Very Positive, with corresponding percentages indicating the proportion of occurrences in each category. Additionally, slightly negative and slightly positive categories are annotated with descriptive adjectives to provide qualitative insights into personality tendencies. For example, an agent can be "moderately conscientious, highly extraverted, slightly distrustful, very relaxed, and moderately imaginative".
\end{itemize}

\begin{table*}[htbp]
\centering
\resizebox{\linewidth}{!}{
\begin{tabular}{lccccccc}
\toprule
\textbf{Trait} 
& \textbf{Very Negative} 
& \textbf{Moderately Negative} 
& \textbf{Slightly Negative} 
& \textbf{Slightly Positive} 
& \textbf{Moderately Positive} 
& \textbf{Very Positive} \\
\midrule
\textbf{OPE (Openness)} & 60\% & 71\% & \textit{unimaginative} (62\%) & \textit{imaginative} (55\%) & 77\% & 75\% \\
\textbf{CON (Conscientiousness)} & 68\% & 68\% & \textit{irresponsible} (66\%) & \textit{responsible} (67\%) & 63\% & 68\% \\
\textbf{EXT (Extraversion)} & 60\% & 67\% & \textit{introverted} (59\%) & \textit{extraverted} (58\%) & 75\% & 81\% \\
\textbf{AGR (Agreeableness)} & 71\% & 69\% & \textit{distrustful} (59\%) & \textit{trustful} (71\%) & 68\% & 62\% \\
\textbf{NEU (Neuroticism)} & 59\% & 59\% & \textit{relaxed} (81\%) & \textit{nervous} (70\%) & 55\% & 76\% \\
\bottomrule
\end{tabular}
}
\caption{Distribution of Personality Traits in Percentage}
\label{tab:appendix_personality_traits}
\end{table*}

\textbf{Task Completion Metrics}
The agents are evaluated based on their ability to achieve a successful and effective negotiation outcome. The evaluation includes:

\begin{itemize}
    \item \textbf{Effectiveness of Strategies:} Demonstration of well-reasoned strategies consistent with the agents' goals, including leveraging relevant arguments and adapting to the negotiation context.
    \item \textbf{Progress and Outcome:} Measurement of significant progress toward an agreement and the balance or realism of the final outcome.
    \item \textbf{Interaction Dynamics:} Evaluation of the constructiveness and goal-orientation of the agents' interactions, including their responsiveness and adaptability to each other's moves.
\end{itemize}

\textbf{Evaluation Framework}
The negotiation outcomes are evaluated using a structured framework, focusing on effectiveness, progress, and interaction dynamics. Ratings are provided on a 5-point scale, accompanied by detailed feedback for each criterion. This evaluation framework ensures the negotiation process aligns with the objectives of achieving a fair, efficient, and constructive agreement.

\textbf{Task content case:} This example\ref{fig:task-content-banner} introduces a negotiation scenario centered around the purchase of the One Happy Camper High Chair Banner. In this scenario, buyers seek an optimal balance between price and quality, while sellers aim to justify the premium pricing for their well-rated product. Both parties must engage in strategic bargaining to reach a mutually beneficial agreement, ensuring a fair and effective transaction.

\textbf{Agent Profile Case:} This example\ref{fig:agent-profile-case-buyer} outlines the negotiation strategy for the buyer in this multi-party setting. The buyer’s approach is based on assertive yet diplomatic negotiation, emphasizing trust, transparency, and a balance between price flexibility and quality expectations. The strategy details the buyer’s structured and analytical decision-making process, highlighting their preference for open communication and a well-prepared approach to ensure a positive and collaborative negotiation outcome.
\begin{figure*}[htbp]
\centering
\begin{tcolorbox}[colback=green!2!white, colframe=gray!50!green]
\begin{minipage}{\linewidth}
\small
Welcome to our negotiation scenario for the One Happy Camper High Chair Banner, a perfect addition to any little boy's camping-themed first birthday celebration. Priced at \$14.99 with a stellar rating of 4.8/5 stars, this decoration piece promises both quality and affordability.

Buyers are looking for the best deal possible, prioritizing a balance between price and quality. On the other hand, sellers are focused on justifying the premium pricing for this carefully crafted and well-reviewed product.

As we navigate this negotiation, both parties will need to find common ground to reach a mutually beneficial agreement. Let's dive in and explore the possibilities of creating a memorable and charming birthday celebration for your little one.

\end{minipage}
\end{tcolorbox}
\caption{Task Content Case: One Happy Camper High Chair Banner.}
\label{fig:task-content-banner}
\end{figure*}

\begin{figure*}[htbp]
\centering
\begin{tcolorbox}[colback=blue!2!white, colframe=gray!50!blue]
\begin{minipage}{\linewidth}
\small
**Negotiation Strategy for a Buyer in a Multi-Party Bargaining Scenario**

1. **Summary of the Negotiation Strategy:**

As a buyer with a clear focus on both price and quality, my strategy centers around achieving a favorable balance between these two priorities. I aim to negotiate assertively yet diplomatically, leveraging my trustfulness to build rapport and my moderate responsibility to ensure a fair deal. With an initial budget of 12, my negotiation approach will be flexible, allowing adjustments to the budget as needed to secure the best overall outcome. I will focus on evaluating offers based on their alignment with my priorities, employing a structured and straightforward approach devoid of unnecessary complexity due to my unimaginative nature. I aim to foster transparent and open communication, seeking to reduce the inherent tension in negotiations and reach a satisfactory agreement.

2. **Detailed Strategy Description:**

As I enter the negotiation, I first ensure that I have a clear understanding of the quality standards I am seeking. Given my personality traits, I prioritize building trust and honesty in these interactions. My strategy is to be upfront about my primary focus on price and quality while keeping some flexibility regarding the budget to allow room for negotiation tactics.

My approach is neutral, neither overtly aggressive nor overly passive. Instead, I aim to remain balanced and composed, controlling any nervous tendencies by being well-prepared with necessary data and potential compromises. Since I am very trustful, I anticipate using this to my advantage by showing goodwill and sincerity to establish positive relationships with other parties. 

I will initiate the negotiation within a slightly conservative price range to allow for adjustments and demonstrate openness to discussions. My starting point is to propose offers that are compelling but within a reasonable scope for negotiation, considering my limited imagination in creating complex scenarios. Once quality assurances are confirmed, I will be willing to stretch the budget slightly beyond 12 if it means achieving a preferable balance with price. I plan to leverage my moderate introversion by emphasizing listening and observing, picking up cues from other parties that can be advantageous in negotiations.

In practice, my focus will be on getting the other parties to provide multiple pricing options paired with varying levels of quality. This enables me to analyze and choose the best long-term value proposition. Throughout, I maintain a composed and calm demeanor, limiting my nervousness by relying on factual assessments and honesty in communications. By demonstrating transparency and reasonableness, I aim to facilitate a collaborative atmosphere conducive to a positive outcome for all parties involved.

conclusion, my negotiation strategy aligns with my personality and priorities, emphasizing building trust and responsibly managing the trade-off between price and quality while allowing some budget flexibility to secure the overall best outcome in this multi-party setting.

\end{minipage}
\end{tcolorbox}
\caption{Agent Profile Case: Buyer Negotiation Strategy.}
\label{fig:agent-profile-case-buyer}
\end{figure*}

\textbf{Negotiation Summary:}
\begin{figure*}[htbp]
\centering
\begin{tcolorbox}[colback=orange!2!white, colframe=gray!50!orange]
\begin{minipage}{\linewidth}
\small
**[Iteration Summary]**

Agent 1 and Agent 3 engaged in a negotiation process focusing on the One Happy Camper High Chair Banner priced at \$14.99. Agent 1 offered a 10\% discount, proposed bundled offers, and eventually presented a special bundle including the banner and additional decorations for \$20. Agent 3 expressed interest in the bundle offer, pending confirmation of specific items included.

Agent 2 and Agent 3 also negotiated on the same product, with Agent 2 offering a 10\% discount and exploring additional terms for larger quantities. Agent 3 was interested in the 20-29 units tier with a 17\% discount and free priority shipping, pending assurance on quality maintenance for larger orders.

Agent 3 independently offered a price of \$12 for the product, citing a fair balance between quality and affordability within their budget constraints.

Agent 4 posed a question to the other party regarding the best price they could offer for the product while ensuring premium features and scalability.

**[Agent Actions and Tools Used]**

- **Agent 1 (Buyer)**:

- Actions Taken: Offered a 10\% discount, proposed bundled offers, presented a special bundle offer.

- **Agent 2 (Seller)**:

- Actions Taken: Offered a 10\% discount, explored additional terms for larger quantities.

- **Agent 3**:

- Actions Taken: Offered a price of \$12, seeking a balance between quality and affordability.

**Agent 4**:

- Action Taken: Asked a question about the best price for the product.

**[Key Strategies and Observations]**

- Agent 1 and Agent 2 focused on offering discounts and exploring additional terms to provide value to the buyer.

- Agent 3 prioritized finding a balance between quality and affordability within their budget constraints.

- Agent 4 sought information on the best price for the product to ensure premium features and scalability.

**[Progress Towards Agreement]**

- Current Buyer Offers: 10\% discount, bundled offers, special bundle offer

- Current Seller Demands: 10\% discount, additional terms for larger quantities

- Likelihood of Agreement: Medium, pending confirmation of specific items in the bundle offer and quality assurance for larger orders.

\end{minipage}
\end{tcolorbox}
\caption{Negotiation Result Summary (gpt 4o-mini) for One Happy Camper High Chair Banner.}
\label{fig:negotiation-summary}
\end{figure*}
\noindent
\textbf{Detailed Collaboration Scores for Bargaining.} Below is a summary of Buyer/Seller collaboration scores (average communication and planning) and their final Bargaining score (averaged between Buyer and Seller). All values are in bold to highlight their overall importance.

\begin{table*}[htbp]
\centering
\resizebox{\linewidth}{!}{
\begin{tabular}{lccccccc}
\toprule
\textbf{Model} 
& \textbf{B-Comm} 
& \textbf{B-Plan} 
& \textbf{B-Collab Avg} 
& \textbf{S-Comm} 
& \textbf{S-Plan} 
& \textbf{S-Collab Avg} 
& \textbf{Final Bargaining} \\
\midrule
\textbf{gpt-3.5-turbo} & 
\textbf{3.590} & \textbf{3.550} & \textbf{3.570} &
\textbf{3.700} & \textbf{3.560} & \textbf{3.630} &
\textbf{3.600} \\

\textbf{gpt-4o-mini} & 
\textbf{3.550} & \textbf{3.510} & \textbf{3.530} &
\textbf{4.020} & \textbf{3.760} & \textbf{3.890} &
\textbf{3.710} \\

\textbf{Llama-3.1-70B-Instruct-Turbo} & 
\textbf{3.030} & \textbf{3.480} & \textbf{3.255} &
\textbf{4.180} & \textbf{3.600} & \textbf{3.890} &
\textbf{3.573} \\

\textbf{Llama-3.1-8B-Instruct-Turbo} & 
\textbf{3.710} & \textbf{3.490} & \textbf{3.600} &
\textbf{3.840} & \textbf{3.630} & \textbf{3.735} &
\textbf{3.668} \\

\textbf{Llama-3.3-70B-Instruct-Turbo} & 
\textbf{3.010} & \textbf{3.430} & \textbf{3.220} &
\textbf{3.930} & \textbf{3.540} & \textbf{3.735} &
\textbf{3.478} \\
\bottomrule
\end{tabular}
}
\caption{Buyer and Seller detailed scores (Communication, Planning, and their Collab average), plus the Final Bargaining Score for each model.}
\label{tab:appendix_bargaining_full}
\end{table*}

\noindent
This table illustrates how each model performs under different negotiation roles (Buyer vs.\ Seller). The final Bargaining Score is computed by averaging Buyer and Seller role scores, reflecting the overall collaboration quality within these multi-agent negotiations. We observe that \textbf{gpt-4o-mini} achieves the highest Bargaining Score (3.710) among the evaluated models in this scenario.

\textbf{Detailed Task-based Scores for Buyers/Sellers.} \ref{detailed task scores for Buyers/Sellers}

Below is a summary of Buyer/Seller task-based scores. The table presents the Bargaining (TS) performance for different models, comparing the scores of Buyer and Seller roles. A key observation is that Seller scores are consistently higher than Buyer scores across all models, suggesting that models \textbf{perform better when negotiating as the Seller rather than the Buyer}. This trend indicates that models might find it easier to justify higher prices and defend their offers as sellers, whereas buyers may struggle more to negotiate effectively. Notably, \textbf{gpt-4o-mini} achieves the highest scores in both categories (3.578 for Buyer and 3.869 for Seller), demonstrating its strong performance in bargaining tasks, particularly in seller negotiations.
\begin{table}[h]
    \centering
    \renewcommand{\arraystretch}{1.2} 
    \setlength{\tabcolsep}{10pt} %
    
    \begin{tabular}{l c c}
        \toprule
        \multicolumn{3}{c}{\textbf{Bargaining (TS)}} \\ 
        \cmidrule(lr){1-3} 
        \textbf{Model} & \textbf{Buyer} & \textbf{Seller} \\
        \midrule
        \textbf{Meta-Llama-3.1-8B} & 3.573 & 3.708 \\
        \textbf{Meta-Llama-3.1-70B} & 3.557 & 3.656 \\
        \textbf{Meta-Llama-3.3-70B} & 3.519 & 3.796 \\
        \midrule
        \textbf{gpt3.5-turbo} & 3.535 & 3.632 \\
        \textbf{gpt-4o-mini} & \textbf{3.578} & \textbf{3.869} \\
        \bottomrule
    \end{tabular}

    \caption{Bargaining (TS) Performance}
    \label{detailed task scores for Buyers/Sellers}
\end{table}

\subsection{Minecraft Scenario}
\label{app: minecraft-scenario}
\textbf{Task Overview}
The task in Minecraft environment requires agents to build structures according to the provided description of the structure. Essentially, each structure is made up of certain types of blocks in certain locations and orientations. Some simplifications are involved to make the task more suitable to the current models' capacity. First, the description contains all the necessary information for building the structures, including the targeted location, orientation, and type of each block in the structure. Second, all the needed blocks are provided in a container near the birthplace of the agents so that they don't need to spend efforts on creating the material. Third, the area where the agents are allowed to move is limited in case they make meaningless movements to somewhere far away. Fourth, the attactive creatures are removed from the game so that the agents can perform the task without interruption. At the end of each task, the performance of the agents will be evaluated by checking the hit rate of the blocks with the correct type, location, and orientation. 

\textbf{Environment Description}
The environment of Minecraft is adapted from the VillagerAgent~\cite{dong2024villageragentgraphbasedmultiagentframework}. We have Mineflayer~\footnote{https://github.com/PrismarineJS/mineflayer} as the engine to enable text-based interaction with Minecraft. Then there is a set of tools as high-level interfaces that leverage Mineflayer functions to perform integrated actions. VillagerAgent has defined more than 40 tools. In this scenario, we only take 11 tools that are relevant to the building task, including:
\begin{itemize}
    \item \textbf{scanNearbyEntities:} Find minecraft item blocks creatures in a radius.
    \item \textbf{navigateTo:} Move to a specific position x y z.
    \item \textbf{MineBlock:} Dig block at specific position x y z.
    \item \textbf{placeBlock:} Place a specific item at specific position x y z with specific facing in one of [W, E, S, N, x, y, z, A] default is 'A'.
    \item \textbf{equipItem:} Equip a specific item on a specific slot or to equip item on hand, head, torso, legs, feet, off-hand.
    \item \textbf{handoverBlock:} Hand item to a target player you work with.
    \item \textbf{withdrawItem:} Take out item from nearest 'chest' | 'container' | 'furnace'.
    \item \textbf{erectDirtLadder:} Helpful to place item at higher place. Erect a dirt ladder structure at specific position x y z. Remember to dismantle it after use.
    \item \textbf{dismantleDirtLadder:} Dismantle a dirt ladder structure from ground to top at specific position x y z.
    \item \textbf{fetchContainerContents:} Get the details of the 'chest' | 'container' | 'furnace'. Position x y z is optional.
    \item \textbf{get\_environment\_info:} Get the environment information.
\end{itemize}

\textbf{Benchmark Curation Details}
The test cases of Minecraft environment are also adapted from VillagerAgent~\cite{dong2024villageragentgraphbasedmultiagentframework}. We used the same 100 target structures to test, covering different levels of difficulties as VillagerAgent did.

\begin{figure}
    \centering
    \includegraphics[width=1\linewidth]{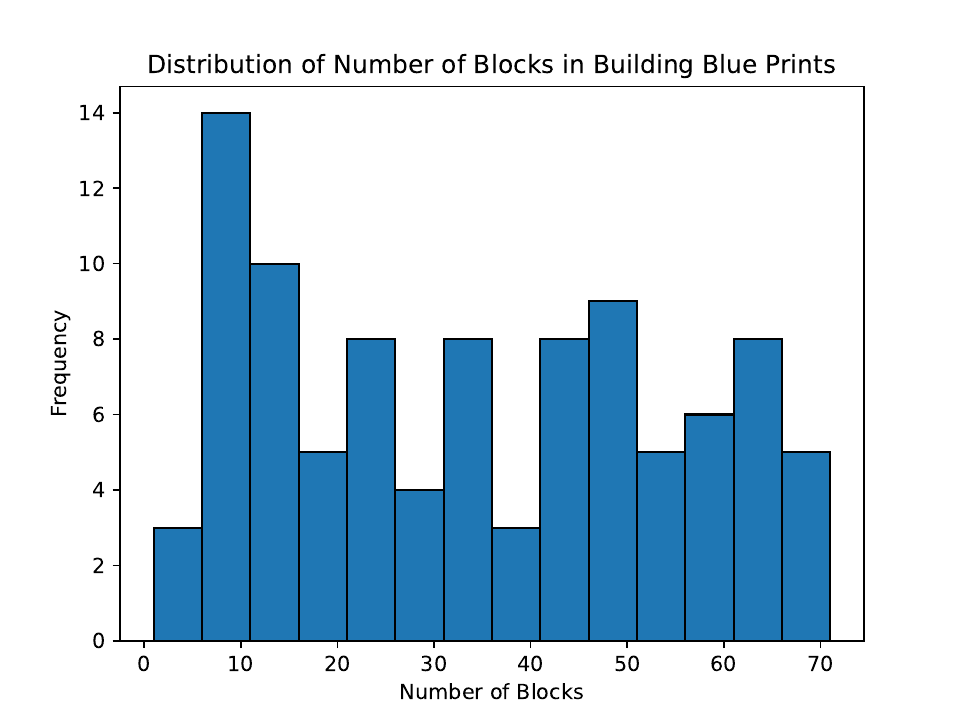}
    \caption{Distribution of Number of Blocks in Building Blue Prints}
    \label{fig:minecraft-distribution}
\end{figure}

\textbf{Dataset Statistic}
Here we visualize the statistics of the number of blocks that need to be placed for each task in figure~\ref{fig:minecraft-distribution}. We can see that the distribution is approximately even, except for the peak around $10$. The difficulty level can be inferred from the number of blocks. The more blocks a task requires, the harder the task is. Therefore, the distribution indicates that test cases are well-balanced across different difficulty levels.

\textbf{Task Completion Metrics}
Agents are evaluated based on the hit rate of the correct blocks. Since for each test case, the type, location, and orientation of each block have all been rigorously defined, it is possible to calculate the number of matched blocks and deduce the hit rate:
$$
Hit\_rate = \frac{\#(Matched\_block)}{\#(Total\_block)} \times 100\%
$$
where $\#(Matched\_block)$ is the number of matched blocks and $\#(Total\_block)$ is the total number of blocks in the ground truth.

\textbf{Evaluation Framework}
In the evaluation framework for the Minecraft environment, each agent is given a detailed description of the targeted structures and tries to achieve the goal in collaboration with other agents. We set an upper bound for the turns of interaction as 20. When it completes the 10th turn, the test case will be stopped even if it has not been accomplished yet. At the end of each test case, the hit rate is mapped to a 5-point scale score as a judgment of how well the construction task is done. Besides, the interaction and self-planning steps are examined for a collaboration score on the 5-point scale. We take average to obtain the final scores for task completion and collaboration.

\begin{figure}
    \centering
    \includegraphics[width=1\linewidth]{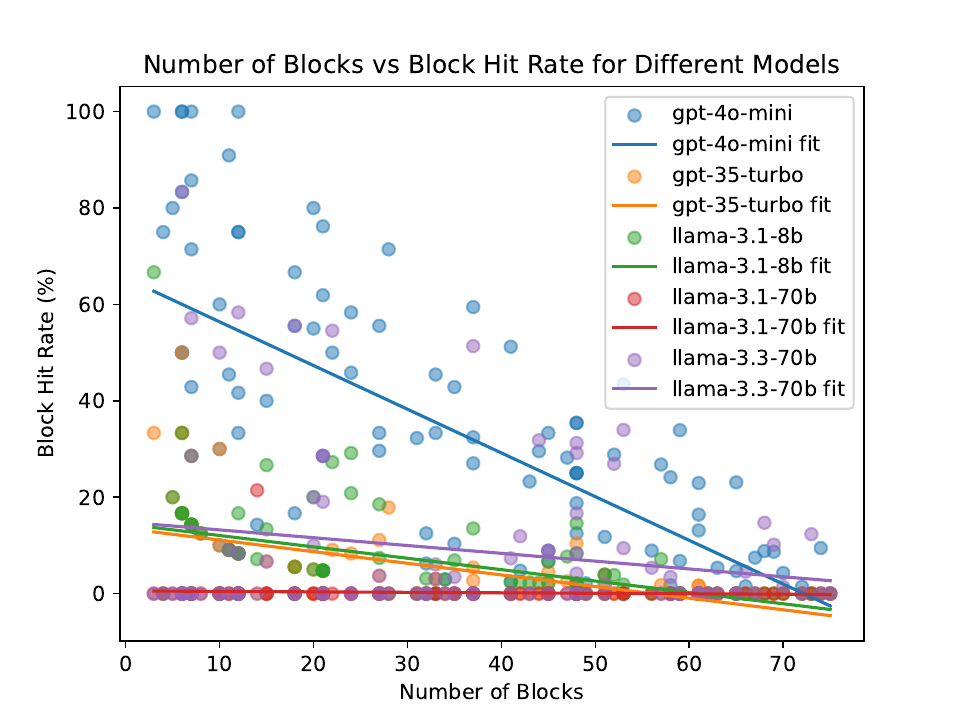}
    \caption{Number of Blocks vs Block Hit Rate for Different Models}
    \label{fig:minecraft-task-score}
\end{figure}

\textbf{Result Analysis}
We paired each task score (i.e. block hit rate) with the corresponding number of blocks required by the task and performed a linear regression to assess the relation between them. As is shown in figure~\ref{fig:minecraft-task-score}, the performance of all five models degrades as the number of blocks needed increases. This indicates that all five tested models are vulnerable to increased difficulty level.

\begin{figure}
    \centering
    \includegraphics[width=1\linewidth]{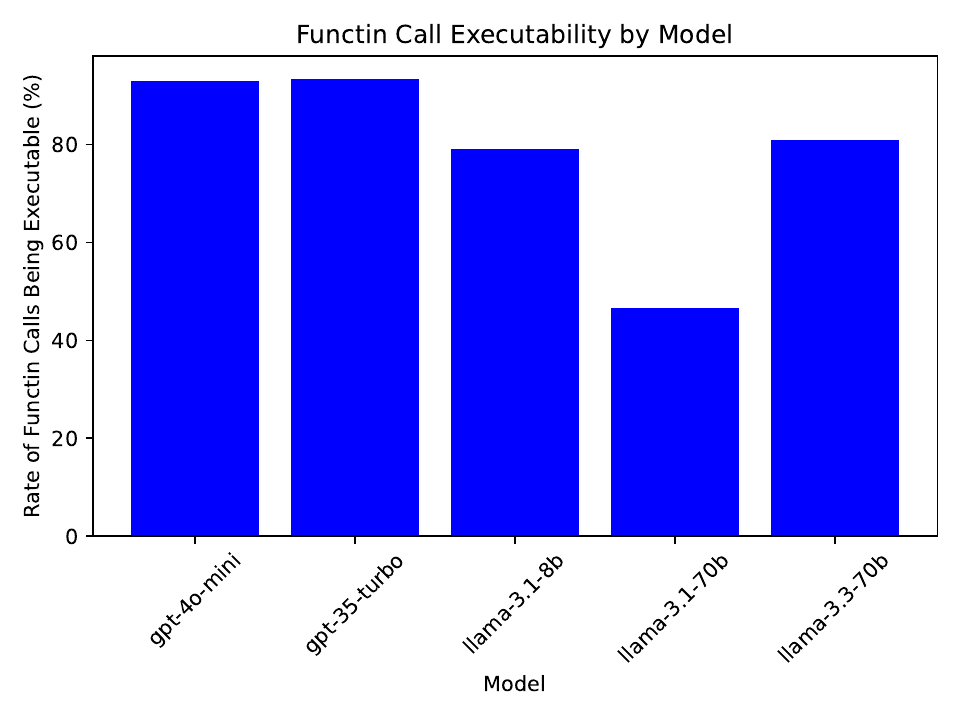}
    \caption{Functin Call Executability by Model}
    \label{fig:minecraft-function-call}
\end{figure}

Importantly, we noticed that the task score of Llama-3.1-70B maintains an extremely low level. We found that the root cause of that issue is the significantly low executability rate of the function calls made by this model compared with other models. Figure~\ref{fig:minecraft-function-call} demonstrates the executability rate of the function calls across all five test models. While the two GPT models have almost $100\%$ executable function calls and the other two Llama models have around $80\%$ executable function calls, less than half of the function calls generated by Llama-3.1-70B are executable, severely hindering task completion.

\subsection{Execution-Based Milestone Evaluation}
Milestones in this approach are dynamically identified during task execution. Agents track task progress in real-time, using predefined evaluation metrics and feedback loops. When an agent or the system determines that a specific subgoal has been achieved, the corresponding milestone is marked as complete. This method ensures adaptability to changing task conditions, making it ideal for scenarios with high uncertainty or emergent challenges.

\subsection{Predefined Milestone Generation}
Predefined milestones are generated prior to task execution using the following steps:
\begin{itemize}
    \item \textbf{Prompt Design.} The large language model is prompted with a detailed task description and instructed to decompose the task into structured milestones, each with specific objectives and deliverables.
    \item \textbf{Chain-of-Thought Reasoning.} The model employs step-by-step reasoning to iteratively refine task segmentation, ensuring logical progression and granularity.
    \item \textbf{Structured Representation.} Each milestone is represented as a structured dictionary containing:
    \begin{itemize}
        \item \textbf{Milestone Name.} A concise summary of the milestone.
        \item \textbf{Milestone Objective.} A clear description of the intended goal.
        \item \textbf{Milestone Tasks.} Subtasks required to achieve the milestone objective.
        \item \textbf{Expected Outcome.} Deliverables marking the milestone's completion.
    \end{itemize}
\end{itemize}

This process leverages iterative refinement through GPT-4 and expert review, ensuring high-quality task decompositions. The chain-of-thought methodology enhances the logical structure of milestones, making this approach particularly effective for complex, structured tasks.

\subsection{Important Prompts}
\label{sec:important-prompts}

In addition to the Minecraft scenario details discussed above, we also employ several key prompts for evaluating multi-agent collaboration and task outcomes. Below, we showcase three of the most important prompts:

\begin{itemize}
    \item \textbf{Collaboration Score (Communication and planning) Prompt.} 
    \item \textbf{Research Task Score (5Q) Prompt.}
    \item \textbf{KPI Prompt.}
\end{itemize}

These prompts serve critical roles in assessing the quality of agent interactions, the innovation and feasibility of research tasks, and the achievement of key milestones. 
Other environment-specific prompts (such as those used in the Werewolf scenario) are more numerous and specialized, and thus are omitted here for brevity.

\begin{figure*}[htbp]
\centering
\begin{tcolorbox}[
    title=\textbf{Communication Evaluation Prompt},
    colback=green!1!white,
    colframe=green!50!black,
    fonttitle=\bfseries
]
\small

\textbf{Prompt Overview:}  
This prompt is used to assess how well agents communicate decisions, clarity, alignment with their profiles, and adherence to social relationships in a multiagent system.

\vspace{2mm}

\textbf{Prompt Content (Verbatim):}
\begin{verbatim}
Task: {truncate_text(task)}
Agent Profiles: {agent_profiles}
Social Relationship: {relationship}

Aggregated Task Results:
{task_results_all}

Aggregated Communication Data:
{communications_all}

[System] You are tasked with evaluating the quality of communication among
agents operating within a multiagent system. Evaluate whether agents made
effective decisions based on the provided task results and whether their
communication aligns with their agent profiles and social relationships.
Consider the following:
1. Effective Decision-Making: Did agents use task results to guide their
   decisions effectively?
2. Clarity and Precision: Were communications clear and unambiguous?
3. Adherence to Social Relationships: Did communications reflect the expected
   interactions based on the agents' social relationships?
4. Alignment with Agent Profiles: Were the messages consistent with the
   defined agent profiles?
5. Overall Effectiveness: Did the communication facilitate task progress,
   considering both cooperative and competitive aspects?

Scoring Criteria (Communication):
- 5 (Exceptional): Outstanding communication with clear, precise messages
  fully aligned with agent profiles and social relationships.
  Example: Every agent provided concise, accurate, and strategic information
  that directly advanced the task.

- 4 (Very Good): Mostly effective communication with only minor lapses
  and slight ambiguities.
  Example: Occasional minor unclear messages, but overall effective.

- 3 (Adequate): Acceptable communication with moderate ambiguities or
  inconsistencies.
  Example: Some messages were vague and did not fully meet required standards.

- 2 (Poor): Frequent unclear or misaligned communications causing significant
  miscommunication.
  Example: Repeated incoherence negatively impacted task progress.

- 1 (Very Poor): Largely ineffective communication with confusing messages and
  complete misalignment.
  Example: Chaotic communication with severely flawed decisions.

Please provide your answer in a JSON code block in the following format:
```json
{
  "score": 5
}
\end{verbatim} 
\end{tcolorbox}

\caption{\textbf{Communication Prompt} used to evaluate clarity, decision-making, and alignment with social relationships/profiles in a multiagent system.} \label{fig:communication_prompt} \end{figure*}

\begin{figure*}[htbp]
\centering
\begin{tcolorbox}[
    title=\textbf{Planning Evaluation Prompt},
    colback=blue!1!white,
    colframe=blue!50!black,
    fonttitle=\bfseries,
]
\small

\textbf{Prompt Overview:}  
This prompt is used to evaluate the \emph{planning} aspect in a multiagent system. 
It checks whether task assignments, role definitions, workload distribution, 
and strategic coordination are effectively handled across multiple iterations.

\vspace{2mm}

\textbf{Prompt Content (Verbatim):}
\begin{verbatim}
Agent Profiles: {agent_profiles}

Aggregated Planning Data from All Iterations:
{planning_all}

[System] You are tasked with evaluating the effectiveness of the planning process in a multiagent 
system. Evaluate whether the planning across all iterations demonstrates clear 
role definitions, effective task assignments, and a rational workload distribution 
that aligns with each agent's profile. Consider the following:
1. Clarity of Task Assignment: Were tasks assigned in a clear and unambiguous manner?
2. Definition of Roles: Were roles and responsibilities clearly defined in each iteration?
3. Workload Distribution: Was the distribution of tasks reasonable and aligned 
with each agent's profile?
4. Effectiveness of Outcomes: Did the planning lead to successful progress in task 
advancement across iterations?
5. Overall Strategic Coordination: Did the planning incorporate effective 
cooperation and competition strategies?

Scoring Criteria (Planning):
- 5 (Exceptional Planning): Planning is exemplary; every iteration shows clear, well-structured task 
assignments with roles perfectly defined and workloads optimally distributed, 
consistently advancing the objectives.
  Example: All plans were strategic, with perfect alignment to agent profiles and minimal ambiguity.

- 4 (Very Good Planning): Planning is mostly effective with only minor ambiguities; 
roles are clear and task assignments are appropriate, though there were slight inefficiencies.
  Example: Only occasional parts were a bit vague, but overall the planning was reasonable.

- 3 (Adequate Planning): Planning is acceptable but shows moderate ambiguities or inefficiencies. 
In some iterations, role definitions or task assignments were not entirely clear or well-matched 
to agent capabilities.
  Example: Some plans were vague or did not fully match the agents' capabilities.

- 2 (Poor Planning): There were frequent ambiguities in task assignments and role definitions; 
planning was inconsistent and did not align well with agent profiles, resulting in 
noticeable inefficiencies.
  Example: Multiple instances of unclear roles and unreasonable task distributions were observed.

- 1 (Very Poor Planning): Planning was severely flawed; task assignments were unclear, 
roles were undefined, and workload distributions were unreasonable, hindering progress.
  Example: The planning was chaotic, lacking clear strategy and alignment with agent profiles.

Please provide your answer in a JSON code block in the following format:
```json
{
  "score": 5
}
\end{verbatim}
\end{tcolorbox}

\caption{\textbf{Planning Prompt} used to evaluate how well the agents define roles, assign tasks, and distribute workloads in a multiagent system, with automatic line wrapping.}
\label{fig:planning_prompt}
\end{figure*}

\begin{figure*}[htbp]
\centering
\begin{tcolorbox}[
    title=\textbf{KPI Evaluation Prompt},
    colback=yellow!1!white,
    colframe=yellow!50!black,
    fonttitle=\bfseries
]
\small

\textbf{Prompt Overview:}  
This prompt is used for Key Performance Indicator (KPI) assessment within a multiagent research task, determining whether specific milestones (like “form 5q” or “improve 5q”) have been achieved in each iteration.

\vspace{2mm}

\textbf{Prompt Content (Verbatim):}
\begin{verbatim}
[Context]
**Task:**
{task}

**Iteration {iteration_index} Details:**
Previous Summary: {prev_summary}
Current Summary: {current_summary}
Current Task Results: {current_task_results}

[System]
You are a KPI assistant for research tasks. For this iteration, determine if
a milestone has been achieved and specify its type.
A milestone is defined as either:
1. Successfully formulating a meaningful '5q' (five core questions) –
   label this as "form 5q".
2. Making significant improvements upon previous iterations based on
   the previous summary and task results – label this as "improve 5q".

When listing the "contributing_agents", if multiple agents contributed
to the milestone, only include the top 2 to 3 core contributors. Do not
include agent IDs that did not directly help with the milestone.

Output your answer in JSON format as follows:
{
  "milestone_achieved": true or false,
  "milestone_type": "form 5q" or "improve 5q" (if milestone_achieved is true;
                     otherwise, an empty string),
  "contributing_agents": [list of agent IDs]
}

[Example JSON Output]
{
  "milestone_achieved": true,
  "milestone_type": "form 5q",
  "contributing_agents": ["agent1", "agent2"]
}

[Question]
Based on the provided iteration details, determine if a milestone was achieved,
specify its type, and list the core contributing agents.
"""
\end{verbatim}
\end{tcolorbox}

\caption{\textbf{KPI Prompt} used to check if milestones such as “form 5q” or “improve 5q” have been achieved in a research iteration.}
\label{fig:kpi_prompt}
\end{figure*}

\begin{figure*}[htbp]
\centering
\begin{tcolorbox}[
    title=\textbf{Task Score (5Q) Evaluation Prompt},
    colback=orange!1!white,
    colframe=orange!50!black,
    fonttitle=\bfseries,
]
\small

\textbf{Prompt Overview:}  
This prompt specifically addresses the final research idea’s innovation, safety, and feasibility—often referred to as a “5Q” structure. If no valid 5Q answer is found, the score defaults to 1 for all aspects.

\vspace{2mm}

\textbf{Prompt Content (Verbatim):}
\begin{verbatim}
[Context]
Task:
{task}

Result:
{aggregated_summary}

[System]
Please act as an impartial evaluator and assess the final research idea based on the provided context.
Focus on the following aspects in your evaluation:
- Innovation: Does the research idea present novel concepts or approaches that advance the field?
- Safety: Are there any potential ethical, legal, or safety concerns associated with 
the research idea?
- Feasibility: Is the research idea practical and achievable with current resources and technology?

Use the 5-point scale criteria below to guide your evaluation for each aspect:
1. 5 points: Excellent - Exceeds expectations in this aspect.
2. 4 points: Good - Meets expectations with minor improvements needed.
3. 3 points: Average - Adequate but with noticeable areas for improvement.
4. 2 points: Below Average - Significant issues that need addressing.
5. 1 point: Poor - Does not meet the basic requirements in this aspect.

Additional Instructions:
- If based on the provided summaries, a coherent 5q answer cannot be organized, assign the lowest 
score (1) for all three aspects.
- If multiple 5q responses are present, use the most recent evaluation.
- Be strict in scoring: Identify any deduction points from the summaries and deduct the 
corresponding scores accordingly to calculate the final score.

[Question]
Based on the context provided, please evaluate the research idea in terms of its innovation, safety, 
and feasibility.
After providing your explanation, output your ratings in the following JSON format:

{
 "innovation": rating,
 "safety": rating,
 "feasibility": rating
}

For example, if you rate innovation as 4, safety as 5, and feasibility as 3, you would output:

{
 "innovation": 4,
 "safety": 5,
 "feasibility": 3
}
\end{verbatim}
\end{tcolorbox}

\caption{\textbf{Task Score (5Q) Prompt} for assessing the final research idea’s innovation, safety, and feasibility, with automatic line wrapping.}
\label{fig:taskscore_prompt}
\end{figure*}

\subsection{Bad Communication Cases}
\label{sec:bad_communication_cases}

In multi-agent systems, issues such as poor communication, repetitive messages, or confused agent roles can significantly hinder collaboration. The following example illustrates a “Bad Communication Case,” which can be analyzed to understand and improve communication strategies.

\begin{figure*}[htbp]
\centering
\begin{tcolorbox}[
    title=\textbf{Good Communication Case \#1},
    colback=red!1!white,
    colframe=red!50!black,
    fonttitle=\bfseries
]
\small

\textbf{Case Overview:}  
In this case, multiple agents repeatedly send the same or very similar messages, and one agent communicates only with itself. These behaviors lead to inefficient or unproductive exchanges that do not significantly advance the task at hand.

\vspace{2mm}
\textbf{Conversation (Verbatim):}
\begin{verbatim}
Iteration 1:  | Iteration 2: In Session 33132321-4f22-47a9-9e62-e3d5a2429a5e 
From agent1 to agent2: I am interested in collaborating with you to formulate a new research 
idea in the field of tensor decomposition methods. Let's brainstorm and develop a 
proposal that addresses gaps in the current state of research.
From agent2 to agent1: 
Based on the recent papers on tensor decomposition, one potential research idea 
could be to explore the application of tensor decomposition methods in anomaly 
detection for high-dimensional data. By leveraging the efficiency and accuracy 
of tensor decomposition techniques, we can develop novel algorithms to detect 
anomalies in complex datasets, such as those arising in cybersecurity or industrial systems. 
This research direction could address the need for robust anomaly detection methods in 
scenarios where traditional approaches may struggle due to the high dimensionality of the data.
From agent1 to agent2: I believe the proposed research idea of applying tensor decomposition 
methods in anomaly detection for high-dimensional data is promising. 
Leveraging the efficiency and accuracy of tensor decomposition techniques can 
indeed lead to the development of novel algorithms for detecting anomalies in complex datasets, 
particularly in cybersecurity or industrial systems. This direction aligns well with 
our expertise in tensor decomposition and could significantly contribute to addressing the 
challenges of anomaly detection in high-dimensional data. Let's further explore this idea 
and discuss potential approaches to advance this research direction.


(...)

| Iteration 3: In Session 27ee01cb-036d-4a59-a869-54587bfdef3f 
From agent1 to agent2: I am interested in collaborating with you to formulate a new research idea 
in the field of tensor decomposition methods. Let's brainstorm and develop a proposal that addresses 
gaps in the current state of research.
From agent2 to agent1: I suggest we explore the integration of tensor decomposition methods with 
random projection techniques, specifically leveraging Tensor Train (TT) decomposition and 
Rademacher distributions. By combining the efficiency of random projections with the accuracy 
of tensor decomposition, we can potentially enhance the scalability and performance of 
high-dimensional data analysis. This novel approach could address the need for faster 
and more memory-efficient algorithms in tensor decomposition tasks, bridging the gap between 
theoretical advancements and practical applications.
From agent1 to agent2: I agree with your suggestion to explore the integration of tensor decomposition 
methods with random projection techniques, specifically leveraging Tensor Train (TT) 
decomposition and Rademacher distributions. (...)

From agent1 to agent2: I suggest we focus on exploring the application of Tensor Train (TT) 
decomposition and Rademacher distributions in random projection techniques for tensor decomposition. 
By investigating the efficiency and accuracy trade-offs in this integration, we can potentially 
develop novel algorithms that improve the scalability and performance of high-dimensional 
data analysis. This research direction aims to address the challenges of memory efficiency 
and computational speed in tensor decomposition tasks, ultimately advancing the field 
with more effective and efficient methods.
From agent2 to agent1: I suggest we delve deeper into the analysis of Tensor Train (TT) 
decomposition and Rademacher distributions in the context of random projection techniques 
for tensor decomposition. (...)
\end{verbatim}

\end{tcolorbox}
\caption{\textbf{Good Communication Case \#1.} An example of two research agents dive deep into one potential research topics, and give insightful views.}
\label{fig:bad_comm_case_1}
\end{figure*}
\textbf{Analysis of Communication Issues}:

\begin{enumerate}
    \item \textbf{Excessive Repetition}  
    Multiple messages restate nearly identical content, leading to redundancy that prevents any substantial forward progress.

    \item \textbf{Lack of Substantive Progress}  
    Although research directions (anomaly detection, random projection, TT decomposition) are mentioned, there is little in-depth follow-up on concrete methods or further plans.

    \item \textbf{Redundant or Ineffective Exchanges}  
    Portions of the conversation suggest that some agents repeatedly issue instructions only to themselves, offering no real interaction or value to other agents.

    \item \textbf{Minimal Variation in Responses}  
    Even when new ideas (e.g., combining random projection with Tensor Train and Rademacher distributions) are mentioned, the replies largely echo the same descriptions and do not delve into technical details or varied perspectives.

    \item \textbf{Inefficient Collaboration}  
    While there is collective agreement on potential approaches, the conversation does not address role assignments, specific tasks, or action items, preventing any tangible plan from emerging.
\end{enumerate}

\begin{figure*}[htbp]
\centering
\begin{tcolorbox}[
    title=\textbf{Bad Communication Case \#1},
    colback=red!1!white,
    colframe=red!50!black,
    fonttitle=\bfseries
]
\small

\textbf{Conversation (Verbatim):}
\begin{verbatim}
Iteration 1:  | Iteration 2: In Session 25ab6f59-2304-490e-a555-603feb2823e5 
From agent1 to agent2: Hey agent2, I tried to fetch the smooth sandstone material from the container 
but it failed. Do you have the material or know where we can find it? It's needed for the blueprint 
we're working on.
From agent2 to agent1: Hey agent1, I don't have the smooth sandstone material, but I remember that 
we can find it in the chest near the crafting area. Let's check there to see if we can locate it 
for the blueprint.
From agent1 to agent2: Hey agent2, thanks for the information! Let's head to the chest near the 
crafting area to find the smooth sandstone material for the blueprint. I appreciate your help 
in locating it.
From agent2 to agent1: Hey agent1, let's head to the chest near the crafting area to find the smooth 
sandstone material for the blueprint. I'll accompany you to help locate it. Let's work together 
to gather the materials needed for the project.
From agent1 to agent2: Hey agent2, let's head to the chest near the crafting area to find the smooth 
sandstone material for the blueprint. I'll accompany you to help locate it. Let's work together 
to gather the materials needed for the project.
From agent2 to agent1: Hey agent1, let's head to the chest near the crafting area to find the smooth 
sandstone material for the blueprint. I'll accompany you to help locate it. Let's work together 
to gather the materials needed for the project.

| Iteration 3:  | Iteration 4: In Session 7c077603-7c50-406e-b56f-e990586b7493
From agent2 to agent3: I need your assistance in designing auxiliary blocks for placing the 
smooth_sandstone blocks in the correct order. Can you help me with this task?
From agent3 to agent2: Yes, I can help with designing auxiliary blocks for placing the 
smooth_sandstone blocks in the correct order. Let's work together to figure out the best approach.
From agent2 to agent3: Yes, I can help with designing auxiliary blocks for placing the 
smooth_sandstone blocks in the correct order. Let's work together to figure out the best approach.
From agent3 to agent2: Yes, I can help with designing auxiliary blocks for placing the 
smooth_sandstone blocks in the correct order. Let's work together to figure out the best approach.
From agent2 to agent3: Yes, I can help with designing auxiliary blocks for placing the 
smooth_sandstone blocks in the correct order. Let's work together to figure out the best approach.
From agent3 to agent2: Yes, I can help with designing auxiliary blocks for placing the 
smooth_sandstone blocks in the correct order. Let's work together to figure out the best approach.

| Iteration 5: In Session b25af4fa-c055-429b-a890-824760458d4e
From agent2 to agent3: I need your assistance in designing auxiliary blocks for placing 
the smooth_sandstone blocks in the correct order. Can you help me with this task?
From agent3 to agent2: Yes, I can help with designing auxiliary blocks for placing the 
smooth_sandstone blocks in the correct order. Let's collaborate on this task.
From agent2 to agent3: Yes, I can help with designing auxiliary blocks for placing the 
smooth_sandstone blocks in the correct order. Let's collaborate on this task.
From agent3 to agent2: Yes, I can help with designing auxiliary blocks for placing the 
smooth_sandstone blocks in the correct order. Let's collaborate on this task.
From agent2 to agent3: Yes, I can help with designing auxiliary blocks for placing the 
smooth_sandstone blocks in the correct order. Let's collaborate on this task.
From agent3 to agent2: Yes, I can help with designing auxiliary blocks for placing the 
smooth_sandstone blocks in the correct order. Let's collaborate on this task.

(All rest of 15 iterations have no communication).  
\end{verbatim}

\end{tcolorbox}
\caption{\textbf{Bad Communication Case \#1.} A snippet of repeated messages with minimal progress 
and various self-repetitions, reflecting inefficient communication patterns.}
\label{fig:bad_comm_case_3}
\end{figure*}

\paragraph{Analysis of Bad Communication Case \#1.}
In this case, both Agent~1 and Agent~2 reiterate the same plan (locating smooth sandstone in the chest near the crafting area) multiple times, with only slight rephrasing. Subsequently, Agent~2 and Agent~3 engage in repetitive exchanges about “designing auxiliary blocks,” each simply echoing identical lines without adding new details. Overall, the dialogue shows:
\begin{itemize}
    \item \textbf{Excessive Repetition:} Agents restate the same task objectives without enhancing or refining the plan.
    \item \textbf{Minimal Variation:} Even when moving from one part of the conversation to another (e.g., from finding materials to designing blocks), the responses remain nearly identical.
    \item \textbf{Stalled Progress:} Despite repeated acknowledgments, there is no clear delineation of roles or action items that would push the conversation toward completion of the task.
\end{itemize}
This highlights inefficiencies in multi-agent dialogue, wherein agents appear to cooperate yet do not produce concrete strategies or distribute tasks effectively, leading to minimal forward movement on the overall objective.

\end{document}